\pdfoutput=1
\documentclass[%
 reprint,
nobibnotes,
 amsmath,amssymb,
 aps,
prb,
]{revtex4-2}

\usepackage{graphicx}
\usepackage{dcolumn}
\usepackage{bm}
\usepackage[backref=none, breaklinks=True]{hyperref} 
\usepackage[usenames,dvipsnames]{xcolor}
\usepackage{color}
\usepackage[framemethod=TikZ]{mdframed}
\newcommand\myshade{85}
\colorlet{mylinkcolor}{violet}
\colorlet{mycitecolor}{YellowOrange}
\colorlet{myurlcolor}{Aquamarine}
\hypersetup{
  linkcolor  = mylinkcolor!\myshade!black,
  citecolor  = mycitecolor!\myshade!black,
  urlcolor   = myurlcolor!\myshade!black,
  colorlinks = true,
}
\usepackage[caption=false]{subfig}
\usepackage{listings}

\definecolor{codegreen}{rgb}{0,0.6,0}
\definecolor{codegray}{rgb}{0.5,0.5,0.5}
\definecolor{codepurple}{rgb}{0.58,0,0.82}
\definecolor{backcolour}{rgb}{0.95,0.95,0.92}

\lstdefinestyle{mystyle}{
    backgroundcolor=\color{backcolour},   
    commentstyle=\color{codegreen},
    keywordstyle=\color{magenta},
    numberstyle=\tiny\color{codegray},
    stringstyle=\color{codepurple},
    basicstyle=\ttfamily\footnotesize,
    breakatwhitespace=false,         
    breaklines=true,                 
    captionpos=b,                    
    keepspaces=true,                 
    numbers=left,                    
    numbersep=5pt,                  
    showspaces=false,                
    showstringspaces=false,
    showtabs=false,                  
    tabsize=2
}

\begin{document}

\preprint{APS/123-QED}

\title{Parallel hybrid quantum-classical machine learning for kernelized time-series classification}

\author{Jack S. Baker$^1$}
\author{Gilchan Park$^2$}
\author{Kwangmin Yu$^2$}
\email{kyu@bnl.gov}
\author{Ara Ghukasyan$^1$}
\author{Oktay Goktas$^1$}%
\author{Santosh Kumar Radha$^1$}%
\email{research@agnostiq.ai}
\affiliation{$^1$Agnostiq Inc., 325 Front St W, Toronto, ON M5V 2Y1}
\affiliation{$^2$Computational Science Initiative, Brookhaven National Laboratory, Upton, New York 11973, USA}

\date{\today}

\begin{abstract}

Supervised time-series classification garners widespread interest because of its applicability throughout a broad application domain including finance, astronomy, biosensors, and many others. In this work, we tackle this problem with hybrid quantum-classical machine learning, deducing pairwise temporal relationships between time-series instances using a time-series Hamiltonian kernel (TSHK). A TSHK is constructed with a sum of inner products generated by quantum states evolved using a parameterized time evolution operator. This sum is then optimally weighted using techniques derived from multiple kernel learning. Because we treat the kernel weighting step as a differentiable convex optimization problem, our method can be regarded as an end-to-end learnable hybrid quantum-classical-convex neural network, or QCC-net, whose output is a data set-generalized kernel function suitable for use in any kernelized machine learning technique such as the support vector machine (SVM). Using our TSHK as input to a SVM, we classify univariate and multivariate time-series using quantum circuit simulators and demonstrate the efficient parallel deployment of the algorithm to 127-qubit superconducting quantum processors using quantum multi-programming.
 
\end{abstract}

\keywords{Quantum multi-programming}

\maketitle

\section{Introduction \label{sec:intro}}

Processes and systems which produce observable characteristics evolving with time are present in topics as disparate as finance, sensor technologies, medicine, astronomy and many others. As a result, new techniques for time-series analysis have become among the most sought after in machine learning (ML) where we seek to learn temporal trends and correlations from time-series data to perform classification \cite{IsmailFawaz2019}, anomaly detection \cite{BlzquezGarca2021, Choi2021}, regression \cite{Clark2020}, forecasting \cite{Deb2017, Torres2021} and to generate synthetic time-series instances \cite{Zhang2018}. Focusing on classification, classical ML has provided a zoo of algorithms where the present state-of-the-art is centered around deep learning. Popular approaches include recurrent neural networks like long-short-term-memory \cite{Hochreiter1997} (and its variations \cite{Melis2019, Nguyen2020}), gated recurrent units and, more recently, transformer networks \cite{Vaswani2017, yang2021, Zerveas2021}. 

While popular for static data (i.e. data without time dependence), kernelized methods like the well-known support vector machine (SVM) \cite{cortes1995support}  see limited use in time-series analysis. Although time-series kernel methods have been proposed \cite{Badiane2018, bailly2018time, fabregues2017forecasting, ruping2001svm} and used as analytical tools for interpreting deep learning models \cite{tino2020dynamical}, none are tailored to encode temporal trends by way of an explicit and learnable time-dependent inner product space. Within classical ML, it is not clear how such a space can be constructed in a non-trivial manner. In this work, we show that time evolution as generated by a parameterized Hamiltonian operator within quantum mechanics is a natural approach for achieving such an objective.  Indeed, by combining inner products evolved to different points of time, we achieve a quantum kernel function adapted for time-series data which we call the Time-Series Hamiltonian Kernel (TSHK). Furthermore, we discuss situations where training a TSHK is useful and provide theoretical tools to analyze the time dependence of the learnt inner product spaces.

The construction of the TSHK is tied to a field known as Multiple Kernel Learning (MKL). In MKL, multiple kernel functions are combined using various tactics \cite{gonen2011multiple, Aiolli2015} to give rise to a combined kernel that is by some measure superior. The TSHK is a combined kernel built using a weighted linear combination of quantum kernel functions each defined at different instances of time $t$. These weights are chosen to maximize the separation between labeled data classes as was proposed in the well-known EasyMKL \cite{Aiolli2015} algorithm. Furthermore, as was originally suggested in \cite{Ghukasayan2023}, the variational parameters of quantum kernels and the kernel weights can be obtained simultaneously by training an end-to-end learnable Quantum-Classical-Convex neural network (QCC-net). The network producing the TSHK can then be regarded as a temporally-aware variation of the original QCC-net. 

Currently in the noisy intermediate-scale quantum (NISQ) era \cite{Bharti2022}, various sources of noise prevent the implementation of deep quantum circuits on real hardware. In light of this, we must ensure that our quantum circuits implementing the TSHK are sufficiently shallow. Conventionally, however, the required time evolution operators are implemented using a potentially high-depth Lie-Suzuki-Trotter expansion \cite{Wiebe2010}. Fortunately, built upon the eigendecomposition approach originally used in the \textit{variational fast forwarding} algorithm \cite{Crstoiu2020}, a family of time evolution operators can be implemented with low depth variational circuits. The space of accessible time evolution operators can controlled by modifying the depth and structure of the circuit Ans{\"a}tz implementing the eigendecomposition. Several works \cite{Radha2021, Horowitz2022, Baker2022, Gibbs2022, Caro2022} have now appeared leveraging this approach to: learn constraint-enforcing mixing operators in the Quantum Approximate Optimization Algorithm \cite{Radha2021}, generate synthetic time-series instances \cite{Horowitz2022} and detect anomalous behaviour in time-series data \cite{Baker2022}. The generalization bounds of this approach have also been studied \cite{Gibbs2022, Caro2022}.

Although the eigendecomposition approach does allow for a shallow circuit implementation of the TSHK, the number of computations  required to compute the kernel matrices scales with the square of the size of the training data set and linearly with the length of the time series. Fortunately, the computations of kernel matrix elements are independent of one another allowing them to be computed in parallel. Indeed, a new approach to efficiently utilize contemporary NISQ computers is by overlapping multiple quantum circuits \cite{das2019case}. This method, called Quantum Multi-Programming (QMP), utilizes NISQ devices by executing multiple quantum circuits concurrently.  In this work, we use QMP to compute the TSHK in parallel. Compared to serial execution, we achieve a notable speed-up (at least $35$ times) with QMP without loss of accuracy using two 127-qubit IBM quantum computers. In light of this large speed-up, this work demonstrates the practical utility of parallelism in hybrid quantum-classical ML workflows.   

The rest of this work is organized as follows: in Sec. \ref{sec:formalism} we describe how we achieve a time-dependent inner product space and show how a weighted sum of inner products can be used to construct the TSHK (Sec. \ref{subsec:time_dep_inner}). Also within this Section, we show how the TSHK can be trained using a QCC-net (Sec. \ref{subsec:training}) and show how the resulting TSHK can be plugged in to a SVM to perform time-series classifications (Sec \ref{subsec:kernel_classifier}). In Sec. \ref{sec:probe_time_dependence}, we describe in detail what is meant by time-dependence in the trained kernel function and show how one can efficiently probe this property. In Sec. \ref{sec:simulator_section}, we demonstrate SVM classification using the TSHK for a synthetic multivariate time-series (Sec. \ref{subsec:syn_demonstration}) and for a real univariate time-series: the well-known gun-point data set \cite{ratanamahatana2005three} (Sec. \ref{subsec:univariate_sim}). Our experiments using QMP on 127 qubit superconducting transmon chips are contained within Sec. \ref{sec:hardware_runs}. The QMP approach is described in detail in Sec. \ref{subsec:qmp_explained} and pertinent QMP design considerations are explained in Sec. \ref{subsec:qmp_design_considerations}. The results of our QMP experiments are shown and discussed in Sec. \ref{subsec:qmp_experiments}. Finally, in Sec. \ref{sec:conclusions}, we conclude and remark on future directions in hybrid quantum-classical ML exploiting parallelism. 

\section{Algorithm description \label{sec:formalism}}
\begin{figure*}
    \centering
    \includegraphics[width=\linewidth]{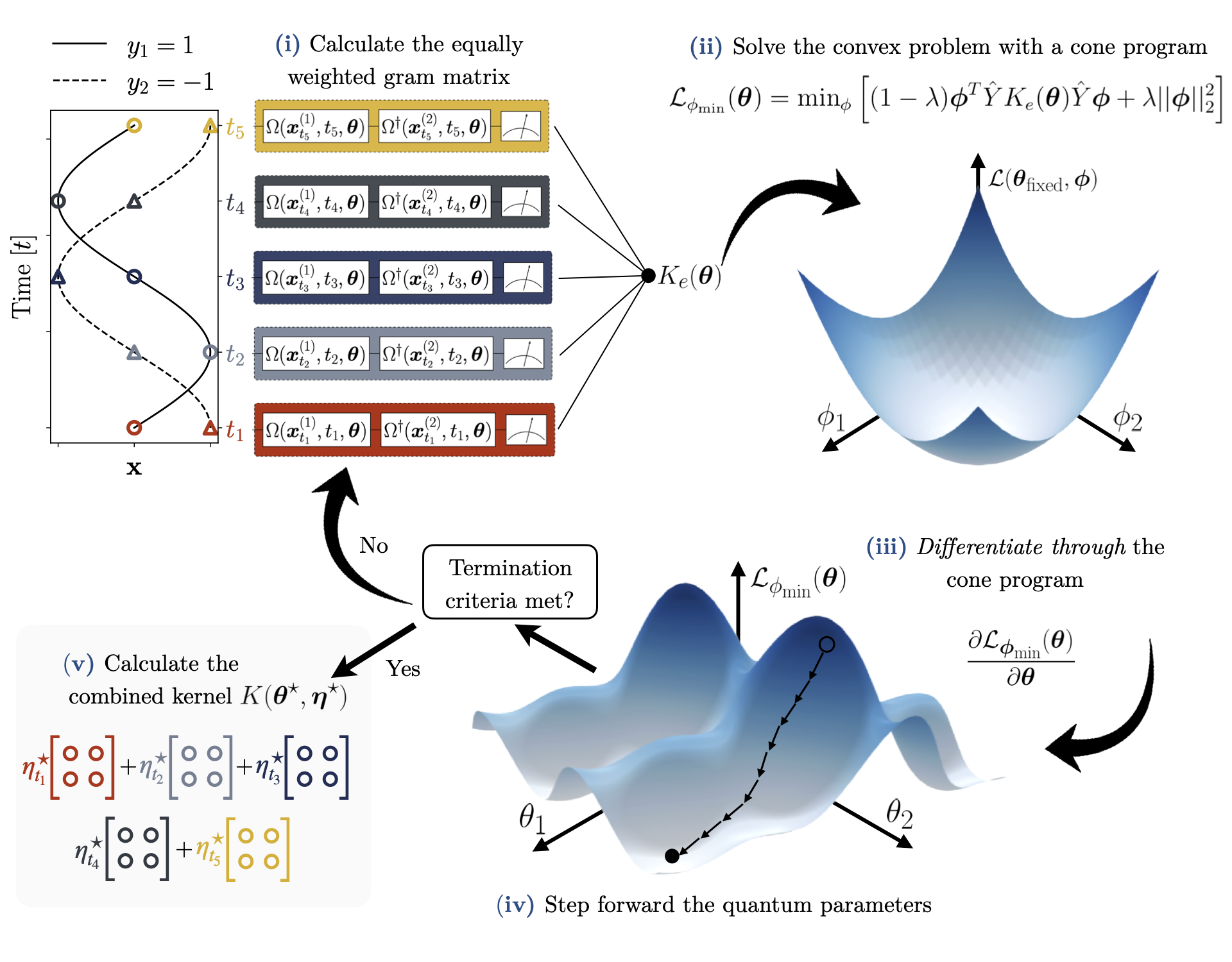}
\caption{Optimizing the combined quantum kernel for a simplified sinusoid versus cosinusoid classification problem with only \textit{one training example} from class $y_i = 1$ [$-\sin(t)$] and class $y_i = -1$ [$-\cos(t)$]. (i) Given some initial $\bm{\theta} = \bm{\theta}_{\text{init}}$, calculate the kernels of Eq. \ref{eq:TSHK} at each time $t_l$ and calculate the equally weighted Gram matrix $K_e(\bm{\theta}) = \sum_{t \in T} K_{t}(\bm{\theta})$. Unitary operator diagrams are color-coded in the Figure and because there is only one unique element (excluding the diagonal which is unity by definition) only one unitary diagram is shown for each kernel computation. (ii) Solve the convex optimization problem with respect to the Lagrangian dual variables $\bm{\phi}$ using a cone program. (iii) Calculate the quantum gradients $\partial \mathcal{L}_{\phi_{\text{min}}}(\bm{\theta})/\partial \bm{\theta}$ by differentiating through the cone program. The surface plot shows a convex optimization landscape spanned by $\phi_1$ and $\phi_2$. (iv) Use the calculated $\partial \mathcal{L}_{\phi_{\text{min}}}(\bm{\theta})/\partial \bm{\theta}$ to step forward a derivative-requiring optimization routine (i.e. gradient descent, \texttt{Adam} etc. ). If the termination criteria (a fixed number of loss iterations, convergence tolerance for $\mathcal{L}_{\phi_{\text{min}}}$ etc.) are met with this new step, advance to (v) else, go back to (i) and enter new loop iteration. Note the negative sign of $\mathcal{L}_{\phi_{\text{min}}}(\bm{\theta})$ in part (iv) of the figure. This is because in practice we maximize the function by minimizing its negative value. (v) Extract the optimal kernel weights $\bm{\eta}^{\star}$ using Eq. \ref{eq:lagrangian_to_weights} to yield the combined kernel function (the TSHK) from Eq. \ref{eq:combined_kernel}. This kernel function can now be used to calculate kernel matrices $K(\bm{\theta}^{\star}, \bm{\eta}^{\star})$ for use in any kernelized learning task including SVM.}
\label{fig:hero_time_kernel}
\end{figure*}

\subsection{Time-dependent inner product space \label{subsec:time_dep_inner}}

We begin with a formal definition of the type of data we are working with: time-series.  A time-series $\bm{x}$ is a sequence of $p \in \mathbb{Z}^+$ observations from a process/system arranged in chronological order
\begin{equation}
\bm{x} := ( \bm{x}_t: t \in T ), \quad T := (t_l : 1 \leq l \leq p)   
\label{eq:time_series_definition}
\end{equation}
where  $\bm{x}_t = [{}^1x_t, {}^2x_t, \ldots {}^dx_t] \in \mathbb{R}^d$, $d \in \mathbb{Z}^+$, $l \in \mathbb{Z}^{+}$ and time $t \in \mathbb{R}^+$.  In what follows, we devise a technique for obtaining a trainable kernel function adapted for  time-series data of the form defined above by (i) formulating a time-dependent quantum inner product space and (ii) using methods from the classical MKL literature \cite{Aiolli2015, Gonen2011} to find an optimally weighted sum of inner products from different points in time. 

The first step towards achieving our goal is to inject time dependence into a quantum state. We achieve this through operating on the $n$-qubit state of zeroes $|0\rangle^{\otimes n}$ using a parameterized unitary time evolution operator $e^{-iH(\bm{\beta}, \bm{\gamma})t^{\prime}}$ where $t^{\prime} \in \mathbb{R}^+$ followed by a parameterized unitary matrix $U(\bm{x}_{t}, \bm{\alpha})$ used to embed $\bm{x}_{t}$ within the quantum state
\begin{equation}
    |\bm{x}_{t}, t^{\prime}, \bm{\theta} \rangle := U(\bm{x}_{t}, \bm{\alpha})e^{-iH(\bm{\beta},\bm{\gamma})t^{\prime}}|0\rangle^{\otimes n}
\label{eq:embedding}
\end{equation}
for real-valued set of free parameter vectors $\bm{\theta} = \{\bm{\alpha}, \bm{\beta} ,\bm{\gamma} \}$ where the embedding unitary matrix $U(\bm{x}_t, \bm{\alpha})$ is permitted to be any quantum feature map \cite{Schuld2019, Havlek2019} implemented using a layered parameterized quantum circuit. It should be stressed that $U(\bm{x}_t, \bm{\alpha})$ is a static map. That is, despite the subscript $t$ bared by the argument $\bm{x}_t$, the resulting unitary matrix is time-independent. Explicitly, should we have $\bm{x}_{t_1} = \bm{x}_{t_2}$ , $t_1 \neq t_2$ the unitary matrices $U(\bm{x}_{t_1}, \bm{\alpha})$ and $U(\bm{x}_{t_2}, \bm{\alpha})$ are indistinguishable. Time-dependence only comes as a result of the time evolution operator $e^{-iH(\bm{\beta}, \bm{\gamma})t^{\prime}}$ which we write as an eigendecomposition
\begin{equation}
V_{t^{\prime}}(\bm{\beta}, \bm{\gamma}) := W^{\dagger}(\bm{\beta})D(\bm{\gamma}, t^{\prime})W(\bm{\beta}) = e^{-iH(\bm{\beta}, \bm{\gamma})t^{\prime}}
\end{equation}
for parameterized unitary matrix of eigenvectors $W(\bm{\beta})$ and parameterized time-encoded diagonal unitary matrix $D(\bm{\gamma}, t^{\prime})$. The above equality with $e^{-iH(\bm{\beta}, \bm{\gamma})t^{\prime}}$ holds following Stone's theorem for strongly continuous one-parameter unitary groups \cite{Stone1932}. The strength of this approach lies in implementing each unitary matrix in the eigendecomposition as a layered parameterized quantum circuit allowing one to control the search space of unitary time evolution operators. Should a small number of layers prove sufficient to optimize the forthcoming loss function (Eq. \ref{eq:cost}), our algorithm becomes suitable for present generation NISQ computers. For notational short-hand, it is convenient to write a combined unitary matrix as a product of the static and time-dependent parts  
\begin{equation}
\Omega(\bm{x}_t, t^{\prime}, \bm{\theta}) := U(\bm{x}_{t}, \bm{\alpha})V_{t^{\prime}}(\bm{\beta}, \bm{\gamma}).
\label{eq:combined_unitary}
\end{equation}
We can now begin to formally derive a time-dependent inner product. To do so, we first must regard the density matrix $\rho_{t^{\prime}}(\bm{x}_t, \bm{\theta}) = |\bm{x}_{t}, t^{\prime}, \bm{\theta} \rangle \langle \bm{x}_{t}, t^{\prime}, \bm{\theta}|$ as the unambiguous feature vector embedding the classical data point $\bm{x}_t$ (regarding $|\bm{x}_{t}, t^{\prime}, \bm{\theta} \rangle$ as the feature vector is ambiguous as the space $\mathbb{C}^{2^n}$ is only physically defined up to a global phase \cite{Havlek2019}) . We now define a kernel by setting $t^{\prime} \rightarrow t$ and taking the Frobenius inner product $\langle \rho_{t}(\bm{x}^{\prime}_t, \bm{\theta}), \rho_{t}(\bm{x}_t, \bm{\theta}) \rangle$. Equivalently, 
\begin{multline}
\kappa_{t}(\bm{x}_{t}, \bm{x}^{\prime}_{t}, \bm{\theta}) := \text{Tr}[\rho_{t}(\bm{x}^{\prime}_t, \bm{\theta}) \rho_{t}(\bm{x}_t, \bm{\theta})] \\ = |\langle \bm{x}^{\prime}_{t}, t, \bm{\theta}|\bm{x}_{t}, t, \bm{\theta} \rangle|^2 
\label{eq:TSHK}
\end{multline}
where  $\bm{x}^{\prime}_{t}$ is an element of another time-series instance $\bm{x}^{\prime}$ defined analogously to Eq. \ref{eq:time_series_definition}. Eq. \ref{eq:TSHK} can be estimated practically by preparing the state $\Omega^{\dagger}(\bm{x}^{\prime}_t, t, \bm{\theta})\Omega(\bm{x}_t, t, \bm{\theta})|0 \rangle^{\otimes n}$ and taking the expectation value of the projector on the state of zeros $P_0 = |0\rangle^{\otimes n} \langle 0|^{\otimes n}$. This requires repeated preparation and measurement of $\Omega^{\dagger}(\bm{x}_t, t, \bm{\theta})\Omega(\bm{x}_t, t, \bm{\theta})|0 \rangle^{\otimes n}$ in the computational basis to estimate the probability of measuring the $n$-bit string of zeros $00\ldots0$ . 

Eq. \ref{eq:TSHK} fulfils the symmetry requirement of a valid kernel function. That is, we have $\kappa_{t}(\bm{x}_{t}, \bm{x}^{\prime}_{t}, \bm{\theta}) = \kappa_{t}(\bm{x}^{\prime}_{t}, \bm{x}_{t}, \bm{\theta})$ for any choice of the two arguments $\bm{x}_{t}$, $\bm{x}^{\prime}_{t} \in \mathbb{R^d}$ or the time argument $t$. Indeed, the achievement of this symmetry motivates the order in which unitary operations are applied in Eq. \ref{eq:embedding}. If we were to switch the order in which we apply of $U(\bm{x}_{t}, \bm{\alpha})$ and $e^{-iH(\bm{\beta},\bm{\gamma})t^{\prime}}$ to $|0\rangle^{\otimes n}$ we yield the state $|\bm{x}_{t}, t^{\prime}, \bm{\theta} \rangle_G := e^{-iH(\bm{\beta},\bm{\gamma})t^{\prime}}U(\bm{x}_{t}, \bm{\alpha})|0\rangle^{\otimes n}$. Computing the inner product of this state at $t^{\prime}=t$ with another state at $t^{\prime}=t^{\prime}$ and multiplying the result by its own complex conjugate, we yield the quantity $\mathbb{G} := |{}_G\langle x_{t}, \bm{\alpha}|e^{-iH(t - t^{\prime})}|x^{\prime}_{t}, \bm{\alpha}\rangle_G|^2$ where we have taken $|x_{t}, \bm{\alpha}\rangle_G := U(\bm{x}_{t}, \bm{\alpha})|0\rangle^{\otimes n}$. We see immediately that for $t \neq t^{\prime}$ this expression is \textit{not} a valid kernel as the symmetry requirement detailed above is violated. It is only at $t = t^{\prime}$ that symmetry is recovered but at the expense of losing time dependence entirely as the time evolution operator becomes the identity. It is worth pointing out that while a time-dependent kernel function is not achievable in this way, the result is otherwise important. That is,  interestingly, $\mathbb{G}$ is equivalent to the product of the Green's function propagator (hence the subscript $G$ in the above) from quantum field theory \cite{bjorken1965relativistic} with its complex conjugate $G_{\bm{\theta}}(\bm{x}_{t}, t; \bm{x}^{\prime}_{t}, t^{\prime})G^{\star}_{\bm{\theta}}(\bm{x}_{t}, t; \bm{x}^{\prime}_{t}, t^{\prime})$. Using parameterized Green's function propagators could prove an interesting avenue for non-kernelized time-series analysis in QML but is beyond the scope of this work.

Returning now to the kernel function defined in Eq. \ref{eq:TSHK}, we compute it $\forall t \in T$ and combine the $p$-terms in a weighted sum to achieve a combined kernel function measuring the similarity between a pair of time-series instances $(\bm{x}, \bm{x}^{\prime})$
\begin{equation}
    \kappa (\bm{x}, \bm{x}^{\prime}, \bm{\theta}, \bm{\eta}) := \sum_{t \in T}  \eta_{t} \cdot \kappa_{t}(\bm{x}_{t}, \bm{x}^{\prime}_{t}, \bm{\theta})
\label{eq:combined_kernel}
\end{equation}
for kernel coefficient vector $\bm{\eta} := [\eta_t| t \in T ] \in \mathbb{R}^p$ subject to the constraint  $\sum_{t \in T} \eta_{t} = 1$.  We call Eq. \ref{eq:combined_kernel} the TSHK. The motivation for such a form of kernel function is to allow the measured similarity between a pair of time-series $(\bm{x}, \bm{x}^{\prime})$ to be influenced by differently weighted contributions at distinct points in time. Should the resulting kernel be used in a classification algorithm like SVM, this allows for ``smoking gun" values of $t$ to discriminate between different classes of time-series. In the next Section, we present a strategy for optimally setting the kernel parameters $\bm{\theta}$ and $\bm{\eta}$ where the exact sense of optimality is to be defined. \par

Before we advance to the next Section, however, now is a good time to discuss what structures of time series data may be learnt well by the kernel function of Eq. \ref{eq:combined_kernel} and for which we may expect performance to be no better than time independent kernels. It is clear from the form of Eq. \ref{eq:combined_kernel} and the time-dependent kernels which construct it (Eq. \ref{eq:TSHK}) that great emphasis is placed on the notion of a ``shared time axis"; a significant inductive bias of our approach. That is, we expect a learnable TSHK when each instance of the training time series data begins at a common initial condition in time. Examples of such data include the temperature in different locations for each day of the year (where we may classify the southern hemisphere versus the northern hemisphere), or traffic (the number of detected vehicles) on roadways given at some frequency on different days (where we may classify weekday versus weekend traffic). In absence of such a shared time axis, we may not expect a learnable TSHK. For example, if we are given ``windowed data" (sub-sequences taken from a larger time series) the initial condition is not guaranteed to be shared among data instances. One example of this is the classification of different genres of music given small audio clips from a longer song. It should be noted that even in this case, it could be possible increase the learnability of the TSHK by preprocessing time series using a time-axis alignment protocol like Dynamic Time Warping \cite{Sakoe1978}. The success of such an approach would have to be assessed on a case-by-case basis.

\subsection{Training the kernel function \label{subsec:training}}

We now present a strategy for training the kernel function of Eq. \ref{eq:combined_kernel} in a supervised learning setting. Throughout this Section, Fig. \ref{fig:hero_time_kernel} should be used as a companion and will be referenced throughout. To begin, we assume to have access to training set $X$ containing $N_X \in \mathbb{Z}^{+}$ time-series instances $\bm{x}^{(i)}$ such that 
\begin{equation}
X := \{\bm{x}^{(i)}: i \in \mathbb{Z}^{+}_{\leq N_X}\}
\label{eq:training_set}
\end{equation}
where each time-series instance $\bm{x}^{(i)}$ has a corresponding binary class label $y_i \in \{1, -1 \}$. Using this labeled data, we define a suitable loss function to be optimized.

For optimizing the kernel coefficients $\bm{\eta}$, it was proposed for the EasyMKL algorithm \cite{Aiolli2015} that the objective function from the Kernel Optimization of the Margin Distribution (KOMD) \cite{Aiolli2008} approach should be used. The KOMD objective represents the separation between positive ($y_i = 1$) and negative ($y_i = -1$) training examples where the kernel weights are represented implicitly with the Lagrangian dual variables $\bm{\phi} := [\phi_1, \phi_2, \ldots \phi_{N_{X}}] \in \mathbb{R}^{N_X}$. That is, $\bm{\eta}$ is derivable using $\bm{\phi}$ as is shown later in Eq.  \ref{eq:lagrangian_to_weights} (see \cite{Aiolli2015} for explicit steps in transforming the optimization problem in $\bm{\eta}$ to that of $\bm{\phi}$ and vice-versa). Recalling that our kernel function is also parameterized by the unitary operator parameters $\bm{\theta}$, the separation between classes as defined in \cite{Aiolli2015} is modified to be written as a function of both parameter sets $\bm{\theta}$ and $\bm{\phi}$:
\begin{equation}
    \mathcal{L}(\bm{\theta}, \bm{\phi}) := (1 - \lambda) \bm{\phi}^T\hat{Y} K_e(\bm{\theta}) \hat{Y} \bm{\phi} + \lambda ||\bm{\phi}||^2_2
\label{eq:cost}
\end{equation}
where $K_e(\bm{\theta}) = \sum_{t \in T} K_{t}(\bm{\theta})$ is an equally weighted sum of Gram matrices with elements $K^{ij}_{t}(\bm{\theta}) = \kappa_{t}(\bm{x}_{t}^{(i)}, \bm{x}_{t}^{(j)}, \bm{\theta})$  from Eq. \ref{eq:TSHK}, $\hat{Y} = \text{diag}(y_1, y_2, \ldots, y_{N_x})$ and $\lambda \in [0, 1]$ is a real-valued penalty scaling factor hyperparameter intended to favor low variance solutions for $\bm{\phi}$ with increasing $\lambda$. As shown in Fig. \ref{fig:hero_time_kernel}(i), $K_e(\bm{\theta})$ is initially computed with some $\bm{\theta}_{\text{init}}$ (chosen randomly or otherwise) which is used to prime the first step of the max-min optimization problem
\begin{equation}
\mathcal{L}^{\star} := \text{max}_{\bm{\theta}}[\text{min}_{\bm{\phi}}\mathcal{L}(\bm{\theta}, \bm{\phi})]
\end{equation}
where $\mathcal{L}^{\star}$ is the optimal loss and the optimal arguments are $\bm{\theta}^{\star}$ and $\bm{\phi}^{\star}$, respectively. Because the internal minimization with respect to $\bm{\phi}$ is a convex objective, as shown in Fig. \ref{fig:hero_time_kernel}(ii), for any given fixed $\bm{\theta} = \bm{\theta}_{\text{fixed}}$, we can efficiently and accurately find $\text{min}_{\bm{\phi}}[\mathcal{L}(\bm{\theta}, \bm{\phi})]$  using a convex solver such as a cone program \cite{Odonoghue2016}. Because we can reasonably assume that the global minimum of a convex problem will be found, we can re-frame the problem as
\begin{equation}
\mathcal{L}^{\star} = \text{max}_{\bm{\theta}}[\mathcal{L}_{\bm{\phi}_{\text{min}}}(\bm{\theta})]
\label{eq:loss_min_phi}
\end{equation}
where $\mathcal{L}_{\bm{\phi}_{\text{min}}}(\bm{\theta}) := \text{min}_{\bm{\phi}}[\mathcal{L}(\bm{\theta}, \bm{\phi})]$ and the subscript $\bm{\phi}_{\text{min}}$ indicates that we are working at the $\bm{\phi}$ which minimizes $\mathcal{L}(\bm{\theta}, \bm{\phi})$ with respect to $\bm{\phi}$ for any given $\bm{\theta}$.  

The maximization with respect to $\bm{\theta}$ can be performed using any general purpose optimization algorithm. Importantly, the gradient $\partial \mathcal{L}_{\bm{\phi}_{\text{min}}}(\bm{\theta})/\partial \bm{\theta}$ is obtainable analytically through an automatic differentiation engine \cite{Bergholm2018} because (i) partial gradients of the kernel function with respect to the quantum circuit parameters $\bm{\theta}$ are known using techniques including the parameter shift rule and its variants \cite{Mitarai2018, Schuld2019grad, Bergholm2018} and (ii) the partial gradients of the \textit{minimum} of the convex optimization problem with respect to the quantum circuit parameters are known because the cone program is differentiable \cite{Agrawal2019, Agrawal2019cone} [Fig. \ref{fig:hero_time_kernel}(iii)]. This allows one to update the quantum parameters $\bm{\theta}$ [Fig. \ref{fig:hero_time_kernel}(iv)] using one of a  whole host of gradient-requiring optimizers available from the wider ML literature \cite{Bottou2018}. Indeed, our method can be interpreted as a hybrid neural network with quantum layers and classical convex optimization layers \cite{Agrawal2019}: a Quantum-Classical-Convex network (QCC-net). 

After meeting the termination criteria of the optimization (which could be a fixed number of iterations or tolerance on changes in the value of Eq. \ref{eq:cost} when the parameters are updated, for example), we arrive at an approximation of $\bm{\theta}^{\star}$ and $\bm{\phi}^{\star}$ where the latter is given as $\text{argmin}_{\bm{\phi}}[\mathcal{L}(\bm{\theta}^{\star}, \bm{\phi})]$. We use both optimal parameter vectors to extract the optimal kernel weights $\eta_{t}^*$
\begin{equation}
\eta_{t}^* = \frac{ \bm{\phi}^{\star T}\hat{Y}K_{t}(\bm{\theta}^{\star})\hat{Y}\bm{\phi^{\star}}}{\sum_{t \in T} \bm{\phi}^{\star T}\hat{Y}K_{t}(\bm{\theta}^{\star})\hat{Y}\bm{\phi^{\star}}}.
\label{eq:lagrangian_to_weights}
\end{equation}
Now, in Eq. \ref{eq:combined_kernel} we set $\bm{\theta} \rightarrow \bm{\theta}^{\star}$ and $\bm{\eta} \rightarrow \bm{\eta}^{\star}$ to obtain our goal: a trained TSHK [Fig. \ref{fig:hero_time_kernel}(v)]. 

It should also be noted that during the training cycle, it is often appropriate to work with mini-batches of the training data. This both reduces the computational effort required in training and is known to improve generalization performance \cite{keskar2017largebatch} in classical ML algorithms. Formally, a mini-batched iteration of the training cycle involves sampling, at random, a training mini-batch $X_{\text{batch}} \subset X$ with cardinality $|X_{\text{batch}}| = N_{\text{batch}}$.  During this cycle, Gram matrices $K_{t}(\bm{\theta})$ and $K_e(\bm{\theta})$ and the diagonal label matrix $\hat{Y}$ have the reduced dimension $\mathbb{R}^{N_{\text{batch}}} \times \mathbb{R}^{N_{\text{batch}}}$ and the Lagrangian parameter vector $\bm{\phi}$ reduces its dimension to $\mathbb{R}^{N_{\text{batch}}}$. It should be clear now that the wall-time for a single mini-batch iteration using a fixed number of shots to evaluate the quantum kernel elements and a fixed number of iterations to solve the convex problem scales quadratically with  $\mathbb{R}^{N_{\text{batch}}}$. When these batches are sampled, we must ensure that each mini-batch contains at least one time-series instance with label $y_i = 1$ and another with $y_i = -1$. Without this, we cannot define a separation between classes and therefore the loss function of Eq. \ref{eq:cost} is undefined. \par

Before moving on, we must consider the trainability of the TSHK. Among the key ingredients of the TSHK are of course quantum kernels. It is known that for static quantum kernels, quantum circuit gradients vanish exponentially with the number of qubits when classical data is embedded using Ans\"{a}tze without sufficient inductive biases \cite{kubler2021inductive}. This is one of the manifestations of the barren plateaus phenomenon \cite{McClean2018}. Within our QCC-net framework, because quantum circuit gradients are obtained by differentiating through convex optimization layers, it is unclear to what extent QCC-nets suffer from barren plateaus. Indeed, not all quantum neural network architectures are found with this problem \cite{Pesah2021}. An interesting topic of future study would be a theoretical analysis of gradient scaling in QCC-nets and how to use convex optimization layers to impart different inductive biases.

\subsection{Using the trained kernel function: classification \label{subsec:kernel_classifier}}
Now equipped with an optimal kernel function, we can use it in any kernelized ML technique like kernelized SVM or kernel ridge regression. In this work we focus on classification using the kernelized soft margin variation of SVM which from now on-wards we refer to as ``SVM". While a full presentation of the formalism describing SVM is beyond the scope of this work (and is discussed elsewhere \cite{cortes1995support}), we now provide a high level explanation. In binary classification, SVM separates data points belonging to two different classes by learning a decision boundary, also known as a hyperplane, from labelled training data. It allows for some misclassifications by introducing a soft margin, where the appetite for misclassfication is controlled by a single hyperparameter $C \in \mathbb{R}$. A kernel function is used to map the data into a higher-dimensional space where the hyperplane can separate the classes more easily.  

For our purposes, SVM can be regarded as a decision function for an unseen data point $\bm{x}$ given four items: $\kappa(\bm{x}, \bm{x}^{\prime}), X, \bm{y}$ and $C$ where each, in order,  is a valid kernel function, a set of training data, the vector of binary class labels for that training data and the misclassification appetite hyperparameter. That is, we have 
\begin{equation}
\mathcal{D}(\bm{x}|\kappa(\bm{x}, \bm{x}^{\prime}), X, \bm{y}, C):= \mathcal{D}^{\star}(\bm{x}) \in \mathbb{R}.
\label{eq:combined_svm_classifer}
\end{equation}
Binary class predictions can now be made using
\begin{equation}
    y_{\text{pred}}(\bm{x}) = \text{sgn}[\mathcal{D}^{\star}(\bm{x})] \in \{-1, 1\}.
    \label{eq:classify_full}
\end{equation}
The most natural way to use this framework to perform time-series classifications is simply to set $\kappa(\bm{x}, \bm{x}^{\prime})$ to a trained kernel function of the form of Eq. \ref{eq:combined_kernel} at the optimal parameters $\bm{\theta}^{\star}$ and $\bm{\eta}^{\star}$. Another way of doing so is the break the problem down into $p$-many time-dependent SVMs. We then have 
\begin{equation}
\mathcal{D}_t(\bm{x}_t|\kappa_t(\bm{x}_t, \bm{x}^{\prime}_{t}, \bm{\theta}^{\star}), X_t, \bm{y}, C):= \mathcal{D}_t^{\star}(\bm{x}_t) \in \mathbb{R}
\label{eq:time_dep_svm}
\end{equation}
where $X_t = \{\bm{x}^1_t, \bm{x}^2_t, \ldots  \bm{x}^{N_X}_t \}$. This function is then used to make a prediction at each $t$ with $y^t_{\text{pred}}(\bm{x}_t) = \text{sgn}[\mathcal{D}_t^{\star}(\bm{x}_t)]$. The final class prediction for the entire time-series $\bm{x}$ is then given by a weighted majority vote from each $t$
\begin{equation}
    y^{\text{vote}}_{\text{pred}}(\bm{x}) = \text{sgn} \left[\sum_{t \in T}\eta_t \cdot y^t_{\text{pred}}(\bm{x}_t) \right] \in \{-1, 1\}.
    \label{eq:classify_by_time}
\end{equation}
While it is not guaranteed that $y_{\text{pred}}(\bm{x})$ will provide more accurate predictions than $y^{\text{vote}}_{\text{pred}}(\bm{x})$ in general, we note that because our training procedure defined in Sec. \ref{subsec:training} refines a combined kernel [used to define $y_{\text{pred}}(\bm{x})$], it is intutive to expect this. Indeed, we find that $y_{\text{{pred}}}(\bm{x})$ is most effective for the data sets treated in Sec. \ref{subsec:syn_demonstration} , \ref{subsec:univariate_sim} and \ref{subsec:qmp_experiments},  so we choose to make predictions using Eq. \ref{eq:classify_full}. However, analyzing $\mathcal{D}_t^{\star}(\bm{x}_t)$ and $y^t_{\text{pred}}(\bm{x}_t)$ can provide valuable insights into the time-dependent nature of a given TSHK function which \textit{are used} in this work.

\section{Probing time dependence \label{sec:probe_time_dependence}}

\begin{figure}
    \centering
    \includegraphics[width=0.8\linewidth]{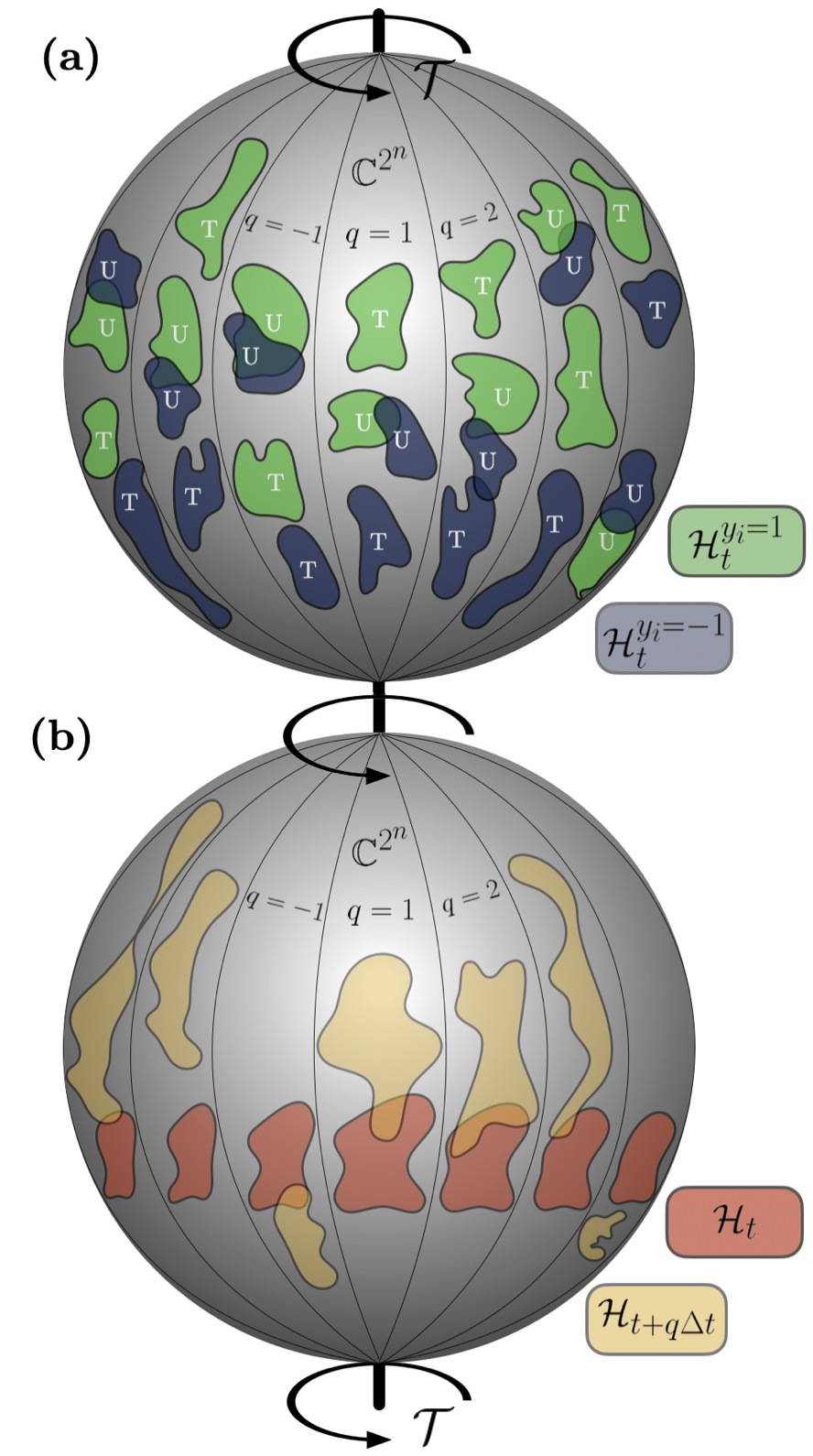}
\caption{Time-evolving Hilbert space illustrations. Spheres are divided into $Q$-many petals each indexed by $q\in \mathbb{Z}^{\neq 0}$. In each petal, the grey area represents the vector space spanning $\mathbb{C}^{2^n}$ such that each point on a petal represents a vector $|v \rangle \in \mathbb{C}^{2^n}$.} A full rotation of the sphere corresponds to the period $\mathcal{T}$ of time evolution operator $e^{-iH(\bm{\beta, \gamma})t}$. We note that a period $\mathcal{T}$ formally exists when the spectrum of $H$ is free of irrational eigenvalues. The $q=0$ case is omitted to retain visual clarity in (b) where colored regions would now (by definition of Eq. \ref{eq:check_time_dependence_naive} and \ref{eq:check_time_dependence_simplified}) maximally overlap. (a) Hilbert space regions accessible by embedded classical data belonging to class 1 or -1 where the kernel function has been Trained (T) or is untrained (or at least not ideally trained; U) at different times $t=t_q$. (b) Hilbert space regions accessible by the entire classical data input space $\bm{\chi} \in \mathbb{R}^{d}$ at time $t$ and time $t + q\Delta t$ where $\Delta t$ is a fixed time step.
\label{fig:probe_time_depend}
\end{figure}

After having achieved a reasonable approximation of $\bm{\theta}^{\star}$ using a classical optimization routine, a relevant quantity to probe is the time dependence of the inner product space. This is an important question to ask because while usage of the TSHK allows for a time dependent inner product space, it does not mean that strong time dependence was achieved when minimizing the cost function of Eq. \ref{eq:cost} which will be strictly dependent on the data set and the choice of parameters/circuit Ans\"{a}tze used to construct the QCC-net. As an extreme case, there can exist some $\bm{\beta}$ and $\bm{\gamma}$ such that $e^{-iH(\bm{\beta}, \bm{\gamma})t} \approx \mathbb{I} \quad \forall t$ thus the resulting TSHK is approximately time-independent. Naively, one may initially think to examine the value of $\kappa_{t}(\bm{x}_{t}, \bm{x}^{\prime}_{t}, \bm{\theta}^{\star})$ as a function of $t$ where $\bm{x}_{t}$ and $ \bm{x}^{\prime}_{t}$ are drawn from a testing data set (or iterate over an entire testing set to calculate an entire kernel matrix). While there is nothing wrong with this approach, we must note that in the ideal case from training, time dependence in this sense \textit{will disappear}. That is, at each point in time $t\in \mathcal{T}$ the kernel function was trained to embed classical data with label $y_i = 1$ in a separate region of Hilbert space to those with label $y_i = -1$. In the ideal case, these two regions do not overlap at all and hence $\kappa_{t}(\bm{x}_{t}, \bm{x}^{\prime}_{t}, \bm{\theta}^{\star}) = 0$ $\forall t$ when   $\bm{x}_{t}$ and $ \bm{x}^{\prime}_{t}$ belong to different classes. Before the kernel function is trained, finite values of $\kappa_{t}(\bm{x}_{t}, \bm{x}^{\prime}_{t}, \bm{\theta})$ are of course permitted under the same conditions for the input. The above discussion is best explained alongside Fig. \ref{fig:probe_time_depend}. Intended for illustrative purposes only, Fig. \ref{fig:probe_time_depend} compresses high-dimensional vectors $|v \rangle \in \mathbb{C}^{2^n}$ into two dimensional regions on the surface of a sphere. Differently colored regions represent the space of feature vectors $| \bm{x}_t, t, \bm{\theta} \rangle$ accessible in different domains defined by the classical data $\bm{x}_t$ (see the figure caption for more information). Looking at Fig. \ref{fig:probe_time_depend}(a), we can see that at any $t$, the time-dependent Hilbert space regions $\mathcal{H}_t^{y_i = 1}$ do not overlap with $\mathcal{H}_t^{y_i = -1}$ when the kernel function is ideally trained (regions marked ``T"). For an untrained kernel (or at least not ideally trained kernel; regions marked ``U"), overlap between these regions are permitted.

To provide a more intuitive notion of time dependence, we must ask another question: ``given a general classical data vector $\bm{\chi} = [{}^1\chi, {}^2\chi, \ldots {}^d\chi] \in \mathbb{R}^d$, does the quantum state embedding of $\bm{\chi}$ given by Eq. \ref{eq:embedding} when we set $\bm{x}_t \rightarrow \bm{\chi}$ change as $t \rightarrow t + \delta t$?". 
The quantity of interest to probe this time dependence is the overlap of the $t$ and $t + \delta t$ quantum embedding spaces integrated over all $\bm{\chi}$. That is
\begin{equation}
 f_{\bm{\theta}}(\delta t) := \int_{^d\chi_{\text{lo}}}^{^d\chi_{ \text{hi}}} \ldots \int_{^1\chi_{\text{lo}}}^{^1\chi_{ \text{hi}}} |\langle \bm{\chi}, t, \bm{\theta}| \bm{\chi}, t + \delta t, \bm{\theta} \rangle|^2 d^1\chi \ldots d^d\chi
    \label{eq:check_time_dependence_naive}
\end{equation}
where the limits ${}^m\chi_{lo}$ and ${}^m\chi_{hi}$ are the minimum and maximum values permitted for classical data embedded into the chosen quantum feature space, respectively. Many quantum feature maps are periodic such that ${}^m\chi \in [{}^m\chi_{\text{lo}}, {}^m\chi_{\text{hi}})$ where ${}^m\chi_{\text{lo}} = 0$ and ${}^m\chi_{\text{hi}} = 2\pi$. Should either limit be unbound, limits should be set to enclose the zone of interest (i.e. some region enclosing all training and testing examples) to the data set. After simplifying the integrand using  Eq. \ref{eq:embedding} (namely, $U^{\dagger}(\bm{\chi}, \bm{\alpha})U(\bm{\chi}, \bm{\alpha}) = \mathbb{I}$ and cancelling $t$ in the exponent of the time evolution operator)  we find
\begin{equation}
 f_{\bm{\beta}, \bm{\gamma}}(\delta t) := \langle a_{\delta t}, \bm{\beta}, \bm{\gamma}|P_0|a_{\delta t}, \bm{\beta}, \bm{\gamma} \rangle \prod_{j=1}^{d}(^j\chi_{hi} - {}^j\chi_{lo})
    \label{eq:check_time_dependence_simplified}
\end{equation}
where  $|a_{\delta t}, \bm{\beta}, \bm{\gamma} \rangle := e^{-iH(\bm{\beta}, \bm{\gamma})\delta t}|0\rangle^{\otimes n}$ and $P_0 = |0\rangle^{\otimes n} \langle 0|^{\otimes n}$. Noting also that the result of the product over the index $j$ in  Eq. \ref{eq:check_time_dependence_simplified} is a real constant scaling factor $k$, it is useful to set $k=1$ choosing to work with the quantity $\mathcal{F}_{\bm{\beta}, \bm{\gamma}}(\delta t) = \langle a_{\delta t}, \bm{\beta}, \bm{\gamma}|P_0|a_{\delta t}, \bm{\beta}, \bm{\gamma} \rangle \in [0, 1]$. Examining $\mathcal{F}_{\bm{\beta}, \bm{\gamma}}(\delta t)$, it is clear that we need only consider the time evolution operator acting on the zero state with no consideration of classical data points. Indeed, much like Eq. \ref{eq:TSHK}, $\mathcal{F}_{\bm{\beta}, \bm{\gamma}}(\delta t)$ can be efficiently computed by estimating the probability of measuring the bit-string of all zeros in the computational basis when a quantum computer is prepared in the state $|a_{\delta t}, \bm{\beta}, \bm{\gamma} \rangle$.

The meaning of Eq. \ref{eq:check_time_dependence_naive} (or, equivalently, Eq. \ref{eq:check_time_dependence_simplified}) is demonstrated visually in Fig.  \ref{fig:probe_time_depend}(b). The sphere is constructed in the same manner as Fig. \ref{fig:probe_time_depend}(a) but colored regions now represent the  Hilbert space region accessible by the embedding of $\bm{\chi}$ at time $t$ ($\mathcal{H}_t$; red regions) and at time $t + q\Delta t$ ($\mathcal{H}_{t + \Delta t}$; yellow regions) where $q \in \mathbb{Z^{+}}$ and $\Delta t \in \mathbb{R}_{>0}$ are fixed constants. The squared modulus of overlapping regions of red and yellow on Fig. \ref{fig:probe_time_depend}. (b) can be interpreted as either  Eq. \ref{eq:check_time_dependence_naive} or \ref{eq:check_time_dependence_simplified}.

\section{Demonstrations with quantum circuit simulators\label{sec:simulator_section}}

\subsection{Building intuition: a synthetic example \label{subsec:syn_demonstration}}

\begin{figure*}[t!]
    \centering
    \includegraphics[width=\linewidth]{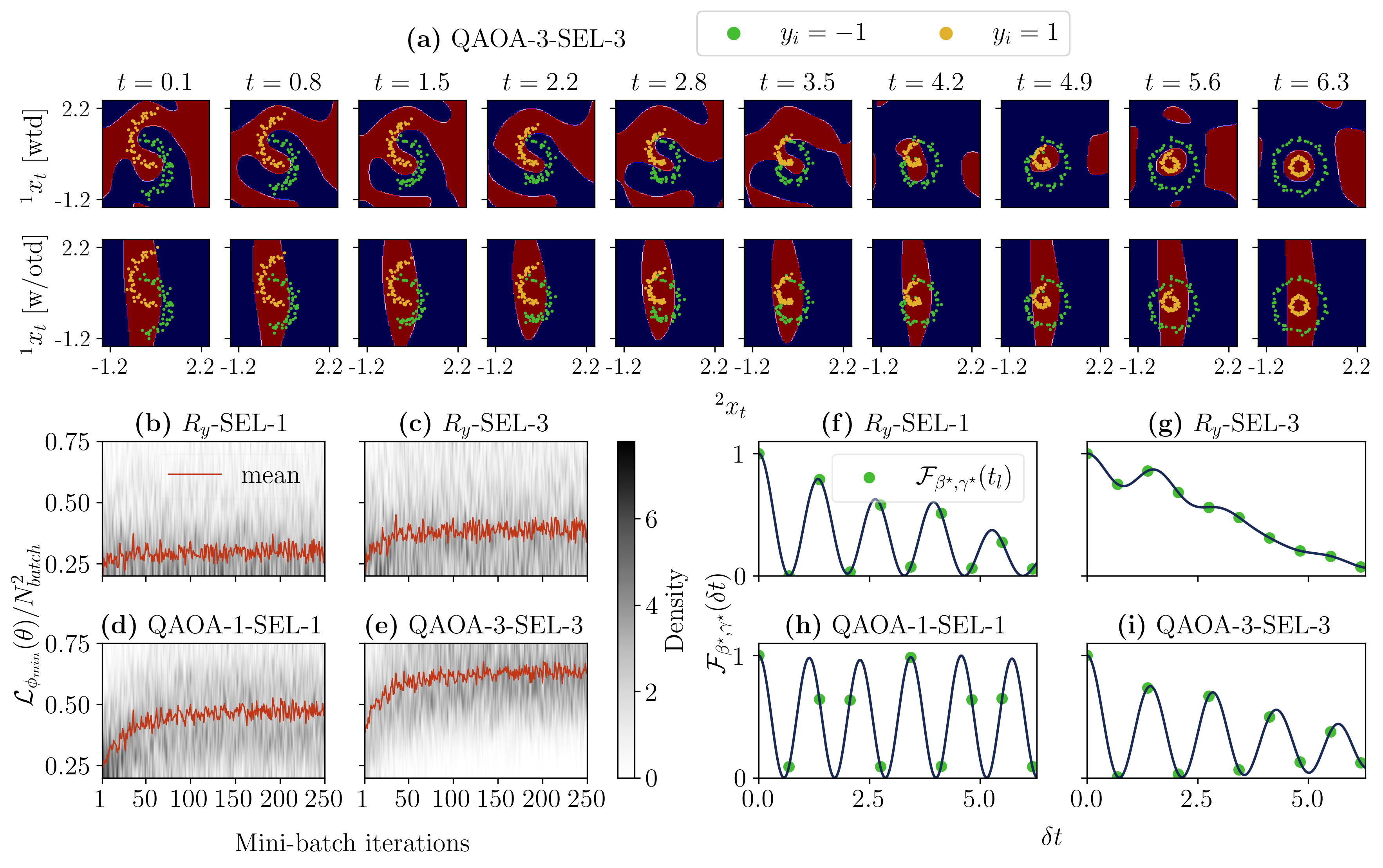}
    \caption{Quantum circuit simulator experiments with the \texttt{moons2circles} data set. (a) The evolution of the time-dependent class prediction $y_{\text{vote}}^t$ as moons (far left panel) gradually become circles (far right panel). Regions predicting $y_i = 1$ are shown in dark red and regions predicting $y_i = -1$ are shown in dark blue. The Testing data set is overlaid and are colored by their truth values as given in the legend. The figure is generated using the best model obtained for the QAOA-3-SEL-3 runs described in the main text. The upper panel displays results with time dependence (wtd) in the inner product space while the lower panel displays results without time dependence in the inner product space (w/otd) (b-e) The normalized loss $\mathcal{L}_{\bm{\phi_{\text{min}}}}(\bm{\theta})/N^2_{\text{batch}}$ as a function of mini-batch iterations for the $R_y$-SEL-1, $R_y$-SEL-3, QAOA-1-SEL-1 and QAOA-3-SEL-3 quantum circuits, respectively. The mean value of 50 independent runs is shown in red and the probability density over these runs is shaded in grey. Recall that according to Eq. \ref{eq:loss_min_phi} we are solving a \textit{maximization} problem, so we expect the loss in \textit{increase} with the number of mini-batch iterations. (f-i) For the same quantum circuits and order defined in (b-e), the time-resolved embedding overlap $\mathcal{F}_{\bm{\beta}^{\star}, \bm{\gamma}^{\star}}(\delta t)$ is plotted using the best model obtained using each circuit. Green markers are inserted at $\mathcal{F}_{\bm{\beta}^{\star}, \bm{\gamma}^{\star}}(t_l)$ to show overlaps of embedding spaces where classical data was seen in training. }
    \label{fig:synthetic_example}
\end{figure*}
As a first example of our time-series classification algorithm, we deal with a didactic case to build intuition. Our goal is to create a synthetic time-series classification problem where (i) data is not linearly separable (thus kernelized SVM is required) and (ii) the shape and position of the decision boundary must evolve with time in order to successfully classify data at different values of $t$. Accordingly, we generate a training set $X_{\text{syn}}$ with $|X_{\text{syn}}| = 100$ two-dimensional ($d=2$) time-series instances. We construct the beginning and endpoints of each time series using the well-known moons and circles data sets, respectively. Intermediate points in the series are generated using a 10-step (i.e. $p=10$) linear interpolation between the start and end points. The result is a gradual transition from moons to circles which we call the \texttt{moons2circles} data set. Testing data from \texttt{moons2circles} is shown as markers overlaid on the panels of Fig. \ref{fig:synthetic_example}(a).  

We use two different varieties of $n=3$ qubit quantum circuit structures to implement the quantum components of the QCC-net . We denote first variety $R_y$-SEL-$g$  and the second QAOA-$g$-SEL-$g$. We implement $g = 1$ and $3$ varieties where $g$ denotes the number of repeated layers of the layered Ans\"{a}tze. $R_y$ means a fixed embedding circuit (i.e. $\bm{\alpha}$ is dropped) of $U(\bm{x}_t) = R_y({}^1x_{t}) \otimes R_y({}^2x_{t}) \otimes I$ is used and QAOA means an embedding circuit inspired by the Quantum Alternating Operator Ans\"atz  \cite{lloyd2020quantum, farhi2014quantum, Hadfield2019} is used to implement $U(\bm{x}_t, \bm{\alpha})$. SEL means a Strongly Entangling Layers Ans\"atz \cite{Schuld2020} is used to implement $W(\bm{\beta})$. For all circuits, $D(\bm{\gamma}, t)$ is implemented using a circuit representing a truncated n-local Walsh-operator expansion as described in \cite{Welch2014}. The kernel function of Eq. \ref{eq:combined_kernel} is then trained using the process described in Sec. \ref{subsec:training} for 250 mini-batch iterations with $N_{\text{batch}} = 4$ using the \texttt{Adam} optimizer \cite{kingma2017adam} with an initial learning rate of 0.05.  For solving the convex problem (i.e, finding the Lagrangian dual variables $\bm{\phi}$) we use the splitting conic solver \cite{ocpb:16, odonoghue:21} which interfaces \texttt{cvxpyayers} \cite{Agrawal2019} for gradient computations \cite{Agrawal2019cone} which are passed to the automatic differentiation in Pennylane \cite{Bergholm2018}. The training is repeated 50 times where each run is seeded with initial parameters $\bm{\theta}_{\text{init}}$ drawn uniformly at random from the range $[-\pi, \pi]$. 

The results of training are shown in Fig. \ref{fig:synthetic_example}(b-e). On average we can see that QAOA-3-SEL-3 produces the largest loss values both initially (i.e. at the first mini-batch iteration) and after training for the 250 mini-batch iterations. While it is intuitive that the loss after training should be largest for  QAOA-3-SEL-3 simply because it has the largest number of quantum circuit parameters $\bm{\theta}$, this does not explain the large initial loss. Indeed, at the first mini-batch iteration, we measure the separation of binary classes for optimally combined random kernels. This means that combined random kernels become better separators as the depth and number of parameters increases for the \texttt{moons2circles} data set. This effect is so pronounced that \textit{untrained} QAOA-3-SEL-3 models, on average, have larger losses than trained $R_y$-SEL-1 and $R_y$-SEL-3 models. Examining the density heat maps [grey shaded areas on Fig. \ref{fig:synthetic_example}(b-e)], we can see that training is highly stochastic, much of which is owed to the small mini-batch size. We do see this stochasticity reduce for the QAOA-$g$-SEL-$g$ models where it is more obvious that an envelope of high density forms around the mean curve showed in dark red.

Using the best (as determined by the largest value of the loss evaluated using the entire testing data set) trained TSHK with the QAOA-3-SEL-3 Ans\"{a}tz, we train a SVM with $C=100$ at each of the 10 time steps and compute $y_{\text{vote}}^t(\bm{x}_t)$ on a fine $100 \times 100$ grid at each $t$. These results are shown on the upper panel of Fig. \ref{fig:synthetic_example}(a) as heat-maps at each point in time. Regions colored in dark red/dark blue show areas where $y_{\text{vote}}^t=1$/ $y_{\text{vote}}^t=-1$ The result is clear: the decision function is time-evolving and adapts to separate data belonging to the two different classes at every time step. Although this is a simple example, we note that the best TSHK for any circuit structure variation achieves 100\% classification  accuracy when passed to a SVM. This is despite the data at intermediate time steps not being perfectly separated (data of different classes overlap). Upon inspection of the kernel weights, we see that the beginning and end-points are most heavily weighted which is intuitive since these points see no overlap between classes. This observation exposes the power of our algorithm: time-series can be classified based on ``smoking-gun" time points as these points are able to be heavily weighted by the convex optimizer. \par

The effects of including a time-dependent inner product space become even clearer should we observe the performance on the \texttt{moons2circles} data set in its absence. This is shown in the lower panel of Fig. \ref{fig:synthetic_example}(a) as achieved by fixing $t=1$ (an arbitrary choice) in the time evolution operator. Other settings are identical to those which produce the results of the upper panel of  Fig. \ref{fig:synthetic_example}(a). While there remains some change in the position of the decision boundary as we advance in time, it is certainly much less malleable than in the case with the time dependent inner product. Now unable to use a different kernel function at each $t$, training finds a single kernel function which strikes a balance between being able to classify moons-like and circles-like data, which, when inserted into a SVM, results in decision boundaries with a striped character. Moreover, classification accuracy with this model slips down to 78\%. From these investigations, we conclude that using the TSHK is most effective for data sets where (i) the correct notion of similarity between pairs of data points changes with time and (ii) there exists a shared time axis in the data as was described in Sec. \ref{subsec:time_dep_inner}.

In Fig. \ref{fig:synthetic_example}(f-g) we probe the time dependence of the best trained models using the approach presented in Sec. \ref{sec:probe_time_dependence}.  With the exception of $R_y$-SEL-3 [Fig. \ref{fig:synthetic_example}(g)] $\mathcal{F}_{\bm{\beta}^{\star}, \bm{\gamma}^{\star}}(\delta t)$ is observed to be highly oscillatory. Importantly, apart from QAOA-1-SEL-1 [Fig. \ref{fig:synthetic_example}(h)], $\mathcal{F}_{\bm{\beta}^{\star}, \bm{\gamma}^{\star}}(\delta t)$ is a broadly decreasing function intuitively suggesting that the quantum embedding space when data is more moons-like becomes different as the data becomes more circles-like. Indeed, even for QAOA-SEL-1, the last point (shown with green markers) of all models shows that $\mathcal{F}_{\bm{\beta}^{\star}, \bm{\gamma}^{\star}}(\delta t)$ commensurate with the time difference between exactly moons [the left-most panel of Fig. \ref{fig:synthetic_example}(a)] and exactly circles [the right-most panel of Fig. \ref{fig:synthetic_example}(a)] is very small which means that the quantum embedding space used for classifying moons is \textit{strongly} different than that used for classifying circles.

\subsection{Univariate time-series: the gun-point data set \label{subsec:univariate_sim}}

In this section, we demonstrate the performance of our algorithm for classification on a real and popular benchmark: the gun-point data set \cite{ratanamahatana2005three}. Although this data set has been described in detail elsewhere, we summarize briefly here.  Two participants, with their hands beginning at their sides, are asked to either (i) point their right finger towards a target or (ii) point a small firearm (a hand gun) at a target and bring their hand back down to the starting position. During this process, the forward motion on their hand is tracked with a sensor at regular intervals and each finger-point or gun-point instance is recorded as a univariate ($d=1$) time-series instance. Labelled gun-point instances belong to class 1 ($y_i = 1$) and finger-point instances belong to class 2 ($y_i = -1$). The training set $X_{tr}$ has size $|X_{tr}|=50$ and the testing set  $X_{te}$ has size $|X_{te}|=150$. Each instance has 150 time stamps ($p=150$). \par
\begin{figure}
    \centering
    \includegraphics[width=\linewidth]{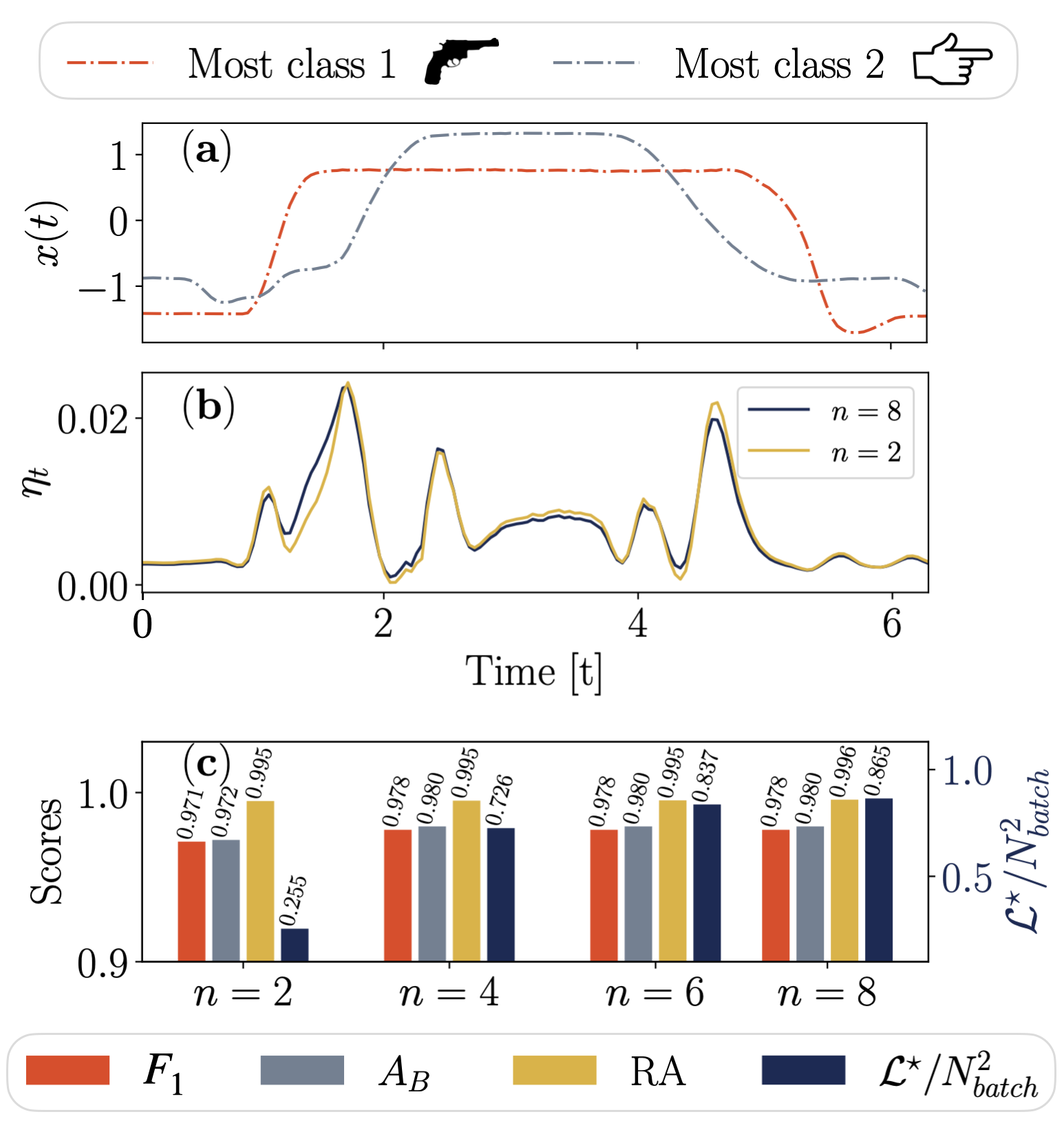}
    \caption{Insights and performance metrics for trained TSHKs passed to SVMs on the gun-point data set. (a) Testing data set time-series instances reported to be the most class 1 (gun-point; orange dotted and dashed line) and most class 2 (finger point; blue dotted and dashed line) as determined by the value of $\mathcal{D}^{\star}(\bm{x})$. (b) The kernels weights $\eta_t$ at each time for the $n=2$ (gold line) and $n=8$ (blue line). (c) $F_1$, $A_B$, and RA scores (left axis) and the normalized optimal loss $\mathcal{L}^{\star}/N_{batch}^2$  (right axis) for the $n=2, 4, 6$ and $8$ models.} 
    \label{fig:simulated_gunpoint}
\end{figure}
Using the QAOA-3-SEL-3 circuit structure described in Sec. \ref{subsec:syn_demonstration}, we train four models with $n=2, 4, 6$ and $8$ qubits. Each model is selected from 20 training attempts each using 500 mini-batch iterations with $N_{\text{batch}} = 4$ with $\bm{\theta}_{\text{init}}$ drawn uniformly at random from $[-\pi, \pi]$. Each model is selected based on the largest value of the loss function evaluated on the testing data set. Before the SVMs are trained, given $\bm{\theta}^{\star}$ each optimal model is passed through the convex optimization step a final time using the entire training data set (i.e, $N_{\text{batch}} = N_X$) to refine the kernel weights $\bm{\eta}$. We now train $C=100$ SVM decision functions $\mathcal{D}^{\star}(\bm{x})$  which we use to classify each time-series instance in $X_{\text{test}}$. With these predictions, we calculate the $F_1$ score, the balanced accuracy score ($A_B$) and the receiver operating characteristic area under the curve score (ROC AUC; abbreviated further to RA for the rest of this work).

Fig. \ref{fig:simulated_gunpoint} shows the results of these experiments. In Fig. \ref{fig:simulated_gunpoint}(a), The time-series instances which were identified as the most class 1 and the most class 2 as determined by having the highest and lowest values of $\mathcal{D}^{\star}(\bm{x})$ are shown. Directly below on the same time axes, in Fig. \ref{fig:simulated_gunpoint}(b), we have the time-resolved kernel weights $\eta_t$ for the $n=2$ and $n=8$ models. It is clear that different points in time are weighted very differently to others. In particular, peaks near the regions where the finger/gun is being lifted and peaks near where the finger/gun is being put back down again are seen. This is in stark contrast to the beginning, end and middle points of the series where finger/gun is approximately stationary. This points to the key discriminating factors between the two cases being a reaction time difference and/or other characteristic differences during the time stamps where the finger/gun is \textit{in motion}. It should also be noted that there is a remarkable symmetry in the profile of $\eta_t$ about the mid-point of time (where the finger/gun is stationary); finger/gun instances can be discriminated equally well from motion on the way up to pointing at the target as they can on the way down. All of the models produce a similar profile for $\eta_t$, which, for $n=2$ and $n=8$ are shown in Fig. \ref{fig:simulated_gunpoint}(a). It can be seen that increases in the qubit size for the model lead only to small changes in the weighting of different time stamps in the classification. Looking now at Fig. \ref{fig:simulated_gunpoint}(c), we see that the balanced accuracy score is already at 97.2\% for the $n=2$ model, and raises a small amount to $98.0\%$ by the time we reach $n=8$. Similar small gains are seen for $F_1$ scores and the ROC AUC score. Remarkably, these scores mean that our hybrid quantum-classical algorithm is competitive with purely classical algorithms for this data set. Specifically, when compared against 9 state-of-the-art deep learning approaches \cite{IsmailFawaz2019}, our algorithm is beaten in terms of accuracy scores by only two of them: fully convolutional neural networks (FCN; $A_B = 1.000$) and residual neural networks (ResNet; $A_B = 0.991$) which, in comparison, require far greater computational resources. \par

\begin{figure}
    \centering
    \includegraphics[width=\linewidth]{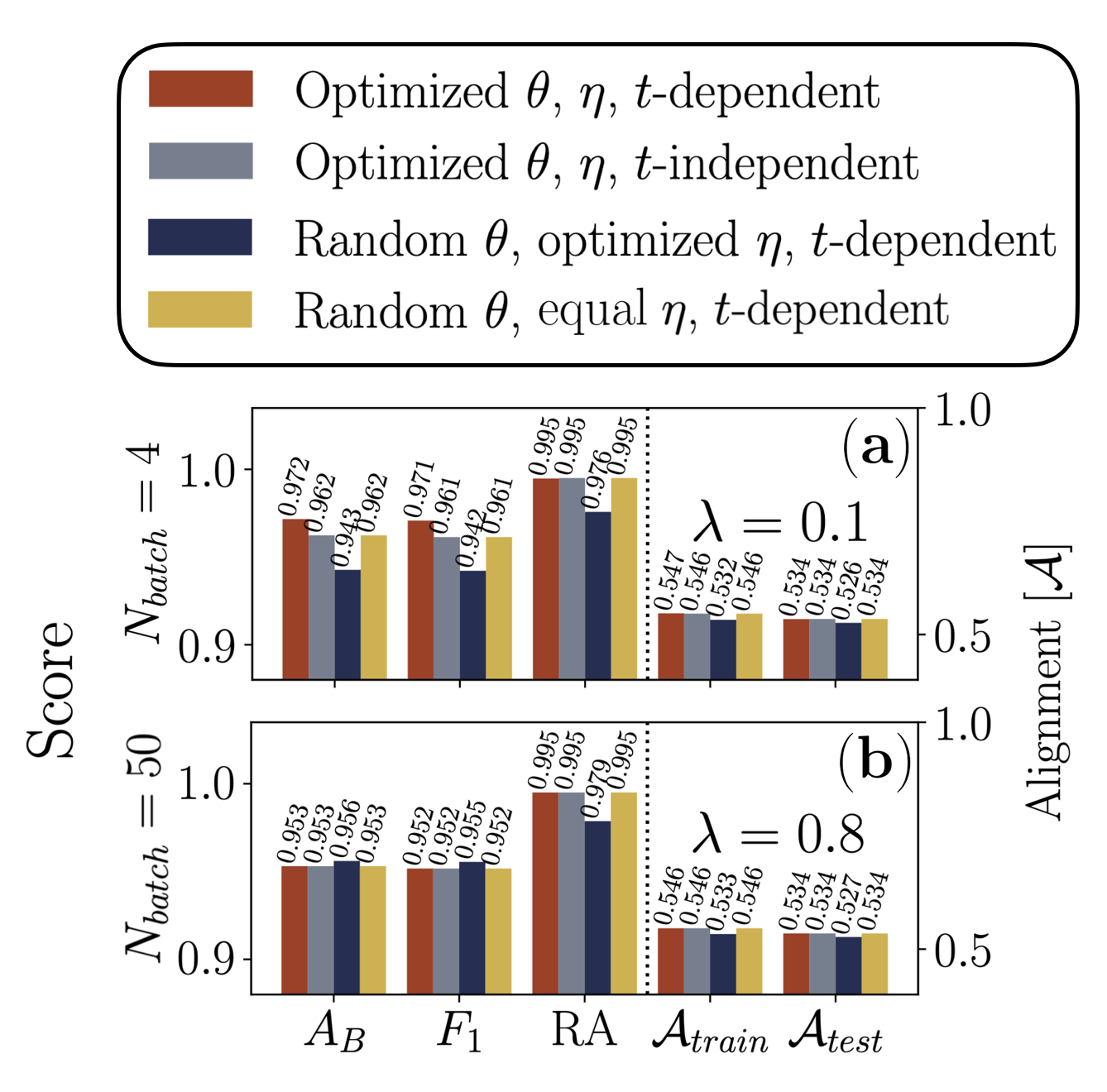}
    \caption{Discerning the impact of different mechanisms defining the TSHK as applied to SVM classification of the gun-point data set. Different approaches include the full TSHK method (red bars), the otherwise full method with time dependence in the inner product space forbidden (a single static quantum kernel is used at each $t$; grey bars), the otherwise full method but with randomly chosen quantum circuit parameters $\bm{\theta}$ (dark blue bars) and an approach using random $\bm{\theta}$, equal kernel weights ($\eta_t = 1/p$ $ \forall t$) and a time-dependent inner product space. $\mathcal{A}_{train/test} = 1 - 1/p^2\sum_{i}^p\sum_{j}^p|K_{train/test, ij} - K_{truth, ij}|$ is the kernel alignment of the training/testing data sets, $K_{train/test, ij}$ are the elements of combined kernels for the training/testing datasets and $K_{truth, ij} = y_i y_j$ are the elements of the truth matrix. (a) $N_{batch} = 4$ and $\lambda = 0.1$. (b) $N_{batch} = 50$ (the entire training data set) and $\lambda = 0.8$. An expanded version of this figure is available in the Supplemental Material \cite{supplement}.} 
    \label{fig:hyperparameter_sweep_summary}
\end{figure}

To quantify which components of our approach are most significant in our experiments and what training hyperparameters yield the best results, we conduct a further study focusing on the $n=2$ qubit case. A summary of the results are shown in Fig \ref{fig:hyperparameter_sweep_summary} and an expanded version of Fig. \ref{fig:hyperparameter_sweep_summary} is available in the Supplemental Material \cite{supplement}. In general, we find that training with small batch sizes and small $\lambda$ is the most effective (Fig. \ref{fig:hyperparameter_sweep_summary}(a)). This effect is enhanced when paired with the full TSHK formalism (red bars) which, for these hyperparameters, yields the largest $A_B$ and $F_1$ scores as well as the largest training kernel alignment $\mathcal{A}_{train}$ (see caption of Fig. \ref{fig:hyperparameter_sweep_summary} for a definition) of all approaches taken in this study. This peak performance must, however, be viewed in the context of the other bars on Fig. \ref{fig:hyperparameter_sweep_summary}(a). Examining the case where the time dependent inner product is forbidden (grey bars), the full approach offers only a 1\% improvement in $A_B$ and $F_1$. Indeed, examining the general level of scores for all other bars on both Fig. \ref{fig:hyperparameter_sweep_summary}(a) and \ref{fig:hyperparameter_sweep_summary}(b), we can see that any of the taken approaches can score in the mid-90\% range. It must therefore be concluded that most of the performance we achieve on this data set is gained from the classical portions of the algorithm (mostly the SVM itself) correctly classifying many instances using any of the combined kernels we supply it. Only a few extra points are gained by using the full TSHK formalism. Given how closely performant the top classical deep learning approaches are for this data set \cite{IsmailFawaz2019}, it is, however, not unreasonable to suggest that these gains are indeed significant. Furthermore, we view the results presented Fig. \ref{fig:simulated_gunpoint} to be sensitive to the choice of circuit Ans{\"a}tz. It is unclear what choice should be made when presented with classical data and is a topic of current research to figure out how to encode inductive biases into classical data encoding Ans{\"a}tze \cite{bowles2023contextuality}.  \par

 Lastly, we remark that while scores only increase modestly as the number of qubits increases [Fig. \ref{fig:simulated_gunpoint}(c)], the loss does gradually get larger. Recalling that this loss represents the separation between positive and negative classes, this means that, within the parameters of our study, higher qubit models are able to find a time-dependent quantum embedding space where classical data is better separated than smaller qubit models. This larger separation is known to give rise to higher generalization performance for SVM  \cite{hastie2009elements}.

\section{Parallel execution on quantum hardware \label{sec:hardware_runs}}

\subsection{Quantum multi-programming \label{subsec:qmp_explained}}

To more efficiently use NISQ hardware, it is possible to run multiple quantum circuits concurrently. Known as Quantum Multi-Programming (QMP), this technique allows NISQ devices to execute several quantum circuits concurrently, even if they differ in structure or complexity. In this Section, we show that computation of TSHKs can be significantly sped up using QMP. 

The primary motivation for QMP is driven by the fact that the number of qubits present in NISQ devices is often significantly higher than their Quantum Volume (QV). Specifically, when compared to trapped ion devices, superconducting quantum computers have restricted qubit connectivity and lower QV. Taking this further, following IBM's quantum volume $V_{Q}$ we have $\displaystyle \log _{2}V_{Q}=\text{argmax}_m\left\{\min \left[m,d(m)\right]\right\}$ where $d(m)$ is the depth of a model circuit with $m$-qubits \cite{cross2019validating}. One depth of the model circuit is composed of a random permutation of the qubits involved in the test, followed by random two-qubit gates. For instance, the QV of the \texttt{ibm\_washington} device (used in this work) is $64$, with $127$ qubits available. This means the model circuit with $6$-qubits runs reliably at six depths of the model circuit on average on $\texttt{ibm\_washington}$.  Other superconducting quantum computers share similar properties such as limited connectivity between qubits, and many more qubits than $\log_2V_Q$. Therefore, it is crucial to use the capacity of modern superconducting quantum devices more efficiently, and QMP has been proposed as a method to achieve this goal. Therefore, QMP fills the gap between the relatively many qubits and the relatively low QV of NISQ devices by executing multiple quantum circuits concurrently, enhancing the throughput and utilization of NISQ devices. 

However, implementing QMP on NISQ devices has several issues to address because QMP accompanies unfavorable impacts on the whole system such as measurement timing of the concurrent circuits \cite{das2019case} and crosstalk between different circuits \cite{ohkura2022simultaneous}. Efficiently mapping qubits between logical and physical states, as well as task scheduling \cite{liu2021qucloud, niu2021enabling, niu2022parallel}, have been studied alongside the aforementioned issues. Despite preliminary investigations into QMP, its integration with quantum algorithms, and especially QML, remains underdeveloped. Nonetheless, there has been recent noteworthy progress in applying QMP to Grover's search algorithm \cite{park2023quantum}, resulting in improved success probabilities compared to previous attempts.

\subsection{QMP implementation considerations} \label{subsec:qmp_design_considerations}
Before proceeding to calculate TSHKs with QMP, we discuss how the typical issues arising out of QMP. Starting with crosstalk, this effect is studied in detail by P. Murali et al \cite{murali2020software}. They suggested several rules to mitigate crosstalk between qubits. The relation between the number of physical buffers and the error rate was studied by Ohkura et al \cite{ohkura2022simultaneous}. They introduced a physical buffer, which is the number of idle qubits between quantum circuits. They tested a different number of controlled-X (CX) gates with a different number of physical buffers. Ohkura et al.~\cite{ohkura2022simultaneous} concluded that only $1$ physical buffer is sufficient until $30$ CX gates. Therefore, we use one physical buffer for our implementations. 

One important issue of QMP is measurement timing. When quantum circuits having different circuit depths run concurrently, the measurement timing influences the shortest circuit. This effect was studied in Ref. \cite{das2019case, ohkura2021crosstalk}. In these studies, they suggested delaying shorter circuits to align the measurement timing. In our case, because the circuits implementing elements of the TSHK each have the same depth, we do not need to make any alterations to measurement timings. 

\begin{figure*}
\centering
\includegraphics[width=\linewidth]{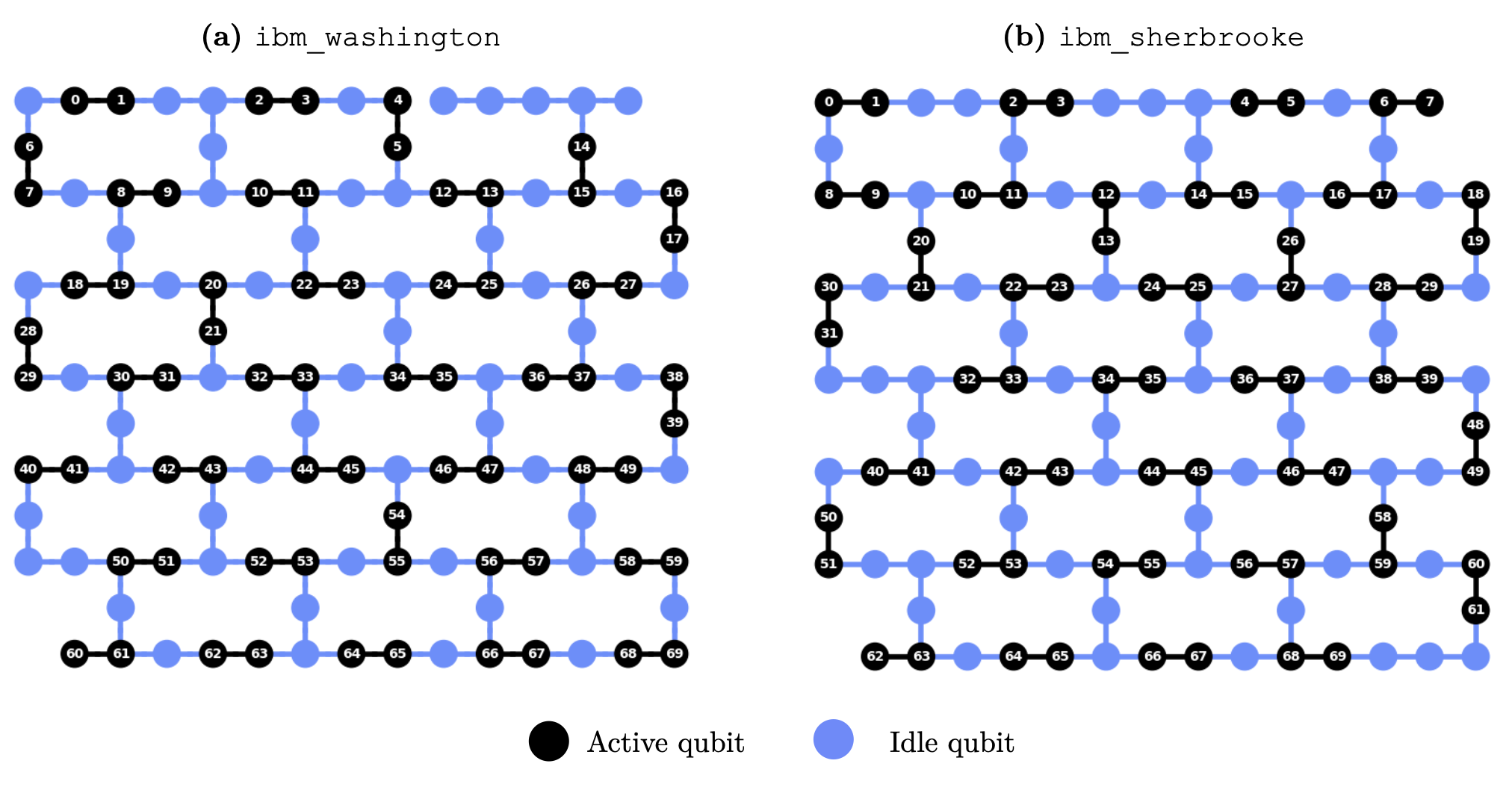}
\caption{The mapping between logical and physical qubits for the QMP circuits used in our experiments on (a)\texttt{ibm\_washington} (\texttt{Eagle r1} processor) and (b) \texttt{ibm\_sherbrooke} (\texttt{Eagle r3} processor). Both machines have 127 physical qubits and have a similar (but not identical) qubit connectivity graph. \texttt{ibm\_washington} has a QV of 64 while  \texttt{ibm\_sherbrooke} has a QV of 32. On \texttt{ibm\_washington}, the native gates are CX, ID, RZ, SX and X and \texttt{ibm\_sherbrooke} uses the same single qubit gates but ECR gates are used for 2-qubit interactions. Idle qubits are shown as blue circles and active qubits are black circles containing integers indexing the logical qubits.}
\label{fig:ibmq_layout}
\end{figure*}

Another crucial issue is efficient physical qubit mapping. Systematic qubit mapping algorithms have been developed \cite{das2019case, liu2021qucloud, niu2022parallel, niu2021enabling}. Even though these qubit mapping algorithms are also important for quantum circuits executed in serial (i.e. without QMP), efficient qubit mapping becomes more challenging and crucial in QMP because practical QMP circuits use more qubits than serial circuits. In our particular circumstance where each quantum thread is composed of two logical qubits, the mapping problem is solved by hand by mapping each thread to two physical qubits with two-qubit gate connectivity subject to the constraint that \textit{at least} one buffer qubit separates individual threads. The QMP layout can be seen in Fig. \ref{fig:ibmq_layout} for \texttt{ibm\_washington} and \texttt{ibm\_sherbrooke}.

Finally, we mention partial measurement in QMP. After the whole measurement of a QMP circuit, we need to extract the measurements from each parallelized circuit. This is called partial measurement and is implemented by post-processing after the whole measurement since the partial measurement probability is the sum of all other uninvolved qubits measurement probability. This is detailed in Ref. \cite{park2023quantum} and in the Supplemental Material \cite{supplement}. Since the example code in Appendix A in Ref. \cite{park2023quantum} iterates over the $2^n$ where $n$ is the number of the measured qubits, the computing time increases exponentially as more qubits are measured. Hence, we re-implement the partial measurement code, which extracts the partial measurement out of the whole measurement, to depend on the number of shots regardless of the number of the measured qubits. Our new implementation is given in the Supplemental Material \cite{supplement}.  

\subsection{QMP experiments with the gun-point data set}
\label{subsec:qmp_experiments}

We conduct QMP experiments using the gun-point data set described in Sec. \ref{subsec:univariate_sim} where the time dimension is decimated by $3 \times$ ($p$ is reduced from 150 to 50) to reduce the total QPU run time to a reasonable level. An optimal simulated (ideal) model for the decimated data set is first obtained using the same method described in Sec. \ref{subsec:univariate_sim}. Using the $\bm{\theta}^{\star}$ derived from the simulated model, we generate the quantum circuits used to calculate the matrix elements of the kernel matrices required to train an SVM. These circuits are deployed on two 127-qubit superconducting QPUs: \texttt{ibm\_washington} (64 QV) and \texttt{ibm\_sherbrooke} (32 QV). The experiments utilize the maximum number of active qubits which can be used in a single QMP run. For both devices, including the buffer qubit constraint,  this number is 70.  Fig.  \ref{fig:ibmq_layout}  shows the qubit layout on \texttt{ibm\_washington} [Fig. \ref{fig:ibmq_layout}(a)] and \texttt{ibm\_sherbrooke} [Fig. \ref{fig:ibmq_layout}(b)] . In addition to mapping considerations already discussed in Sec. \ref{subsec:qmp_design_considerations},  we manually choose the qubits with low error ratios (read assignment error, etc.). Considering each circuit has two qubits, we can run 35 circuits in parallel with a single QMP run. The Trial Reduction Factor (TRF) \cite{das2019case} describes the ratio of the number of trials (shots) that executes individually (baseline) to the number of trials (shots) of the QMP circuit. This factor represents the efficiency of the QMP in terms of trials (shots). In our case, the TRF is $35$. The QMP used $8192$ shot numbers to measure $35$ circuits whereas non-QMP implementations need $8192 \times 35$ shots to measure $35$ circuits. 

Recalling the total size of the training and testing data sets in the decimated gun-point data set, a naive serial QPU implementation (without QMP) requires $436,250$ individual calls to a quantum device to calculate all of the relevant kernel matrices to train and test an SVM. By applying QMP, the total number of calls to the QPU is \textit{significantly} reduced by a factor of 35 to $12,465$ because the QMP circuit parallelizes $35$ quantum circuits. As a preliminary step, we first verified that our QMP implementation produced consistent outputs with serial execution by comparing the output of both using a quantum circuit simulator. The outputs for serial and parallel are in perfect agreement. The details of these tests, and others, are given in the Supplemental Material \cite{supplement}. Importantly, in practice, QMP reduces the queuing time of the circuit. Because of the much-reduced number of calls to the QPU, there are far fewer circuits in the queuing system of an individual device.  The advantage of this is two-fold as (i) the wait time is reduced for the user and other users and (ii) schedulers often use a ``fair-share" allowance on the number of jobs submitted by a given user. The fewer number of circuits sent means that job priority levels administered by the scheduler remain high. Finally, we note that because of our choice to use two separate 127 qubit machines, circuits were sent to both in parallel thus, two queues were being occupied at all times which itself leads to a practical speedup. 
\begin{figure}
    \centering
    \includegraphics[width=\linewidth]{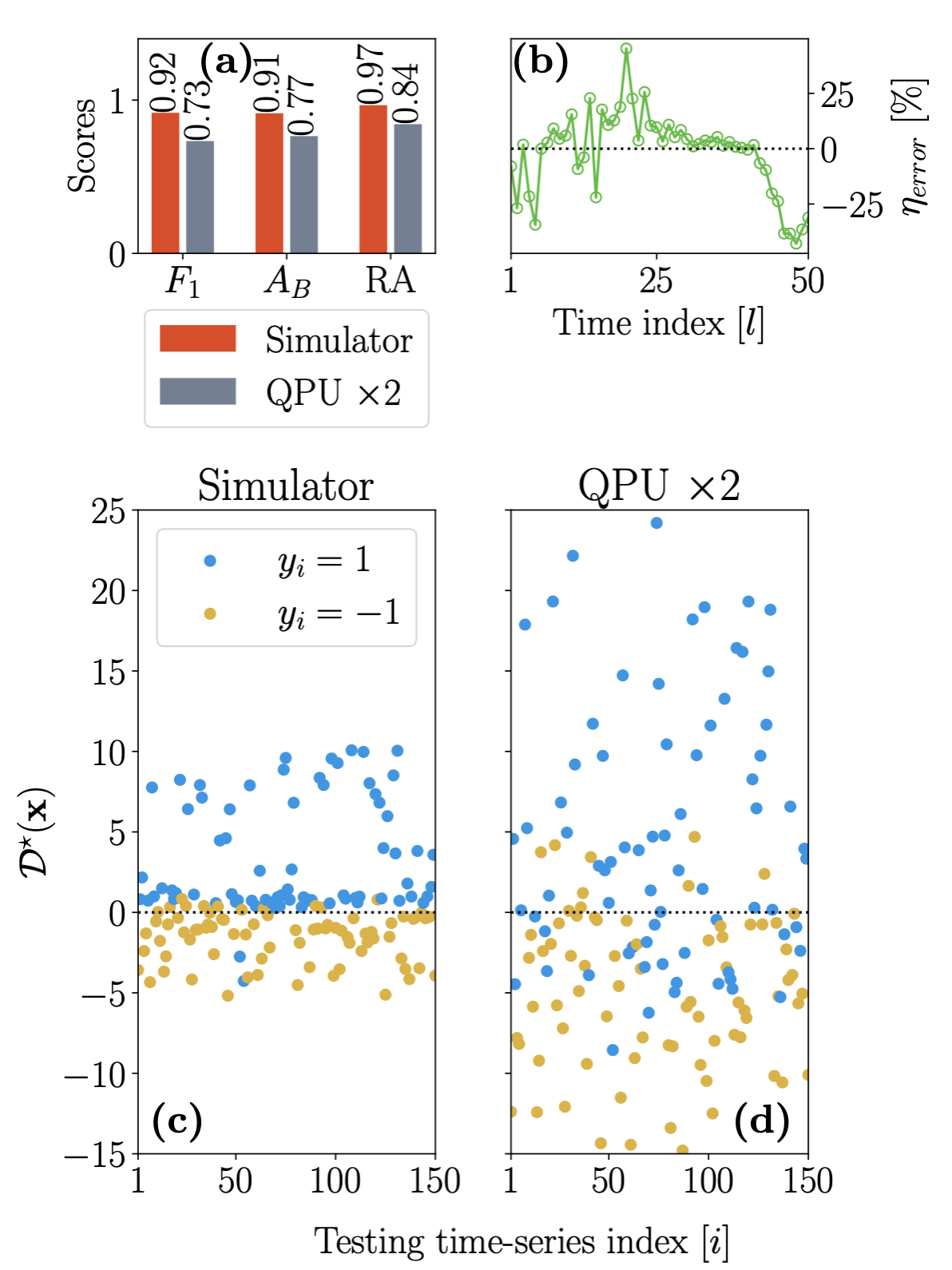}
    \caption{Simulator versus device results on the sub-sampled gun-point data set. (a) The $F_1$, $A_B$ (balanced accuracy) and RA (ROC AUC) scores for the simulator (red) and device (grey; QPU $\times 2$) results. (b) The percentage error in the kernel weights from the device results vs the simulator given as $\bm{\eta}_{error} = 100\cdot(\bm{\eta}_{device} - \bm{\eta}_{ideal})/\bm{\eta}_{ideal}$ plotted as a function of time index $l$. (c-d) The SVM decision function $\mathcal{D}^{\star}(\bm{x})$ for each of the 150 testing time-series using the simulated kernel matrices (c) or the device kernel matrices (d). Points are colored by their truth label. Blue points above the black dotted line are correctly classified while those below it are incorrectly classified. The reverse is true for gold points.}
    \label{fig:device_versus_simulator}
\end{figure}
The results of our QMP experiments are shown in Fig. \ref{fig:device_versus_simulator}. Fig. \ref{fig:device_versus_simulator}(a) compares $F_1$, $A_B$  and RA scores for the simulated results (red bars) and the QMP runs executed on \texttt{ibm\_washington} and \texttt{ibm\_sherbrooke} (QPU $\times 2$; grey bars). Compared with the simulator results from Sec. \ref{subsec:univariate_sim}, we see that MKL SVM on decimated gun-point data set gives rise to lower scores for all three metrics. This is intuitive as the algorithm is being shown fewer data points (100 fewer time points) in training. When run on $2 \times$ QPU, scores fall $\approx$ 10-20\% across the board as a result of all of the various types of hardware noise. We should note that besides the use of buffer qubits and selection of low-error qubits on the machines, no other optimizations/error mitigation techniques were used so the results presented in Fig. \ref{fig:device_versus_simulator} can be considered a baseline. In reality, more highly optimized transpiler passes could be used alongside a whole host of dedicated error mitigation techniques including (but not limited to) Pauli twirling, dynamical decoupling and various flavors of read-out mitigation. Fig. \ref{fig:device_versus_simulator}(b) shows the error in the kernel weights $\bm{\eta}$ as a function of time (i.e. $\eta_t$ is shown). Errors exceeding 25\% occur at low and high $t$ while errors are very small at intermediate $t$. Interestingly (although not shown), areas where the weights themselves are larger (at intermediate $t$) have the smallest errors which means despite a significant amount of noise entering our kernel matrix computations, the kernel weight values are quite robust. Fig. \ref{fig:device_versus_simulator}(c) and \ref{fig:device_versus_simulator}(d) show the SVM decision function $\mathcal{D}^{\star}(\bm{x})$ (discussed in Sec. \ref{subsec:kernel_classifier}) evaluated using all 150 of the testing time-series instances using where the TSHK was either computed with an ideal quantum circuit simulator [Fig. \ref{fig:device_versus_simulator}(c)] or calculated on real devices using QMP  [Fig. \ref{fig:device_versus_simulator} (d)]. Recall that $\text{sgn}[\mathcal{D}^{\star}(\bm{x})]$ distinguishes the predicted binary class label such that points above the black dotted line at $\mathcal{D}^{\star}(\bm{x}) = 0$ are predicted as class 1 while those below it are predicted as class -1. Because each point is colored by its truth label $y_i$, it is possible to see from Fig. \ref{fig:device_versus_simulator}(c) and \ref{fig:device_versus_simulator}(d) which points are correctly and incorrectly classified. Looking first at the simulator results [Fig. \ref{fig:device_versus_simulator}(c) ] it can be seen that many points sit \textit{very close} to $\mathcal{D}^{\star}(\bm{x}) = 0$ meaning their classification is a \textit{close call}. This finding is important as it provides some insight into why performance metrics decrease for the QPU results.  That is, we can see from Fig. \ref{fig:device_versus_simulator}(d) that those \textit{close call} points which were sitting close to $\mathcal{D}^{\star}(\bm{x}) = 0$ for the simulated results have been smeared across the decision boundary by a non-trivial propagation of error incurred by (i) training the SVM with a noisy (and only approximately positive semi-definite) Gram matrix and (ii) making predictions using already noisy trained SVM model using a similarly noisy kernel function. We observe also that those points which were sitting further from the decision boundary in the simulator results now even further from it. \par

Following the approach taken in Ref. \cite{Hubgregson2022}, errors incurred by the processes described in points (i) and (ii) above can be mitigated against. Using the Tikhonov regularization scheme \cite{Hubgregson2022}, Gram matrices $K$ can regularized (be made positive semi-definite) with a simple correction:
\begin{equation}
      K^{\text{reg}} =
    \begin{cases}
      K - \epsilon_{\text{min}} I & \text{if } \epsilon_{\text{min}} < 0\\
      K & \text{otherwise}
    \end{cases}  
    \label{eq: tik_reg}
\end{equation}
for regularized gram matrix $K^{\text{reg}}$, smallest eigenvalue in the spectrum of $K$, $\epsilon_{\text{min}}$ and identity matrix $I$. We apply the correction of Eq. \ref{eq: tik_reg} to each of the noisy time-dependent Gram matrices $K_t$ and use the regularized output to find the optimal kernel weights $\eta_t^{\star}$ given in Eq. \ref{eq:lagrangian_to_weights}. Finally, we build a regularized combined kernel using Eq. \ref{eq:combined_kernel}. We now use the regularized combined kernel as input to a SVM to classify the gun-point data set. We find that all considered metrics are improved: $A_B = 0.81$, $F_1 = 0.82$ and $RA = 0.90$. This shows the success of even simple regularization techniques and supports the need for future works assessing the best possible way to use kernel methods on noisy QPUs.

\section{Conclusions and Perspectives\label{sec:conclusions}}

In this work, we proposed and demonstrated a hybrid quantum-classical supervised ML algorithm for time-series classification. The algorithm works by utilizing time-dependent inner product spaces as generated by a common time-evolution operator and combining the inner products using classical MKL techniques. Together, the components of the algorithm create a temporally-aware QCC-net able to tailor kernel functions for arbitrary time-series data. When these tailored kernel functions were used with kernelized SVM, we found classification performance to be comparable with purely classical methods for the well-known gun-point data set, although, a great deal of the performance stemmed from the classical portions of the algorithm. Through conducting experiments on the synthetic \texttt{moons2circles} data set, we verified that the TSHK is most effective in situations where the correct notion of similarity between pairs of data points evolves with time and there exists a shared time axis in the data as was described in Sec. \ref{subsec:time_dep_inner}.\par

Furthermore, we developed a method to study the resulting time-dependence present within a trained time-series kernel function. We must stress that algorithms trained to explicitly learn time-dependent inner product spaces are novel even within the classical ML literature and indeed, we do not know of an existing classical approach that can precisely emulate the method proposed in this work. \par 

We also found that the total run time of our method can be greatly reduced by utilizing the state-of-the-art method, QMP. That is, because the computation of quantum kernel matrix elements is an \textit{embarrassingly parallel} problem, many matrix elements can be computed in parallel by distributing multiple quantum threads across large QPUs. We demonstrated this parallelism using two 127-qubit superconducting quantum processors to perform SVM classification on a decimated version of the gun-point data set. With this method, we were able to reduce the total number of required QPU calls by $35 \times$, which, assuming constant job queuing times for both 127-qubit QPUs, leads to a $70 \times$ speed-up compared to serial execution on a single QPU.  We have therefore demonstrated, for the first time, that QMP is a valuable tool for significantly speeding up QML algorithms on present and near future NISQ machines. With recent advancements in low-latency CPU-QPU interactions by \texttt{Qiskit Runtime} and others, in the near future, we expect to see the emergence of real-time training of error-mitigated QML where the quantum components of the algorithm are accelerated by QMP and the classical parts are accelerated using conventional parallel processing paradigms like MPI and OpenMP. This work, therefore, makes strides along the path toward the much-anticipated intersection of quantum computation and classical high performance computing.

\section*{Acknowledgments}
We acknowledge the use of IBM Quantum services for this work. The views expressed are those of the authors and do not reflect the official policy or position of IBM or the IBM Quantum team.

\section*{Data and code availability}
The data used to produce Fig. \ref{fig:synthetic_example}, \ref{fig:simulated_gunpoint} and \ref{fig:device_versus_simulator} can be found at \href{https://doi.org/10.5281/zenodo.7996534}{https://doi.org/10.5281/zenodo.7996534} alongside a code example implementing a QCC-net for the Sine vs Cosine classification shown in Fig. \ref{fig:hero_time_kernel}.

\section*{Competing interests}
The authors declare no competing personal or financial interests.

\section*{Funding}
This research used quantum computing resources of the Oak Ridge Leadership Computing Facility, which is a DOE Office of Science User Facility supported under Contract DE-AC05-00OR22725. 
This research used resources of the National Energy Research Scientific Computing Center, a DOE Office of Science User Facility supported by the Office of Science of the U.S. Department of Energy under Contract No. DE-AC02-05CH11231 using NERSC award DDR-ERCAP0024165. 

\section*{Author's contributions}
Jack S. Baker made the figures, conducted the numerical experiments on quantum circuit simulators, wrote the manuscript and helped develop the theoretical method. Gilchan Park conducted the QMP experiments and contributed to the manuscript. Kwangmin Yu oversaw the QML experiments, contributed to the manuscript and the supplemental information. Ara Ghukasayan helped to develop the theoretical method and reviewed the manuscript. Oktay Goktas oversaw the numerical experiments and reviewed the manuscript. Santosh Kumar Radha seeded the idea of using time evolution operators to create  a time dependent quantum kernel function and reviewed the manuscript.

\bibliography{ms.bib}

\clearpage
\newpage
\mbox{~}
\onecolumngrid
\setcounter{equation}{0}
\setcounter{figure}{0}
\setcounter{table}{0}
\setcounter{page}{1}
\setcounter{secnumdepth}{2}

\makeatletter
\renewcommand{\theequation}{S\arabic{equation}}
\renewcommand{\thefigure}{S\arabic{figure}}
\renewcommand{\bibnumfmt}[1]{[S#1]}
\renewcommand{\citenumfont}[1]{S#1}
\vspace*{\fill}
\begin{center}
\textbf{\large Supplemental Material: Parallel hybrid quantum-classical machine learning for kernelized time-series classification}
\end{center}

\section*{Overview}
This document contains supplemental material pertaining to the article "Massively parallel hybrid quantum-classical machine learning for kernelized time-series classification". We begin by presenting a preliminary study regarding the gun-point data set, examining the effects of hyperparameter tuning on model performance and the effects of turning on or off optimization over the parameters of the quantum kernels and the coefficients of the kernels in the multiple kernel combination. We then provide a didactic example for how quantum multiprocessing is conducted with our study and show agreement between our method and standard serial quantum computations.

\section*{Hyperparameter tuning: gun-point data set}

\begin{figure*}[t!]
\centering
\includegraphics[width=\textwidth]{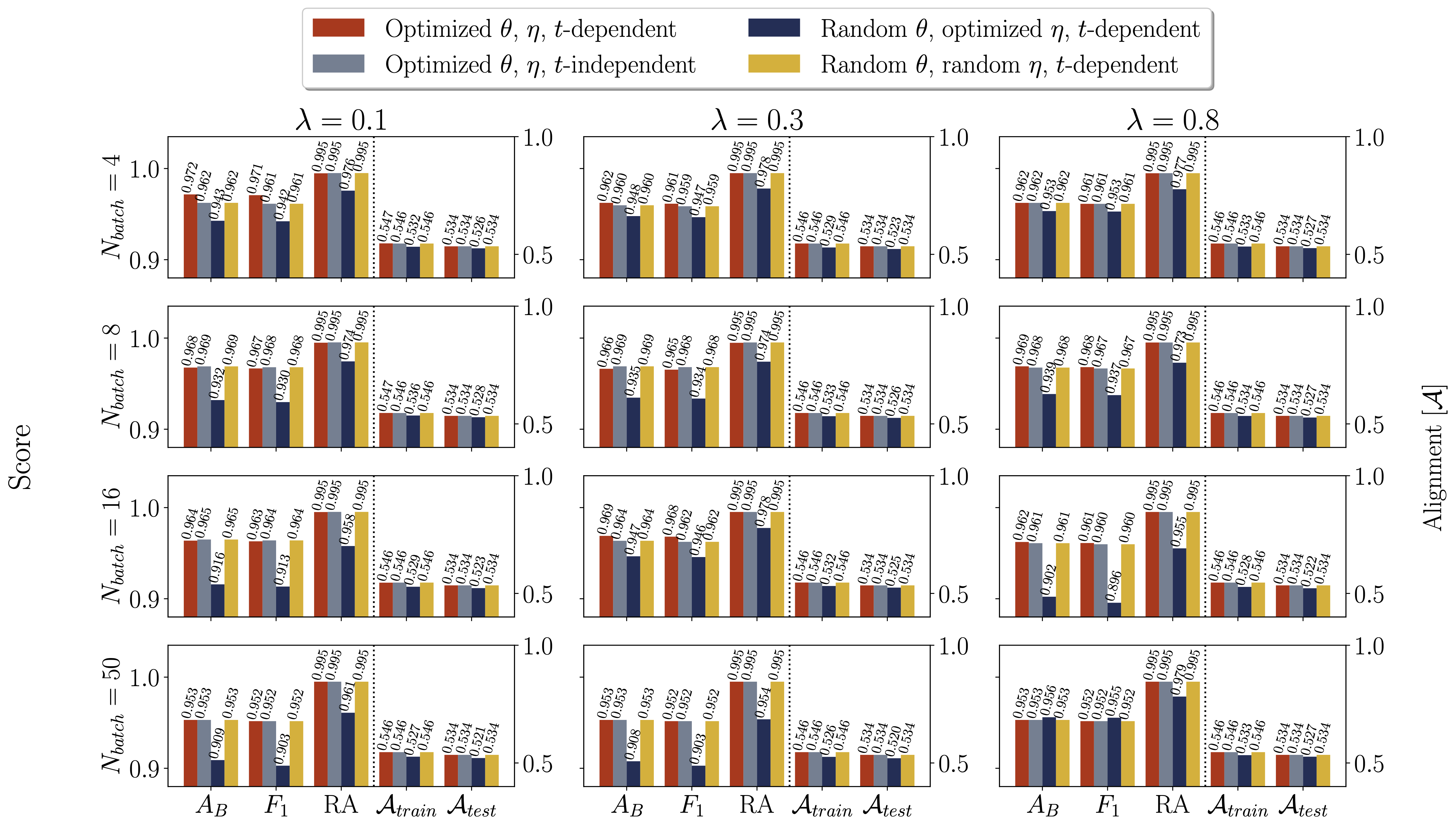}
\caption{A study on the performance of the time-resolved QCC-net for different values of the $\lambda$ and batch size $N_{batch}$ hyperparameters. The study also includes the effects of removing time dependence from the inner product space and from turning on or off various parts of the optimization procedure. }
\label{fig:hyperparam_tune}
\end{figure*}

While the benefits of using our temporally aware QCC-net are clear in the context of the synthetic data set discussed in the main article, realistic data requires more careful assessment to see if (i) the allowance for time dependent inner product spaces or (ii) the optimization over kernel parameters and coefficients has a marked effect when it comes to performance. We conducted such a study and the results are shown in Fig. \ref{fig:hyperparam_tune}. 

The study is structured as follows. Using the gun-point data set described in the main text, we performed an SVM classification on the testing time-series instances given (1) a value of the penalty scaling hyperparameter $\lambda \in \{0.1, 0.3, 0.8\}$, (2) a value for the batch size $N_{batch} \in \{4, 8, 16, 50\}$ and (3) a method for obtaining a combined kernel function used in the SVM classification. Looking further into point (3), we look at four cases: (i) The case where the variational parameters $\bm{\theta}$ and the kernel coefficients $\bm{\eta}$ are optimized and the inner product space is permitted to be time-dependent (red bars on Fig. \ref{fig:hyperparam_tune}); the full form of the QCC net proposed in the main article. (ii) The same optimization as (i) but time-dependence is forbidden in the inner product space so time-dependence is only retained classically (i.e. through the time-dependent kernel weights; grey bars on Fig. \ref{fig:hyperparam_tune}). (iii) Kernels are generated with a random value of $\bm{\theta}$, $\bm{\eta}$ is optimized and time-dependence is allowed in the inner product space (dark blue bars on Fig. \ref{fig:hyperparam_tune}). (iv) Kernels are generated with a random value of $\bm{\theta}$, the kernel weights are $\bm{\eta}$ are fixed as $1/p$ and time-dependence is allowed in the inner product space (gold bars on Fig. \ref{fig:hyperparam_tune}). After each run, we record the balanced accuracy score $A_B$, the $F_1$ scores, the ROC AUC score (RA in shorthand) and the kernel alignments $\mathcal{A}_{train/test}$ given by $1 - 1/N\sum_{i}\sum_{j}|K_{train/test, ij} - K_{truth, ij}|$ where N is the total number of matrix elements, $K_{train/test, ij}$ are the elements of the kernel matrix obtained using the given combination of (1), (2) and (3) on either the training or testing data set and $K_{truth, ij}$ are the truth matrix elements given by $y_i y_j$ for labels $y_i, y_j \in \{1, -1\}$. Each result is given as the average of five random initializations of the parameters $\bm{\theta}$. Those runs where $\bm{\theta}$ is optimized are trained for 500 mini-batch iterations. 

We see that of all the runs, the scenario achieving the best overall performance in all of the considered metrics is low $\lambda$, low $N_{batch}$ and the full optimization over all parameters in a time-dependent QCC-net. Specifically, best performance is seen at $\lambda = 0.1$ and $N_{batch} = 4$ thus motivating the settings used in the main article. It should be noted that gains are small over the other methods and for intermediate $\lambda$ and $N_{batch}$ are often seen at all. There are also some cases where the full time-aware QCC-net is slightly outperformed by the version with the time-independent inner product space (grey bars on Fig. \ref{fig:hyperparam_tune}). It should be noted also that despite all of these variations in $A_B$, $F_1$ and $RA$, we only see very slight variations in $\mathcal{A}_{train}$ and $\mathcal{A}_{test}$. Indeed, this can be rationalized by examining Fig. 6(c) and 6(d) from the main article by examining how many time-series instances sit so close to the decision boundary. It is quite possible that very marginal changes in the kernel alignment bring close call time-series instances over the decision boundary to becoming correctly classified.

\section*{Quantum Multiprocessing}

In the past few years, companies such as Google, IBM, IonQ, Quantinuum, and Rigetti have achieved remarkable advancements in the development of quantum computers.
A number of companies have now launched quantum computers with around 50 qubits, with IBM leading the way with a 127-qubit quantum computer - the largest number of qubits available at the time of writing \cite{IBM_127q}.
Despite the fact that the accuracy of quantum computers is dependent on the implementation technique of the Noisy-Intermediate Scale Quantum (NISQ) \cite{Preskill2018quantumcomputingin} hardware, such as superconducting or trapped ion, regardless of their architecture, they are prone to various noises, errors, and decoherence.
Contemporary quantum devices are restricted in their ability to produce dependable results for practical computing problems. Despite vendors releasing more advanced quantum computers with higher Quantum Volume \cite{moll2018quantum, cross2019validating}, which is a means of quantifying a quantum device's computational power, these devices are still a long way from achieving quantum supremacy for practical problems.
Moreover, the attainment of fault-tolerant quantum computers is currently unfeasible and may remain so for several decades. Consequently, it is imperative to maximize the utilization of NISQ devices to gain a quantum advantage on such devices.

To optimize the utilization of NISQ devices, various methods have been proposed.
Variational Quantum Algorithms (VQAs) \cite{cerezo2021variational}, including the Quantum Approximate Optimization Algorithm (QAOA) \cite{farhi2014quantum} and the Variational Quantum Eigensolver (VQE) \cite{peruzzo2014variational}, have gained significant attention as potential approaches to achieve quantum advantage on Noisy Intermediate-Scale Quantum (NISQ) devices.
An alternative strategy involves enhancing classical-quantum algorithms, such as Quantum Amplitude Estimation (QAE) \cite{aaronson2020quantum, suzuki2020amplitude, yu2020practical,  grinko2021iterative, rao2020quantum} and Grover's search \cite{zhang2020depth, zhang2021implementation,  zhang2022quantum}, so that they can be implemented on NISQ devices.
These methods are aimed at formulating quantum algorithms that are compatible with NISQ devices, but face challenges with respect to long quantum circuit depth. Although these algorithms offer better precision and fidelity due to reduced circuit depth, the use of classical computers in the process weakens the potential for quantum speedup.
\begin{figure*}[t!]
\centering
\includegraphics[width=0.85\textwidth]{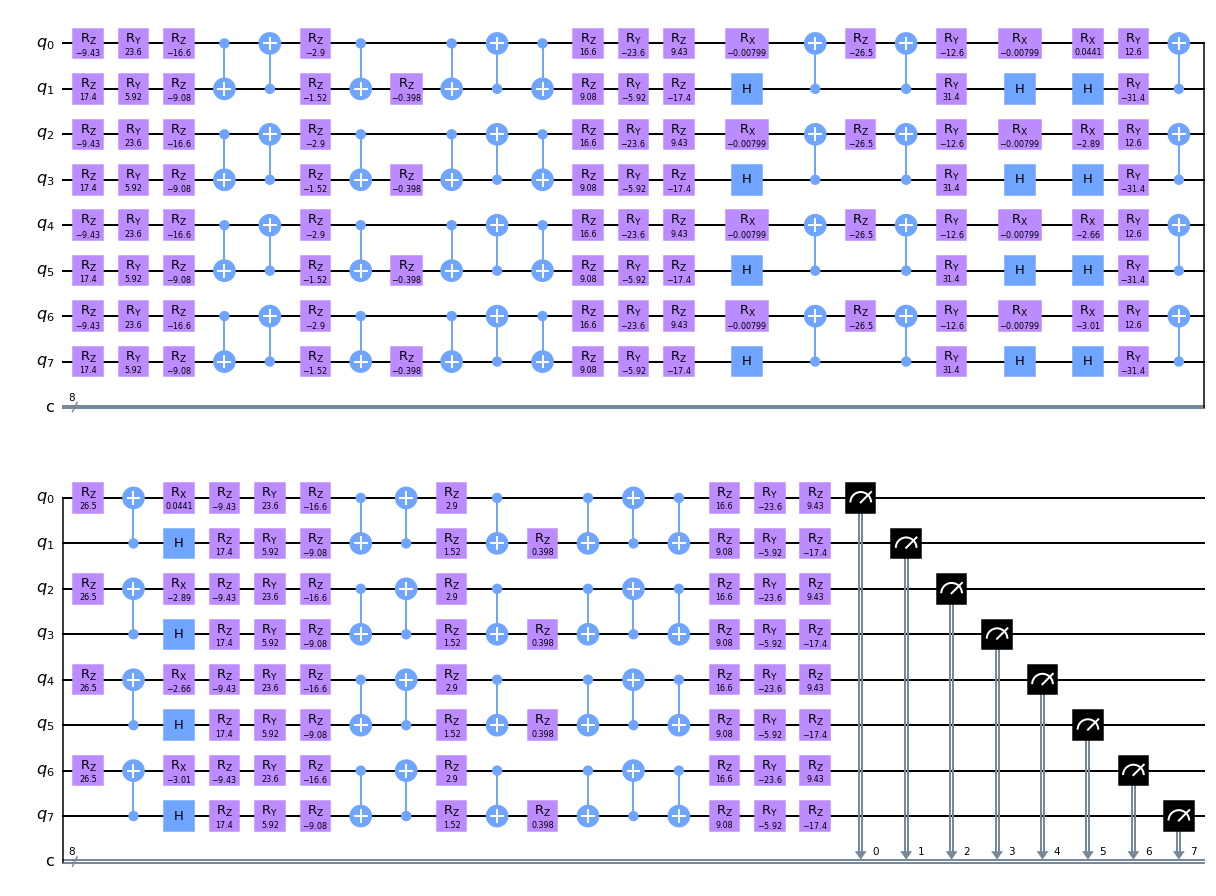}
\caption{Quantum circuit diagram of the QMP. This diagram shows the quantum circuit diagram of the QMP overlapping four independent circuits.}
\label{fig:qmp_circuit}
\end{figure*}

\begin{figure*}[t!]
\centering
  \subfloat[Whole measurement]{%
    \includegraphics[width=0.45\textwidth]{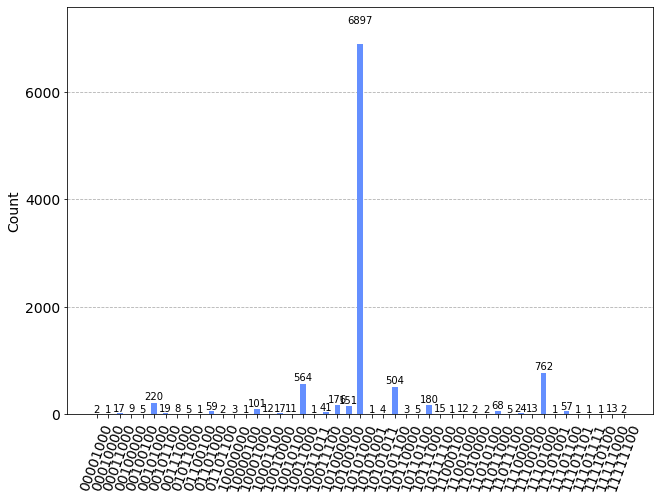}}
    ~ 
  \subfloat[Partial measurement]{%
    \includegraphics[width=0.52\textwidth]{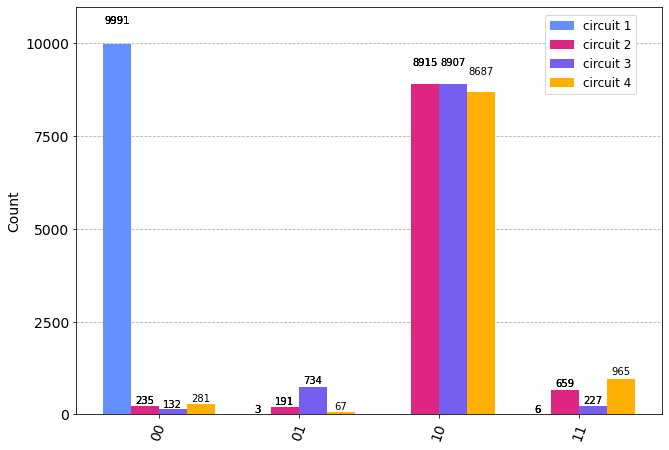}}
  \caption{The measurement histogram of the circuit of Fig. \ref{fig:qmp_circuit} with $10000$ trials (shots). Figure (a) shows the result of the whole measurement of the circuit in Fig. \ref{fig:qmp_circuit}. The result of `circuit 1' is the partial measurement of qubit $q_0$ and $q_1$. This partial measurement in Fig. (b) is extracted by the partial measurement algorithm in Appendix A in Ref. \cite{park2023quantum} based on the whole measurement shown in Fig. (a).}
  \label{fig:qmp_sim_measurement}
\end{figure*}

\begin{figure*}[t!]
\centering
\includegraphics[width=0.52\textwidth]{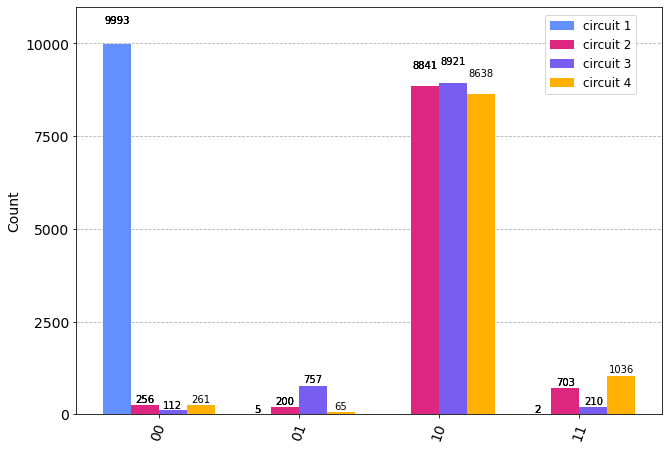}
\caption{The measurement histogram of each circuit (2 qubits) in the circuit diagram in Fig. \ref{fig:qmp_circuit}. This histogram is the results of each circuit consisting of the QMP circuit in Fig. \ref{fig:qmp_circuit}. This histogram is presented to compare with the QMP result shown in Fig. \ref{fig:qmp_sim_measurement} (b). }
\label{fig:sim_measurement_each}
\end{figure*}

\subsection{Example}
\label{sec:example}

In this section, we explain the QMP with an example from the Qiskit simulator. In the bit ordering convention, the rightmost and the uppermost qubits are the least significant bit (LSB) in the bra-ket notation and the quantum circuit, respectively.

Figure \ref{fig:qmp_circuit} shows a QMP quantum circuit diagram of four quantum circuits for the training for the kernelized time-series classification with the gun-point data.
The QMP circuit parallelizes four independent quantum circuits.
The first quantum circuit (`circuit 1') is placed on the qubit $q_0$ and $q_1$, and the second quantum circuit (`circuit 2') is placed on the qubit $q_2$ and $q_3$, respectively.

From the sampling result of $10,000$ trials (shots) shown in Fig. \ref{fig:qmp_sim_measurement} (a), we extract the partial measurement of each circuit. We reimplement the partial measurement code based on the code in Appendix A in Ref. \cite{park2023quantum}.
The partial measurement results of each circuit of the QMP are shown in Fig. \ref{fig:qmp_sim_measurement} (b).
The results of `circuit 1' and `circuit 2' are the partial measurements of qubit $q_0$, $q_1$ and $q_2$, $q_3$, respectively.

The Trial Reduction Factor (TRF) \cite{das2019case} describes the ratio of the number of trials (shots) that executes individually (baseline) to the number of trials (shots) of the QMP circuit.
This factor represents the efficiency of the QMP in terms of trials (shots).
In this example, the TRF is $4$ because each circuit has $10,000$ shots and the total shot number is $40,000$ when the circuits are executed in serial.

\subsection{Result Fidelity}

Result fidelity is a benchmark metric introduced in Ref. \cite{lubinski2021application}. It quantifies the distance between the output probabilistic distribution $P_\text{out}$ with the ideal $P_\text{ideal}$ with the normalization in terms of the uniform distribution $P_\text{uni}$. 
The definition is 
    \begin{equation}
    \label{eq:def_fidelity_1}
         F(P_\text{out},P_\text{ideal}) = \max \{ F_{raw}(P_\text{out},P_\text{ideal}), 0 \}
    \end{equation}
    where
    \begin{equation}
    \label{eq:def_fidelity_2}
         F_{raw}(P_\text{out},P_\text{ideal}) = \frac{f(P_\text{out},P_\text{ideal})-f(P_\text{uni},P_\text{ideal})}{1-f(P_\text{uni},P_\text{ideal})}
    \end{equation}
    with the distance
    \begin{equation}
        f(P_1,P_2) = \left(\sum_j \sqrt{P_1(j)P_2(j)}\right)^2.
    \end{equation}
Note that we have $F(P_\text{ideal},P_\text{ideal})=1$, $F(P_\text{uni},P_\text{ideal}) = 0$, and $F(P_\text{out},P_\text{ideal})\leq 1$.
The result fidelity quantifies the degree of degradation due to the noises.
Hence, we quantify the difference between the QMP output and the outputs of each circuit (2 qubit) execution (comparison between Fig. \ref{fig:qmp_sim_measurement} (b) and Fig. \ref{fig:sim_measurement_each}).
We set the outputs of each circuit execution as the ideal result distribution and compare the QMP outputs.
In this setting, we have $0.99999$, $0.99889$, $0.99986$, and $0.99979$ result fidelity of `circuit 1', `circuit 2', `circuit 3', and `circuit 4', respectively.
When we consider the finite sampling number of the distribution, this result fidelity shows the QMP result is consistent with the independent circuit execution result.

The gun-point data consists of 50 training and 50 test time-series instances. The training data (matrices) for the SVM has symmetrical $50\times50$ dimensions, and the test data has $150\times50$ dimensions. The number of circuits to compute for each training sample is $1275 = (50\times51)/2$, and $7500 = 150\times50$ for each test sample. By applying the QMP, the number of device runs is significantly reduced to 37 and 215 times for each training and test matrix. As a preliminary step, we first verified that the QMP produced consistent outputs with the conventional method (non-QMP) by comparing both performances on an IBM simulator and a device with 35 test circuits. Due to the limit of the maximum number of qubits on a simulator, we used 12 circuits, whereas all 35 circuits were run on a device for the QMP at a time. 
Table \ref{table:fid-scores} demonstrates that QMP and non-QMP generated almost identical outcomes with over 0.99 fidelity on both simulation (NQ\_S vs Q\_S) and real quantum machine (NQ\_D vs Q\_D) environments. The fidelity between a simulator and a device for the QMP (Q\_S vs Q\_D) also showed over 0.95, which is even higher than the non-QMP score, 0.91 (NQ\_S vs NQ\_D).
Since both the simulator and device comparison between QMP and non-QMP show more than 0.99 fidelity, we conclude that the QMP implementation does not have accuracy degradation with respect to the canonical (non-QMP) implementation in the experiments.

\begin{table}[t!]
\caption{\label{table:fid-scores}
Result fidelity \cite{lubinski2021application} scores between Non-QMP (NQ), QMP (Q) on the simulator (S) and device (D).
}
\setlength{\tabcolsep}{10pt} 
\renewcommand{\arraystretch}{1.5} 
\centering
\begin{tabular}{lcccc}
\hline
 & NQ\_S & Q\_S & NQ\_D & Q\_D\\
\hline
NQ\_S    & \textit{NA} & 0.9997 & 0.9134 & 0.91769 \\
Q\_S     & 0.9997 & \textit{NA} & 0.9446 & 0.9508 \\
NQ\_D    & 0.9134 & 0.9446 & \textit{NA} & 0.9909 \\
Q\_D     & 0.9177 & 0.9508 & 0.9909 & \textit{NA} \\
\hline
\end{tabular}
\end{table}

\subsection{Partial Measurement Code}
\begin{lstlisting}[language=Python, caption={A sample code for partial measurement extraction from the whole measurement}, label={list:after}]
import numpy as np
    
def get_partial_measurement(count, least, n):
    """
    parameters:
    - count: return of result.get_counts()
             the total number of counts is the number shots (e.g., 8192)
    - least: the least significant bit index of the partial measurement
    - n    : the number of bits for partial measurement
    """
    cnt = {}
    for i in range(2**n):
        key = f'{{0:0>{n}b}}'.format(i)
        cnt[key] = 0
        
        for m, c in count.items(): # measurement keys, counts
            start_idx = len(m) - (least+n)
            end_idx = len(m) - least

            if key == m[start_idx:end_idx]:
                cnt[key] += c
    return cnt
\end{lstlisting}

\begin{@fileswfalse}
\bibliography{ms.bib}

\begin{thebibliography}{84}%
\makeatletter
\providecommand \@ifxundefined [1]{%
 \@ifx{#1\undefined}
}%
\providecommand \@ifnum [1]{%
 \ifnum #1\expandafter \@firstoftwo
 \else \expandafter \@secondoftwo
 \fi
}%
\providecommand \@ifx [1]{%
 \ifx #1\expandafter \@firstoftwo
 \else \expandafter \@secondoftwo
 \fi
}%
\providecommand \natexlab [1]{#1}%
\providecommand \enquote  [1]{``#1''}%
\providecommand \bibnamefont  [1]{#1}%
\providecommand \bibfnamefont [1]{#1}%
\providecommand \citenamefont [1]{#1}%
\providecommand \href@noop [0]{\@secondoftwo}%
\providecommand \href [0]{\begingroup \@sanitize@url \@href}%
\providecommand \@href[1]{\@@startlink{#1}\@@href}%
\providecommand \@@href[1]{\endgroup#1\@@endlink}%
\providecommand \@sanitize@url [0]{\catcode `\\12\catcode `\$12\catcode
  `\&12\catcode `\#12\catcode `\^12\catcode `\_12\catcode `\%12\relax}%
\providecommand \@@startlink[1]{}%
\providecommand \@@endlink[0]{}%
\providecommand \url  [0]{\begingroup\@sanitize@url \@url }%
\providecommand \@url [1]{\endgroup\@href {#1}{\urlprefix }}%
\providecommand \urlprefix  [0]{URL }%
\providecommand \Eprint [0]{\href }%
\providecommand \doibase [0]{https://doi.org/}%
\providecommand \selectlanguage [0]{\@gobble}%
\providecommand \bibinfo  [0]{\@secondoftwo}%
\providecommand \bibfield  [0]{\@secondoftwo}%
\providecommand \translation [1]{[#1]}%
\providecommand \BibitemOpen [0]{}%
\providecommand \bibitemStop [0]{}%
\providecommand \bibitemNoStop [0]{.\EOS\space}%
\providecommand \EOS [0]{\spacefactor3000\relax}%
\providecommand \BibitemShut  [1]{\csname bibitem#1\endcsname}%
\let\auto@bib@innerbib\@empty
\bibitem [{\citenamefont {Fawaz}\ \emph {et~al.}(2019)\citenamefont {Fawaz},
  \citenamefont {Forestier}, \citenamefont {Weber}, \citenamefont {Idoumghar},\
  and\ \citenamefont {Muller}}]{IsmailFawaz2019}%
  \BibitemOpen
  \bibfield  {author} {\bibinfo {author} {\bibfnamefont {H.~I.}\ \bibnamefont
  {Fawaz}}, \bibinfo {author} {\bibfnamefont {G.}~\bibnamefont {Forestier}},
  \bibinfo {author} {\bibfnamefont {J.}~\bibnamefont {Weber}}, \bibinfo
  {author} {\bibfnamefont {L.}~\bibnamefont {Idoumghar}},\ and\ \bibinfo
  {author} {\bibfnamefont {P.-A.}\ \bibnamefont {Muller}},\ }\bibfield  {title}
  {\bibinfo {title} {Deep learning for time series classification: a review},\
  }\href {https://doi.org/10.1007/s10618-019-00619-1} {\bibfield  {journal}
  {\bibinfo  {journal} {Data Mining and Knowledge Discovery}\ }\textbf
  {\bibinfo {volume} {33}},\ \bibinfo {pages} {917} (\bibinfo {year}
  {2019})}\BibitemShut {NoStop}%
\bibitem [{\citenamefont {Bl{\'{a}}zquez-Garc{\'{\i}}a}\ \emph
  {et~al.}(2021)\citenamefont {Bl{\'{a}}zquez-Garc{\'{\i}}a}, \citenamefont
  {Conde}, \citenamefont {Mori},\ and\ \citenamefont
  {Lozano}}]{BlzquezGarca2021}%
  \BibitemOpen
  \bibfield  {author} {\bibinfo {author} {\bibfnamefont {A.}~\bibnamefont
  {Bl{\'{a}}zquez-Garc{\'{\i}}a}}, \bibinfo {author} {\bibfnamefont
  {A.}~\bibnamefont {Conde}}, \bibinfo {author} {\bibfnamefont
  {U.}~\bibnamefont {Mori}},\ and\ \bibinfo {author} {\bibfnamefont {J.~A.}\
  \bibnamefont {Lozano}},\ }\bibfield  {title} {\bibinfo {title} {A review on
  outlier/anomaly detection in time series data},\ }\href
  {https://doi.org/10.1145/3444690} {\bibfield  {journal} {\bibinfo  {journal}
  {{ACM} Computing Surveys}\ }\textbf {\bibinfo {volume} {54}},\ \bibinfo
  {pages} {1} (\bibinfo {year} {2021})}\BibitemShut {NoStop}%
\bibitem [{\citenamefont {Choi}\ \emph {et~al.}(2021)\citenamefont {Choi},
  \citenamefont {Yi}, \citenamefont {Park},\ and\ \citenamefont
  {Yoon}}]{Choi2021}%
  \BibitemOpen
  \bibfield  {author} {\bibinfo {author} {\bibfnamefont {K.}~\bibnamefont
  {Choi}}, \bibinfo {author} {\bibfnamefont {J.}~\bibnamefont {Yi}}, \bibinfo
  {author} {\bibfnamefont {C.}~\bibnamefont {Park}},\ and\ \bibinfo {author}
  {\bibfnamefont {S.}~\bibnamefont {Yoon}},\ }\bibfield  {title} {\bibinfo
  {title} {Deep learning for anomaly detection in time-series data: Review,
  analysis, and guidelines},\ }\href
  {https://doi.org/10.1109/access.2021.3107975} {\bibfield  {journal} {\bibinfo
   {journal} {{IEEE} Access}\ }\textbf {\bibinfo {volume} {9}},\ \bibinfo
  {pages} {120043} (\bibinfo {year} {2021})}\BibitemShut {NoStop}%
\bibitem [{\citenamefont {Clark}\ \emph {et~al.}(2020)\citenamefont {Clark},
  \citenamefont {Hyndman}, \citenamefont {Pagendam},\ and\ \citenamefont
  {Ryan}}]{Clark2020}%
  \BibitemOpen
  \bibfield  {author} {\bibinfo {author} {\bibfnamefont {S.}~\bibnamefont
  {Clark}}, \bibinfo {author} {\bibfnamefont {R.~J.}\ \bibnamefont {Hyndman}},
  \bibinfo {author} {\bibfnamefont {D.}~\bibnamefont {Pagendam}},\ and\
  \bibinfo {author} {\bibfnamefont {L.~M.}\ \bibnamefont {Ryan}},\ }\bibfield
  {title} {\bibinfo {title} {Modern strategies for time series regression},\
  }\bibfield  {journal} {\bibinfo  {journal} {International Statistical
  Review}\ }\textbf {\bibinfo {volume} {88}},\ \href
  {https://doi.org/10.1111/insr.12432} {10.1111/insr.12432} (\bibinfo {year}
  {2020})\BibitemShut {NoStop}%
\bibitem [{\citenamefont {Deb}\ \emph {et~al.}(2017)\citenamefont {Deb},
  \citenamefont {Zhang}, \citenamefont {Yang}, \citenamefont {Lee},\ and\
  \citenamefont {Shah}}]{Deb2017}%
  \BibitemOpen
  \bibfield  {author} {\bibinfo {author} {\bibfnamefont {C.}~\bibnamefont
  {Deb}}, \bibinfo {author} {\bibfnamefont {F.}~\bibnamefont {Zhang}}, \bibinfo
  {author} {\bibfnamefont {J.}~\bibnamefont {Yang}}, \bibinfo {author}
  {\bibfnamefont {S.~E.}\ \bibnamefont {Lee}},\ and\ \bibinfo {author}
  {\bibfnamefont {K.~W.}\ \bibnamefont {Shah}},\ }\bibfield  {title} {\bibinfo
  {title} {A review on time series forecasting techniques for building energy
  consumption},\ }\href {https://doi.org/10.1016/j.rser.2017.02.085} {\bibfield
   {journal} {\bibinfo  {journal} {Renewable and Sustainable Energy Reviews}\
  }\textbf {\bibinfo {volume} {74}},\ \bibinfo {pages} {902} (\bibinfo {year}
  {2017})}\BibitemShut {NoStop}%
\bibitem [{\citenamefont {Torres}\ \emph {et~al.}(2021)\citenamefont {Torres},
  \citenamefont {Hadjout}, \citenamefont {Sebaa}, \citenamefont
  {Mart{\'{\i}}nez-{\'{A}}lvarez},\ and\ \citenamefont
  {Troncoso}}]{Torres2021}%
  \BibitemOpen
  \bibfield  {author} {\bibinfo {author} {\bibfnamefont {J.~F.}\ \bibnamefont
  {Torres}}, \bibinfo {author} {\bibfnamefont {D.}~\bibnamefont {Hadjout}},
  \bibinfo {author} {\bibfnamefont {A.}~\bibnamefont {Sebaa}}, \bibinfo
  {author} {\bibfnamefont {F.}~\bibnamefont {Mart{\'{\i}}nez-{\'{A}}lvarez}},\
  and\ \bibinfo {author} {\bibfnamefont {A.}~\bibnamefont {Troncoso}},\
  }\bibfield  {title} {\bibinfo {title} {Deep learning for time series
  forecasting: A survey},\ }\href {https://doi.org/10.1089/big.2020.0159}
  {\bibfield  {journal} {\bibinfo  {journal} {Big Data}\ }\textbf {\bibinfo
  {volume} {9}},\ \bibinfo {pages} {3} (\bibinfo {year} {2021})}\BibitemShut
  {NoStop}%
\bibitem [{\citenamefont {Zhang}\ \emph {et~al.}(2018)\citenamefont {Zhang},
  \citenamefont {Kuppannagari}, \citenamefont {Kannan},\ and\ \citenamefont
  {Prasanna}}]{Zhang2018}%
  \BibitemOpen
  \bibfield  {author} {\bibinfo {author} {\bibfnamefont {C.}~\bibnamefont
  {Zhang}}, \bibinfo {author} {\bibfnamefont {S.~R.}\ \bibnamefont
  {Kuppannagari}}, \bibinfo {author} {\bibfnamefont {R.}~\bibnamefont
  {Kannan}},\ and\ \bibinfo {author} {\bibfnamefont {V.~K.}\ \bibnamefont
  {Prasanna}},\ }\bibfield  {title} {\bibinfo {title} {Generative adversarial
  network for synthetic time series data generation in smart grids},\ }in\
  \href {https://doi.org/10.1109/smartgridcomm.2018.8587464} {\emph {\bibinfo
  {booktitle} {2018 {IEEE} International Conference on Communications, Control,
  and Computing Technologies for Smart Grids ({SmartGridComm})}}}\ (\bibinfo
  {publisher} {{IEEE}},\ \bibinfo {year} {2018})\BibitemShut {NoStop}%
\bibitem [{\citenamefont {Hochreiter}\ and\ \citenamefont
  {Schmidhuber}(1997)}]{Hochreiter1997}%
  \BibitemOpen
  \bibfield  {author} {\bibinfo {author} {\bibfnamefont {S.}~\bibnamefont
  {Hochreiter}}\ and\ \bibinfo {author} {\bibfnamefont {J.}~\bibnamefont
  {Schmidhuber}},\ }\bibfield  {title} {\bibinfo {title} {Long short-term
  memory},\ }\href {https://doi.org/10.1162/neco.1997.9.8.1735} {\bibfield
  {journal} {\bibinfo  {journal} {Neural Computation}\ }\textbf {\bibinfo
  {volume} {9}},\ \bibinfo {pages} {1735} (\bibinfo {year} {1997})}\BibitemShut
  {NoStop}%
\bibitem [{\citenamefont {Melis}\ \emph {et~al.}(2019)\citenamefont {Melis},
  \citenamefont {Kočiský},\ and\ \citenamefont {Blunsom}}]{Melis2019}%
  \BibitemOpen
  \bibfield  {author} {\bibinfo {author} {\bibfnamefont {G.}~\bibnamefont
  {Melis}}, \bibinfo {author} {\bibfnamefont {T.}~\bibnamefont {Kočiský}},\
  and\ \bibinfo {author} {\bibfnamefont {P.}~\bibnamefont {Blunsom}},\ }\href
  {https://doi.org/10.48550/ARXIV.1909.01792} {\bibinfo {title} {Mogrifier
  lstm}} (\bibinfo {year} {2019})\BibitemShut {NoStop}%
\bibitem [{\citenamefont {Nguyen}\ \emph {et~al.}(2020)\citenamefont {Nguyen},
  \citenamefont {Baraniuk}, \citenamefont {Bertozzi}, \citenamefont {Osher},\
  and\ \citenamefont {Wang}}]{Nguyen2020}%
  \BibitemOpen
  \bibfield  {author} {\bibinfo {author} {\bibfnamefont {T.}~\bibnamefont
  {Nguyen}}, \bibinfo {author} {\bibfnamefont {R.}~\bibnamefont {Baraniuk}},
  \bibinfo {author} {\bibfnamefont {A.}~\bibnamefont {Bertozzi}}, \bibinfo
  {author} {\bibfnamefont {S.}~\bibnamefont {Osher}},\ and\ \bibinfo {author}
  {\bibfnamefont {B.}~\bibnamefont {Wang}},\ }\bibfield  {title} {\bibinfo
  {title} {Momentumrnn: Integrating momentum into recurrent neural networks},\
  }in\ \href@noop {} {\emph {\bibinfo {booktitle} {Advances in Neural
  Information Processing Systems}}},\ Vol.~\bibinfo {volume} {33},\ \bibinfo
  {editor} {edited by\ \bibinfo {editor} {\bibfnamefont {H.}~\bibnamefont
  {Larochelle}}, \bibinfo {editor} {\bibfnamefont {M.}~\bibnamefont {Ranzato}},
  \bibinfo {editor} {\bibfnamefont {R.}~\bibnamefont {Hadsell}}, \bibinfo
  {editor} {\bibfnamefont {M.}~\bibnamefont {Balcan}},\ and\ \bibinfo {editor}
  {\bibfnamefont {H.}~\bibnamefont {Lin}}}\ (\bibinfo  {publisher} {Curran
  Associates, Inc.},\ \bibinfo {year} {2020})\ pp.\ \bibinfo {pages}
  {1924--1936}\BibitemShut {NoStop}%
\bibitem [{\citenamefont {Vaswani}\ \emph {et~al.}(2017)\citenamefont
  {Vaswani}, \citenamefont {Shazeer}, \citenamefont {Parmar}, \citenamefont
  {Uszkoreit}, \citenamefont {Jones}, \citenamefont {Gomez}, \citenamefont
  {Kaiser},\ and\ \citenamefont {Polosukhin}}]{Vaswani2017}%
  \BibitemOpen
  \bibfield  {author} {\bibinfo {author} {\bibfnamefont {A.}~\bibnamefont
  {Vaswani}}, \bibinfo {author} {\bibfnamefont {N.}~\bibnamefont {Shazeer}},
  \bibinfo {author} {\bibfnamefont {N.}~\bibnamefont {Parmar}}, \bibinfo
  {author} {\bibfnamefont {J.}~\bibnamefont {Uszkoreit}}, \bibinfo {author}
  {\bibfnamefont {L.}~\bibnamefont {Jones}}, \bibinfo {author} {\bibfnamefont
  {A.~N.}\ \bibnamefont {Gomez}}, \bibinfo {author} {\bibfnamefont {L.~u.}\
  \bibnamefont {Kaiser}},\ and\ \bibinfo {author} {\bibfnamefont
  {I.}~\bibnamefont {Polosukhin}},\ }\bibfield  {title} {\bibinfo {title}
  {Attention is all you need},\ }in\ \href@noop {} {\emph {\bibinfo {booktitle}
  {Advances in Neural Information Processing Systems}}},\ Vol.~\bibinfo
  {volume} {30},\ \bibinfo {editor} {edited by\ \bibinfo {editor}
  {\bibfnamefont {I.}~\bibnamefont {Guyon}}, \bibinfo {editor} {\bibfnamefont
  {U.~V.}\ \bibnamefont {Luxburg}}, \bibinfo {editor} {\bibfnamefont
  {S.}~\bibnamefont {Bengio}}, \bibinfo {editor} {\bibfnamefont
  {H.}~\bibnamefont {Wallach}}, \bibinfo {editor} {\bibfnamefont
  {R.}~\bibnamefont {Fergus}}, \bibinfo {editor} {\bibfnamefont
  {S.}~\bibnamefont {Vishwanathan}},\ and\ \bibinfo {editor} {\bibfnamefont
  {R.}~\bibnamefont {Garnett}}}\ (\bibinfo  {publisher} {Curran Associates,
  Inc.},\ \bibinfo {year} {2017})\BibitemShut {NoStop}%
\bibitem [{\citenamefont {Yang}\ \emph {et~al.}(2021)\citenamefont {Yang},
  \citenamefont {Tsai},\ and\ \citenamefont {Chen}}]{yang2021}%
  \BibitemOpen
  \bibfield  {author} {\bibinfo {author} {\bibfnamefont {C.-H.~H.}\
  \bibnamefont {Yang}}, \bibinfo {author} {\bibfnamefont {Y.-Y.}\ \bibnamefont
  {Tsai}},\ and\ \bibinfo {author} {\bibfnamefont {P.-Y.}\ \bibnamefont
  {Chen}},\ }\bibfield  {title} {\bibinfo {title} {Voice2series: Reprogramming
  acoustic models for time series classification},\ }in\ \href@noop {} {\emph
  {\bibinfo {booktitle} {International Conference on Machine Learning}}}\
  (\bibinfo {organization} {PMLR},\ \bibinfo {year} {2021})\ pp.\ \bibinfo
  {pages} {11808--11819}\BibitemShut {NoStop}%
\bibitem [{\citenamefont {Zerveas}\ \emph {et~al.}(2021)\citenamefont
  {Zerveas}, \citenamefont {Jayaraman}, \citenamefont {Patel}, \citenamefont
  {Bhamidipaty},\ and\ \citenamefont {Eickhoff}}]{Zerveas2021}%
  \BibitemOpen
  \bibfield  {author} {\bibinfo {author} {\bibfnamefont {G.}~\bibnamefont
  {Zerveas}}, \bibinfo {author} {\bibfnamefont {S.}~\bibnamefont {Jayaraman}},
  \bibinfo {author} {\bibfnamefont {D.}~\bibnamefont {Patel}}, \bibinfo
  {author} {\bibfnamefont {A.}~\bibnamefont {Bhamidipaty}},\ and\ \bibinfo
  {author} {\bibfnamefont {C.}~\bibnamefont {Eickhoff}},\ }\bibfield  {title}
  {\bibinfo {title} {A transformer-based framework for multivariate time series
  representation learning},\ }in\ \href
  {https://doi.org/10.1145/3447548.3467401} {\emph {\bibinfo {booktitle}
  {Proceedings of the 27th {ACM} {SIGKDD} Conference on Knowledge Discovery;
  Data Mining}}}\ (\bibinfo  {publisher} {{ACM}},\ \bibinfo {year}
  {2021})\BibitemShut {NoStop}%
\bibitem [{\citenamefont {Cortes}\ and\ \citenamefont
  {Vapnik}(1995)}]{cortes1995support}%
  \BibitemOpen
  \bibfield  {author} {\bibinfo {author} {\bibfnamefont {C.}~\bibnamefont
  {Cortes}}\ and\ \bibinfo {author} {\bibfnamefont {V.}~\bibnamefont
  {Vapnik}},\ }\bibfield  {title} {\bibinfo {title} {Support-vector networks},\
  }\href@noop {} {\bibfield  {journal} {\bibinfo  {journal} {Machine learning}\
  }\textbf {\bibinfo {volume} {20}},\ \bibinfo {pages} {273} (\bibinfo {year}
  {1995})}\BibitemShut {NoStop}%
\bibitem [{\citenamefont {Badiane}\ \emph {et~al.}(2018)\citenamefont
  {Badiane}, \citenamefont {O'Reilly},\ and\ \citenamefont
  {Cunningham}}]{Badiane2018}%
  \BibitemOpen
  \bibfield  {author} {\bibinfo {author} {\bibfnamefont {M.}~\bibnamefont
  {Badiane}}, \bibinfo {author} {\bibfnamefont {M.}~\bibnamefont {O'Reilly}},\
  and\ \bibinfo {author} {\bibfnamefont {P.}~\bibnamefont {Cunningham}},\
  }\bibfield  {title} {\bibinfo {title} {Kernel methods for time series
  classification and regression.},\ }in\ \href@noop {} {\emph {\bibinfo
  {booktitle} {AICS}}}\ (\bibinfo {year} {2018})\ pp.\ \bibinfo {pages}
  {54--65}\BibitemShut {NoStop}%
\bibitem [{\citenamefont {Bailly}(2018)}]{bailly2018time}%
  \BibitemOpen
  \bibfield  {author} {\bibinfo {author} {\bibfnamefont {A.}~\bibnamefont
  {Bailly}},\ }\emph {\bibinfo {title} {Time series classification algorithms
  with applications in remote sensing}},\ \href@noop {} {Ph.D. thesis},\
  \bibinfo  {school} {Universit{\'e} Rennes 2} (\bibinfo {year}
  {2018})\BibitemShut {NoStop}%
\bibitem [{\citenamefont {F{\'a}bregues de~los
  Santos}(2017)}]{fabregues2017forecasting}%
  \BibitemOpen
  \bibfield  {author} {\bibinfo {author} {\bibfnamefont {L.}~\bibnamefont
  {F{\'a}bregues de~los Santos}},\ }\emph {\bibinfo {title} {Forecasting
  financial time series using multiple Kernel Learning}},\ \href@noop {}
  {Master's thesis},\ \bibinfo  {school} {Universitat Polit{\`e}cnica de
  Catalunya} (\bibinfo {year} {2017})\BibitemShut {NoStop}%
\bibitem [{\citenamefont {R{\"u}ping}(2001)}]{ruping2001svm}%
  \BibitemOpen
  \bibfield  {author} {\bibinfo {author} {\bibfnamefont {S.}~\bibnamefont
  {R{\"u}ping}},\ }\href@noop {} {\emph {\bibinfo {title} {SVM kernels for time
  series analysis}}},\ \bibinfo {type} {Tech. Rep.}\ (\bibinfo  {institution}
  {Technical report},\ \bibinfo {year} {2001})\BibitemShut {NoStop}%
\bibitem [{\citenamefont {Tino}(2020)}]{tino2020dynamical}%
  \BibitemOpen
  \bibfield  {author} {\bibinfo {author} {\bibfnamefont {P.}~\bibnamefont
  {Tino}},\ }\bibfield  {title} {\bibinfo {title} {Dynamical systems as
  temporal feature spaces},\ }\href@noop {} {\bibfield  {journal} {\bibinfo
  {journal} {The Journal of Machine Learning Research}\ }\textbf {\bibinfo
  {volume} {21}},\ \bibinfo {pages} {1649} (\bibinfo {year}
  {2020})}\BibitemShut {NoStop}%
\bibitem [{\citenamefont {G{\"o}nen}\ and\ \citenamefont
  {Alpayd{\i}n}(2011{\natexlab{a}})}]{gonen2011multiple}%
  \BibitemOpen
  \bibfield  {author} {\bibinfo {author} {\bibfnamefont {M.}~\bibnamefont
  {G{\"o}nen}}\ and\ \bibinfo {author} {\bibfnamefont {E.}~\bibnamefont
  {Alpayd{\i}n}},\ }\bibfield  {title} {\bibinfo {title} {Multiple kernel
  learning algorithms},\ }\href@noop {} {\bibfield  {journal} {\bibinfo
  {journal} {The Journal of Machine Learning Research}\ }\textbf {\bibinfo
  {volume} {12}},\ \bibinfo {pages} {2211} (\bibinfo {year}
  {2011}{\natexlab{a}})}\BibitemShut {NoStop}%
\bibitem [{\citenamefont {Aiolli}\ and\ \citenamefont
  {Donini}(2015)}]{Aiolli2015}%
  \BibitemOpen
  \bibfield  {author} {\bibinfo {author} {\bibfnamefont {F.}~\bibnamefont
  {Aiolli}}\ and\ \bibinfo {author} {\bibfnamefont {M.}~\bibnamefont
  {Donini}},\ }\bibfield  {title} {\bibinfo {title} {{EasyMKL}: a scalable
  multiple kernel learning algorithm},\ }\href
  {https://doi.org/10.1016/j.neucom.2014.11.078} {\bibfield  {journal}
  {\bibinfo  {journal} {Neurocomputing}\ }\textbf {\bibinfo {volume} {169}},\
  \bibinfo {pages} {215} (\bibinfo {year} {2015})}\BibitemShut {NoStop}%
\bibitem [{\citenamefont {Ghukasyan}\ \emph {et~al.}(2023)\citenamefont
  {Ghukasyan}, \citenamefont {Baker}, \citenamefont {Goktas}, \citenamefont
  {Carrasquilla},\ and\ \citenamefont {Radha}}]{Ghukasayan2023}%
  \BibitemOpen
  \bibfield  {author} {\bibinfo {author} {\bibfnamefont {A.}~\bibnamefont
  {Ghukasyan}}, \bibinfo {author} {\bibfnamefont {J.~S.}\ \bibnamefont
  {Baker}}, \bibinfo {author} {\bibfnamefont {O.}~\bibnamefont {Goktas}},
  \bibinfo {author} {\bibfnamefont {J.}~\bibnamefont {Carrasquilla}},\ and\
  \bibinfo {author} {\bibfnamefont {S.~K.}\ \bibnamefont {Radha}},\ }\href@noop
  {} {\bibinfo {title} {Quantum-classical multiple kernel learning}} (\bibinfo
  {year} {2023}),\ \Eprint {https://arxiv.org/abs/2305.17707} {arXiv:2305.17707
  [quant-ph]} \BibitemShut {NoStop}%
\bibitem [{\citenamefont {Bharti}\ \emph {et~al.}(2022)\citenamefont {Bharti},
  \citenamefont {Cervera-Lierta}, \citenamefont {Kyaw}, \citenamefont {Haug},
  \citenamefont {Alperin-Lea}, \citenamefont {Anand}, \citenamefont {Degroote},
  \citenamefont {Heimonen}, \citenamefont {Kottmann}, \citenamefont {Menke},
  \citenamefont {Mok}, \citenamefont {Sim}, \citenamefont {Kwek},\ and\
  \citenamefont {Aspuru-Guzik}}]{Bharti2022}%
  \BibitemOpen
  \bibfield  {author} {\bibinfo {author} {\bibfnamefont {K.}~\bibnamefont
  {Bharti}}, \bibinfo {author} {\bibfnamefont {A.}~\bibnamefont
  {Cervera-Lierta}}, \bibinfo {author} {\bibfnamefont {T.~H.}\ \bibnamefont
  {Kyaw}}, \bibinfo {author} {\bibfnamefont {T.}~\bibnamefont {Haug}}, \bibinfo
  {author} {\bibfnamefont {S.}~\bibnamefont {Alperin-Lea}}, \bibinfo {author}
  {\bibfnamefont {A.}~\bibnamefont {Anand}}, \bibinfo {author} {\bibfnamefont
  {M.}~\bibnamefont {Degroote}}, \bibinfo {author} {\bibfnamefont
  {H.}~\bibnamefont {Heimonen}}, \bibinfo {author} {\bibfnamefont {J.~S.}\
  \bibnamefont {Kottmann}}, \bibinfo {author} {\bibfnamefont {T.}~\bibnamefont
  {Menke}}, \bibinfo {author} {\bibfnamefont {W.-K.}\ \bibnamefont {Mok}},
  \bibinfo {author} {\bibfnamefont {S.}~\bibnamefont {Sim}}, \bibinfo {author}
  {\bibfnamefont {L.-C.}\ \bibnamefont {Kwek}},\ and\ \bibinfo {author}
  {\bibfnamefont {A.}~\bibnamefont {Aspuru-Guzik}},\ }\bibfield  {title}
  {\bibinfo {title} {Noisy intermediate-scale quantum algorithms},\ }\href
  {https://doi.org/10.1103/RevModPhys.94.015004} {\bibfield  {journal}
  {\bibinfo  {journal} {Rev. Mod. Phys.}\ }\textbf {\bibinfo {volume} {94}},\
  \bibinfo {pages} {015004} (\bibinfo {year} {2022})}\BibitemShut {NoStop}%
\bibitem [{\citenamefont {Wiebe}\ \emph {et~al.}(2010)\citenamefont {Wiebe},
  \citenamefont {Berry}, \citenamefont {H{\o}yer},\ and\ \citenamefont
  {Sanders}}]{Wiebe2010}%
  \BibitemOpen
  \bibfield  {author} {\bibinfo {author} {\bibfnamefont {N.}~\bibnamefont
  {Wiebe}}, \bibinfo {author} {\bibfnamefont {D.}~\bibnamefont {Berry}},
  \bibinfo {author} {\bibfnamefont {P.}~\bibnamefont {H{\o}yer}},\ and\
  \bibinfo {author} {\bibfnamefont {B.~C.}\ \bibnamefont {Sanders}},\
  }\bibfield  {title} {\bibinfo {title} {Higher order decompositions of ordered
  operator exponentials},\ }\href
  {https://doi.org/10.1088/1751-8113/43/6/065203} {\bibfield  {journal}
  {\bibinfo  {journal} {Journal of Physics A: Mathematical and Theoretical}\
  }\textbf {\bibinfo {volume} {43}},\ \bibinfo {pages} {065203} (\bibinfo
  {year} {2010})}\BibitemShut {NoStop}%
\bibitem [{\citenamefont {C{\^{\i}}rstoiu}\ \emph {et~al.}(2020)\citenamefont
  {C{\^{\i}}rstoiu}, \citenamefont {Holmes}, \citenamefont {Iosue},
  \citenamefont {Cincio}, \citenamefont {Coles},\ and\ \citenamefont
  {Sornborger}}]{Crstoiu2020}%
  \BibitemOpen
  \bibfield  {author} {\bibinfo {author} {\bibfnamefont {C.}~\bibnamefont
  {C{\^{\i}}rstoiu}}, \bibinfo {author} {\bibfnamefont {Z.}~\bibnamefont
  {Holmes}}, \bibinfo {author} {\bibfnamefont {J.}~\bibnamefont {Iosue}},
  \bibinfo {author} {\bibfnamefont {L.}~\bibnamefont {Cincio}}, \bibinfo
  {author} {\bibfnamefont {P.~J.}\ \bibnamefont {Coles}},\ and\ \bibinfo
  {author} {\bibfnamefont {A.}~\bibnamefont {Sornborger}},\ }\bibfield  {title}
  {\bibinfo {title} {Variational fast forwarding for quantum simulation beyond
  the coherence time},\ }\bibfield  {journal} {\bibinfo  {journal} {npj Quantum
  Information}\ }\textbf {\bibinfo {volume} {6}},\ \href
  {https://doi.org/10.1038/s41534-020-00302-0} {10.1038/s41534-020-00302-0}
  (\bibinfo {year} {2020})\BibitemShut {NoStop}%
\bibitem [{\citenamefont {Radha}(2021)}]{Radha2021}%
  \BibitemOpen
  \bibfield  {author} {\bibinfo {author} {\bibfnamefont {S.~K.}\ \bibnamefont
  {Radha}},\ }\href {https://doi.org/10.48550/ARXIV.2105.06770} {\bibinfo
  {title} {Quantum constraint learning for quantum approximate optimization
  algorithm}} (\bibinfo {year} {2021})\BibitemShut {NoStop}%
\bibitem [{\citenamefont {Horowitz}\ \emph {et~al.}(2022)\citenamefont
  {Horowitz}, \citenamefont {Rao},\ and\ \citenamefont {Radha}}]{Horowitz2022}%
  \BibitemOpen
  \bibfield  {author} {\bibinfo {author} {\bibfnamefont {H.}~\bibnamefont
  {Horowitz}}, \bibinfo {author} {\bibfnamefont {P.}~\bibnamefont {Rao}},\ and\
  \bibinfo {author} {\bibfnamefont {S.~K.}\ \bibnamefont {Radha}},\ }\href
  {https://doi.org/10.48550/ARXIV.2204.06150} {\bibinfo {title} {A quantum
  generative model for multi-dimensional time series using hamiltonian
  learning}} (\bibinfo {year} {2022})\BibitemShut {NoStop}%
\bibitem [{\citenamefont {Baker}\ \emph {et~al.}(2022)\citenamefont {Baker},
  \citenamefont {Horowitz}, \citenamefont {Radha}, \citenamefont {Fernandes},
  \citenamefont {Jones}, \citenamefont {Noorani}, \citenamefont {Skavysh},
  \citenamefont {Lamontangne},\ and\ \citenamefont {Sanders}}]{Baker2022}%
  \BibitemOpen
  \bibfield  {author} {\bibinfo {author} {\bibfnamefont {J.~S.}\ \bibnamefont
  {Baker}}, \bibinfo {author} {\bibfnamefont {H.}~\bibnamefont {Horowitz}},
  \bibinfo {author} {\bibfnamefont {S.~K.}\ \bibnamefont {Radha}}, \bibinfo
  {author} {\bibfnamefont {S.}~\bibnamefont {Fernandes}}, \bibinfo {author}
  {\bibfnamefont {C.}~\bibnamefont {Jones}}, \bibinfo {author} {\bibfnamefont
  {N.}~\bibnamefont {Noorani}}, \bibinfo {author} {\bibfnamefont
  {V.}~\bibnamefont {Skavysh}}, \bibinfo {author} {\bibfnamefont
  {P.}~\bibnamefont {Lamontangne}},\ and\ \bibinfo {author} {\bibfnamefont
  {B.~C.}\ \bibnamefont {Sanders}},\ }\href
  {https://doi.org/10.48550/ARXIV.2210.16438} {\bibinfo {title} {Quantum
  variational rewinding for time series anomaly detection}} (\bibinfo {year}
  {2022})\BibitemShut {NoStop}%
\bibitem [{\citenamefont {Gibbs}\ \emph {et~al.}(2022)\citenamefont {Gibbs},
  \citenamefont {Holmes}, \citenamefont {Caro}, \citenamefont {Ezzell},
  \citenamefont {Huang}, \citenamefont {Cincio}, \citenamefont {Sornborger},\
  and\ \citenamefont {Coles}}]{Gibbs2022}%
  \BibitemOpen
  \bibfield  {author} {\bibinfo {author} {\bibfnamefont {J.}~\bibnamefont
  {Gibbs}}, \bibinfo {author} {\bibfnamefont {Z.}~\bibnamefont {Holmes}},
  \bibinfo {author} {\bibfnamefont {M.~C.}\ \bibnamefont {Caro}}, \bibinfo
  {author} {\bibfnamefont {N.}~\bibnamefont {Ezzell}}, \bibinfo {author}
  {\bibfnamefont {H.-Y.}\ \bibnamefont {Huang}}, \bibinfo {author}
  {\bibfnamefont {L.}~\bibnamefont {Cincio}}, \bibinfo {author} {\bibfnamefont
  {A.~T.}\ \bibnamefont {Sornborger}},\ and\ \bibinfo {author} {\bibfnamefont
  {P.~J.}\ \bibnamefont {Coles}},\ }\href
  {https://doi.org/10.48550/ARXIV.2204.10269} {\bibinfo {title} {Dynamical
  simulation via quantum machine learning with provable generalization}}
  (\bibinfo {year} {2022})\BibitemShut {NoStop}%
\bibitem [{\citenamefont {Caro}\ \emph {et~al.}(2022)\citenamefont {Caro},
  \citenamefont {Huang}, \citenamefont {Ezzell}, \citenamefont {Gibbs},
  \citenamefont {Sornborger}, \citenamefont {Cincio}, \citenamefont {Coles},\
  and\ \citenamefont {Holmes}}]{Caro2022}%
  \BibitemOpen
  \bibfield  {author} {\bibinfo {author} {\bibfnamefont {M.~C.}\ \bibnamefont
  {Caro}}, \bibinfo {author} {\bibfnamefont {H.-Y.}\ \bibnamefont {Huang}},
  \bibinfo {author} {\bibfnamefont {N.}~\bibnamefont {Ezzell}}, \bibinfo
  {author} {\bibfnamefont {J.}~\bibnamefont {Gibbs}}, \bibinfo {author}
  {\bibfnamefont {A.~T.}\ \bibnamefont {Sornborger}}, \bibinfo {author}
  {\bibfnamefont {L.}~\bibnamefont {Cincio}}, \bibinfo {author} {\bibfnamefont
  {P.~J.}\ \bibnamefont {Coles}},\ and\ \bibinfo {author} {\bibfnamefont
  {Z.}~\bibnamefont {Holmes}},\ }\href
  {https://doi.org/10.48550/ARXIV.2204.10268} {\bibinfo {title}
  {Out-of-distribution generalization for learning quantum dynamics}} (\bibinfo
  {year} {2022})\BibitemShut {NoStop}%
\bibitem [{\citenamefont {Das}\ \emph {et~al.}(2019)\citenamefont {Das},
  \citenamefont {Tannu}, \citenamefont {Nair},\ and\ \citenamefont
  {Qureshi}}]{das2019case}%
  \BibitemOpen
  \bibfield  {author} {\bibinfo {author} {\bibfnamefont {P.}~\bibnamefont
  {Das}}, \bibinfo {author} {\bibfnamefont {S.~S.}\ \bibnamefont {Tannu}},
  \bibinfo {author} {\bibfnamefont {P.~J.}\ \bibnamefont {Nair}},\ and\
  \bibinfo {author} {\bibfnamefont {M.}~\bibnamefont {Qureshi}},\ }\bibfield
  {title} {\bibinfo {title} {A case for multi-programming quantum computers},\
  }in\ \href@noop {} {\emph {\bibinfo {booktitle} {Proceedings of the 52nd
  Annual IEEE/ACM International Symposium on Microarchitecture}}}\ (\bibinfo
  {year} {2019})\ pp.\ \bibinfo {pages} {291--303}\BibitemShut {NoStop}%
\bibitem [{\citenamefont {Ratanamahatana}\ and\ \citenamefont
  {Keogh}(2005)}]{ratanamahatana2005three}%
  \BibitemOpen
  \bibfield  {author} {\bibinfo {author} {\bibfnamefont {C.~A.}\ \bibnamefont
  {Ratanamahatana}}\ and\ \bibinfo {author} {\bibfnamefont {E.}~\bibnamefont
  {Keogh}},\ }\bibfield  {title} {\bibinfo {title} {Three myths about dynamic
  time warping data mining},\ }in\ \href@noop {} {\emph {\bibinfo {booktitle}
  {Proceedings of the 2005 SIAM international conference on data mining}}}\
  (\bibinfo {organization} {SIAM},\ \bibinfo {year} {2005})\ pp.\ \bibinfo
  {pages} {506--510}\BibitemShut {NoStop}%
\bibitem [{\citenamefont {G{\"o}nen}\ and\ \citenamefont
  {Alpayd{\i}n}(2011{\natexlab{b}})}]{Gonen2011}%
  \BibitemOpen
  \bibfield  {author} {\bibinfo {author} {\bibfnamefont {M.}~\bibnamefont
  {G{\"o}nen}}\ and\ \bibinfo {author} {\bibfnamefont {E.}~\bibnamefont
  {Alpayd{\i}n}},\ }\bibfield  {title} {\bibinfo {title} {Multiple kernel
  learning algorithms},\ }\href@noop {} {\bibfield  {journal} {\bibinfo
  {journal} {The Journal of Machine Learning Research}\ }\textbf {\bibinfo
  {volume} {12}},\ \bibinfo {pages} {2211} (\bibinfo {year}
  {2011}{\natexlab{b}})}\BibitemShut {NoStop}%
\bibitem [{\citenamefont {Schuld}\ and\ \citenamefont
  {Killoran}(2019)}]{Schuld2019}%
  \BibitemOpen
  \bibfield  {author} {\bibinfo {author} {\bibfnamefont {M.}~\bibnamefont
  {Schuld}}\ and\ \bibinfo {author} {\bibfnamefont {N.}~\bibnamefont
  {Killoran}},\ }\bibfield  {title} {\bibinfo {title} {Quantum machine learning
  in feature hilbert spaces},\ }\bibfield  {journal} {\bibinfo  {journal}
  {Physical Review Letters}\ }\textbf {\bibinfo {volume} {122}},\ \href
  {https://doi.org/10.1103/physrevlett.122.040504}
  {10.1103/physrevlett.122.040504} (\bibinfo {year} {2019})\BibitemShut
  {NoStop}%
\bibitem [{\citenamefont {Havl{\'{\i}}{\v{c}}ek}\ \emph
  {et~al.}(2019)\citenamefont {Havl{\'{\i}}{\v{c}}ek}, \citenamefont
  {C{\'{o}}rcoles}, \citenamefont {Temme}, \citenamefont {Harrow},
  \citenamefont {Kandala}, \citenamefont {Chow},\ and\ \citenamefont
  {Gambetta}}]{Havlek2019}%
  \BibitemOpen
  \bibfield  {author} {\bibinfo {author} {\bibfnamefont {V.}~\bibnamefont
  {Havl{\'{\i}}{\v{c}}ek}}, \bibinfo {author} {\bibfnamefont {A.~D.}\
  \bibnamefont {C{\'{o}}rcoles}}, \bibinfo {author} {\bibfnamefont
  {K.}~\bibnamefont {Temme}}, \bibinfo {author} {\bibfnamefont {A.~W.}\
  \bibnamefont {Harrow}}, \bibinfo {author} {\bibfnamefont {A.}~\bibnamefont
  {Kandala}}, \bibinfo {author} {\bibfnamefont {J.~M.}\ \bibnamefont {Chow}},\
  and\ \bibinfo {author} {\bibfnamefont {J.~M.}\ \bibnamefont {Gambetta}},\
  }\bibfield  {title} {\bibinfo {title} {Supervised learning with
  quantum-enhanced feature spaces},\ }\href
  {https://doi.org/10.1038/s41586-019-0980-2} {\bibfield  {journal} {\bibinfo
  {journal} {Nature}\ }\textbf {\bibinfo {volume} {567}},\ \bibinfo {pages}
  {209} (\bibinfo {year} {2019})}\BibitemShut {NoStop}%
\bibitem [{\citenamefont {Stone}(1932)}]{Stone1932}%
  \BibitemOpen
  \bibfield  {author} {\bibinfo {author} {\bibfnamefont {M.~H.}\ \bibnamefont
  {Stone}},\ }\bibfield  {title} {\bibinfo {title} {On one-parameter unitary
  groups in hilbert space},\ }\href {https://doi.org/10.2307/1968538}
  {\bibfield  {journal} {\bibinfo  {journal} {The Annals of Mathematics}\
  }\textbf {\bibinfo {volume} {33}},\ \bibinfo {pages} {643} (\bibinfo {year}
  {1932})}\BibitemShut {NoStop}%
\bibitem [{\citenamefont {Bjorken}\ and\ \citenamefont
  {Drell}(1965)}]{bjorken1965relativistic}%
  \BibitemOpen
  \bibfield  {author} {\bibinfo {author} {\bibfnamefont {J.~D.}\ \bibnamefont
  {Bjorken}}\ and\ \bibinfo {author} {\bibfnamefont {S.~D.}\ \bibnamefont
  {Drell}},\ }\href@noop {} {\emph {\bibinfo {title} {Relativistic quantum
  fields}}}\ (\bibinfo  {publisher} {McGraw-Hill},\ \bibinfo {year}
  {1965})\BibitemShut {NoStop}%
\bibitem [{\citenamefont {Sakoe}\ and\ \citenamefont
  {Chiba}(1978)}]{Sakoe1978}%
  \BibitemOpen
  \bibfield  {author} {\bibinfo {author} {\bibfnamefont {H.}~\bibnamefont
  {Sakoe}}\ and\ \bibinfo {author} {\bibfnamefont {S.}~\bibnamefont {Chiba}},\
  }\bibfield  {title} {\bibinfo {title} {Dynamic programming algorithm
  optimization for spoken word recognition},\ }\href
  {https://doi.org/10.1109/TASSP.1978.1163055} {\bibfield  {journal} {\bibinfo
  {journal} {IEEE Transactions on Acoustics, Speech, and Signal Processing}\
  }\textbf {\bibinfo {volume} {26}},\ \bibinfo {pages} {43} (\bibinfo {year}
  {1978})}\BibitemShut {NoStop}%
\bibitem [{\citenamefont {Aiolli}\ \emph {et~al.}(2008)\citenamefont {Aiolli},
  \citenamefont {Martino},\ and\ \citenamefont {Sperduti}}]{Aiolli2008}%
  \BibitemOpen
  \bibfield  {author} {\bibinfo {author} {\bibfnamefont {F.}~\bibnamefont
  {Aiolli}}, \bibinfo {author} {\bibfnamefont {G.~D.~S.}\ \bibnamefont
  {Martino}},\ and\ \bibinfo {author} {\bibfnamefont {A.}~\bibnamefont
  {Sperduti}},\ }\bibfield  {title} {\bibinfo {title} {A kernel method for the
  optimization of the margin distribution},\ }in\ \href
  {https://doi.org/10.1007/978-3-540-87536-9_32} {\emph {\bibinfo {booktitle}
  {Artificial Neural Networks - {ICANN} 2008}}}\ (\bibinfo  {publisher}
  {Springer Berlin Heidelberg},\ \bibinfo {year} {2008})\ pp.\ \bibinfo {pages}
  {305--314}\BibitemShut {NoStop}%
\bibitem [{\citenamefont {O'Donoghue}\ \emph
  {et~al.}(2016{\natexlab{a}})\citenamefont {O'Donoghue}, \citenamefont {Chu},
  \citenamefont {Parikh},\ and\ \citenamefont {Boyd}}]{Odonoghue2016}%
  \BibitemOpen
  \bibfield  {author} {\bibinfo {author} {\bibfnamefont {B.}~\bibnamefont
  {O'Donoghue}}, \bibinfo {author} {\bibfnamefont {E.}~\bibnamefont {Chu}},
  \bibinfo {author} {\bibfnamefont {N.}~\bibnamefont {Parikh}},\ and\ \bibinfo
  {author} {\bibfnamefont {S.}~\bibnamefont {Boyd}},\ }\bibfield  {title}
  {\bibinfo {title} {Conic optimization via operator splitting and homogeneous
  self-dual embedding},\ }\href {http://stanford.edu/~boyd/papers/scs.html}
  {\bibfield  {journal} {\bibinfo  {journal} {Journal of Optimization Theory
  and Applications}\ }\textbf {\bibinfo {volume} {169}},\ \bibinfo {pages}
  {1042} (\bibinfo {year} {2016}{\natexlab{a}})}\BibitemShut {NoStop}%
\bibitem [{\citenamefont {Bergholm}\ \emph {et~al.}(2018)\citenamefont
  {Bergholm}, \citenamefont {Izaac}, \citenamefont {Schuld}, \citenamefont
  {Gogolin}, \citenamefont {Ahmed}, \citenamefont {Ajith}, \citenamefont
  {Alam}, \citenamefont {Alonso-Linaje}, \citenamefont {AkashNarayanan},
  \citenamefont {Asadi}, \citenamefont {Arrazola}, \citenamefont {Azad},
  \citenamefont {Banning}, \citenamefont {Blank}, \citenamefont {Bromley},
  \citenamefont {Cordier}, \citenamefont {Ceroni}, \citenamefont {Delgado},
  \citenamefont {Di~Matteo}, \citenamefont {Dusko}, \citenamefont {Garg},
  \citenamefont {Guala}, \citenamefont {Hayes}, \citenamefont {Hill},
  \citenamefont {Ijaz}, \citenamefont {Isacsson}, \citenamefont {Ittah},
  \citenamefont {Jahangiri}, \citenamefont {Jain}, \citenamefont {Jiang},
  \citenamefont {Khandelwal}, \citenamefont {Kottmann}, \citenamefont {Lang},
  \citenamefont {Lee}, \citenamefont {Loke}, \citenamefont {Lowe},
  \citenamefont {McKiernan}, \citenamefont {Meyer}, \citenamefont
  {Montañez-Barrera}, \citenamefont {Moyard}, \citenamefont {Niu},
  \citenamefont {O'Riordan}, \citenamefont {Oud}, \citenamefont {Panigrahi},
  \citenamefont {Park}, \citenamefont {Polatajko}, \citenamefont {Quesada},
  \citenamefont {Roberts}, \citenamefont {Sá}, \citenamefont {Schoch},
  \citenamefont {Shi}, \citenamefont {Shu}, \citenamefont {Sim}, \citenamefont
  {Singh}, \citenamefont {Strandberg}, \citenamefont {Soni}, \citenamefont
  {Száva}, \citenamefont {Thabet}, \citenamefont {Vargas-Hernández},
  \citenamefont {Vincent}, \citenamefont {Vitucci}, \citenamefont {Weber},
  \citenamefont {Wierichs}, \citenamefont {Wiersema}, \citenamefont {Willmann},
  \citenamefont {Wong}, \citenamefont {Zhang},\ and\ \citenamefont
  {Killoran}}]{Bergholm2018}%
  \BibitemOpen
  \bibfield  {author} {\bibinfo {author} {\bibfnamefont {V.}~\bibnamefont
  {Bergholm}}, \bibinfo {author} {\bibfnamefont {J.}~\bibnamefont {Izaac}},
  \bibinfo {author} {\bibfnamefont {M.}~\bibnamefont {Schuld}}, \bibinfo
  {author} {\bibfnamefont {C.}~\bibnamefont {Gogolin}}, \bibinfo {author}
  {\bibfnamefont {S.}~\bibnamefont {Ahmed}}, \bibinfo {author} {\bibfnamefont
  {V.}~\bibnamefont {Ajith}}, \bibinfo {author} {\bibfnamefont {M.~S.}\
  \bibnamefont {Alam}}, \bibinfo {author} {\bibfnamefont {G.}~\bibnamefont
  {Alonso-Linaje}}, \bibinfo {author} {\bibfnamefont {B.}~\bibnamefont
  {AkashNarayanan}}, \bibinfo {author} {\bibfnamefont {A.}~\bibnamefont
  {Asadi}}, \bibinfo {author} {\bibfnamefont {J.~M.}\ \bibnamefont {Arrazola}},
  \bibinfo {author} {\bibfnamefont {U.}~\bibnamefont {Azad}}, \bibinfo {author}
  {\bibfnamefont {S.}~\bibnamefont {Banning}}, \bibinfo {author} {\bibfnamefont
  {C.}~\bibnamefont {Blank}}, \bibinfo {author} {\bibfnamefont {T.~R.}\
  \bibnamefont {Bromley}}, \bibinfo {author} {\bibfnamefont {B.~A.}\
  \bibnamefont {Cordier}}, \bibinfo {author} {\bibfnamefont {J.}~\bibnamefont
  {Ceroni}}, \bibinfo {author} {\bibfnamefont {A.}~\bibnamefont {Delgado}},
  \bibinfo {author} {\bibfnamefont {O.}~\bibnamefont {Di~Matteo}}, \bibinfo
  {author} {\bibfnamefont {A.}~\bibnamefont {Dusko}}, \bibinfo {author}
  {\bibfnamefont {T.}~\bibnamefont {Garg}}, \bibinfo {author} {\bibfnamefont
  {D.}~\bibnamefont {Guala}}, \bibinfo {author} {\bibfnamefont
  {A.}~\bibnamefont {Hayes}}, \bibinfo {author} {\bibfnamefont
  {R.}~\bibnamefont {Hill}}, \bibinfo {author} {\bibfnamefont {A.}~\bibnamefont
  {Ijaz}}, \bibinfo {author} {\bibfnamefont {T.}~\bibnamefont {Isacsson}},
  \bibinfo {author} {\bibfnamefont {D.}~\bibnamefont {Ittah}}, \bibinfo
  {author} {\bibfnamefont {S.}~\bibnamefont {Jahangiri}}, \bibinfo {author}
  {\bibfnamefont {P.}~\bibnamefont {Jain}}, \bibinfo {author} {\bibfnamefont
  {E.}~\bibnamefont {Jiang}}, \bibinfo {author} {\bibfnamefont
  {A.}~\bibnamefont {Khandelwal}}, \bibinfo {author} {\bibfnamefont
  {K.}~\bibnamefont {Kottmann}}, \bibinfo {author} {\bibfnamefont {R.~A.}\
  \bibnamefont {Lang}}, \bibinfo {author} {\bibfnamefont {C.}~\bibnamefont
  {Lee}}, \bibinfo {author} {\bibfnamefont {T.}~\bibnamefont {Loke}}, \bibinfo
  {author} {\bibfnamefont {A.}~\bibnamefont {Lowe}}, \bibinfo {author}
  {\bibfnamefont {K.}~\bibnamefont {McKiernan}}, \bibinfo {author}
  {\bibfnamefont {J.~J.}\ \bibnamefont {Meyer}}, \bibinfo {author}
  {\bibfnamefont {J.~A.}\ \bibnamefont {Montañez-Barrera}}, \bibinfo {author}
  {\bibfnamefont {R.}~\bibnamefont {Moyard}}, \bibinfo {author} {\bibfnamefont
  {Z.}~\bibnamefont {Niu}}, \bibinfo {author} {\bibfnamefont {L.~J.}\
  \bibnamefont {O'Riordan}}, \bibinfo {author} {\bibfnamefont {S.}~\bibnamefont
  {Oud}}, \bibinfo {author} {\bibfnamefont {A.}~\bibnamefont {Panigrahi}},
  \bibinfo {author} {\bibfnamefont {C.-Y.}\ \bibnamefont {Park}}, \bibinfo
  {author} {\bibfnamefont {D.}~\bibnamefont {Polatajko}}, \bibinfo {author}
  {\bibfnamefont {N.}~\bibnamefont {Quesada}}, \bibinfo {author} {\bibfnamefont
  {C.}~\bibnamefont {Roberts}}, \bibinfo {author} {\bibfnamefont
  {N.}~\bibnamefont {Sá}}, \bibinfo {author} {\bibfnamefont {I.}~\bibnamefont
  {Schoch}}, \bibinfo {author} {\bibfnamefont {B.}~\bibnamefont {Shi}},
  \bibinfo {author} {\bibfnamefont {S.}~\bibnamefont {Shu}}, \bibinfo {author}
  {\bibfnamefont {S.}~\bibnamefont {Sim}}, \bibinfo {author} {\bibfnamefont
  {A.}~\bibnamefont {Singh}}, \bibinfo {author} {\bibfnamefont
  {I.}~\bibnamefont {Strandberg}}, \bibinfo {author} {\bibfnamefont
  {J.}~\bibnamefont {Soni}}, \bibinfo {author} {\bibfnamefont {A.}~\bibnamefont
  {Száva}}, \bibinfo {author} {\bibfnamefont {S.}~\bibnamefont {Thabet}},
  \bibinfo {author} {\bibfnamefont {R.~A.}\ \bibnamefont {Vargas-Hernández}},
  \bibinfo {author} {\bibfnamefont {T.}~\bibnamefont {Vincent}}, \bibinfo
  {author} {\bibfnamefont {N.}~\bibnamefont {Vitucci}}, \bibinfo {author}
  {\bibfnamefont {M.}~\bibnamefont {Weber}}, \bibinfo {author} {\bibfnamefont
  {D.}~\bibnamefont {Wierichs}}, \bibinfo {author} {\bibfnamefont
  {R.}~\bibnamefont {Wiersema}}, \bibinfo {author} {\bibfnamefont
  {M.}~\bibnamefont {Willmann}}, \bibinfo {author} {\bibfnamefont
  {V.}~\bibnamefont {Wong}}, \bibinfo {author} {\bibfnamefont {S.}~\bibnamefont
  {Zhang}},\ and\ \bibinfo {author} {\bibfnamefont {N.}~\bibnamefont
  {Killoran}},\ }\href {https://doi.org/10.48550/ARXIV.1811.04968} {\bibinfo
  {title} {Pennylane: Automatic differentiation of hybrid quantum-classical
  computations}} (\bibinfo {year} {2018})\BibitemShut {NoStop}%
\bibitem [{\citenamefont {Mitarai}\ \emph {et~al.}(2018)\citenamefont
  {Mitarai}, \citenamefont {Negoro}, \citenamefont {Kitagawa},\ and\
  \citenamefont {Fujii}}]{Mitarai2018}%
  \BibitemOpen
  \bibfield  {author} {\bibinfo {author} {\bibfnamefont {K.}~\bibnamefont
  {Mitarai}}, \bibinfo {author} {\bibfnamefont {M.}~\bibnamefont {Negoro}},
  \bibinfo {author} {\bibfnamefont {M.}~\bibnamefont {Kitagawa}},\ and\
  \bibinfo {author} {\bibfnamefont {K.}~\bibnamefont {Fujii}},\ }\bibfield
  {title} {\bibinfo {title} {Quantum circuit learning},\ }\bibfield  {journal}
  {\bibinfo  {journal} {Physical Review A}\ }\textbf {\bibinfo {volume} {98}},\
  \href {https://doi.org/10.1103/physreva.98.032309}
  {10.1103/physreva.98.032309} (\bibinfo {year} {2018})\BibitemShut {NoStop}%
\bibitem [{\citenamefont {Schuld}\ \emph {et~al.}(2019)\citenamefont {Schuld},
  \citenamefont {Bergholm}, \citenamefont {Gogolin}, \citenamefont {Izaac},\
  and\ \citenamefont {Killoran}}]{Schuld2019grad}%
  \BibitemOpen
  \bibfield  {author} {\bibinfo {author} {\bibfnamefont {M.}~\bibnamefont
  {Schuld}}, \bibinfo {author} {\bibfnamefont {V.}~\bibnamefont {Bergholm}},
  \bibinfo {author} {\bibfnamefont {C.}~\bibnamefont {Gogolin}}, \bibinfo
  {author} {\bibfnamefont {J.}~\bibnamefont {Izaac}},\ and\ \bibinfo {author}
  {\bibfnamefont {N.}~\bibnamefont {Killoran}},\ }\bibfield  {title} {\bibinfo
  {title} {Evaluating analytic gradients on quantum hardware},\ }\bibfield
  {journal} {\bibinfo  {journal} {Physical Review A}\ }\textbf {\bibinfo
  {volume} {99}},\ \href {https://doi.org/10.1103/physreva.99.032331}
  {10.1103/physreva.99.032331} (\bibinfo {year} {2019})\BibitemShut {NoStop}%
\bibitem [{\citenamefont {Agrawal}\ \emph
  {et~al.}(2019{\natexlab{a}})\citenamefont {Agrawal}, \citenamefont {Amos},
  \citenamefont {Barratt}, \citenamefont {Boyd}, \citenamefont {Diamond},\ and\
  \citenamefont {Kolter}}]{Agrawal2019}%
  \BibitemOpen
  \bibfield  {author} {\bibinfo {author} {\bibfnamefont {A.}~\bibnamefont
  {Agrawal}}, \bibinfo {author} {\bibfnamefont {B.}~\bibnamefont {Amos}},
  \bibinfo {author} {\bibfnamefont {S.}~\bibnamefont {Barratt}}, \bibinfo
  {author} {\bibfnamefont {S.}~\bibnamefont {Boyd}}, \bibinfo {author}
  {\bibfnamefont {S.}~\bibnamefont {Diamond}},\ and\ \bibinfo {author}
  {\bibfnamefont {J.~Z.}\ \bibnamefont {Kolter}},\ }\bibfield  {title}
  {\bibinfo {title} {Differentiable convex optimization layers},\ }in\
  \href@noop {} {\emph {\bibinfo {booktitle} {Advances in Neural Information
  Processing Systems}}},\ Vol.~\bibinfo {volume} {32},\ \bibinfo {editor}
  {edited by\ \bibinfo {editor} {\bibfnamefont {H.}~\bibnamefont {Wallach}},
  \bibinfo {editor} {\bibfnamefont {H.}~\bibnamefont {Larochelle}}, \bibinfo
  {editor} {\bibfnamefont {A.}~\bibnamefont {Beygelzimer}}, \bibinfo {editor}
  {\bibfnamefont {F.}~\bibnamefont {d\textquotesingle Alch\'{e}-Buc}}, \bibinfo
  {editor} {\bibfnamefont {E.}~\bibnamefont {Fox}},\ and\ \bibinfo {editor}
  {\bibfnamefont {R.}~\bibnamefont {Garnett}}}\ (\bibinfo  {publisher} {Curran
  Associates, Inc.},\ \bibinfo {year} {2019})\BibitemShut {NoStop}%
\bibitem [{\citenamefont {Agrawal}\ \emph
  {et~al.}(2019{\natexlab{b}})\citenamefont {Agrawal}, \citenamefont {Barratt},
  \citenamefont {Boyd}, \citenamefont {Busseti},\ and\ \citenamefont
  {Moursi}}]{Agrawal2019cone}%
  \BibitemOpen
  \bibfield  {author} {\bibinfo {author} {\bibfnamefont {A.}~\bibnamefont
  {Agrawal}}, \bibinfo {author} {\bibfnamefont {S.}~\bibnamefont {Barratt}},
  \bibinfo {author} {\bibfnamefont {S.}~\bibnamefont {Boyd}}, \bibinfo {author}
  {\bibfnamefont {E.}~\bibnamefont {Busseti}},\ and\ \bibinfo {author}
  {\bibfnamefont {W.~M.}\ \bibnamefont {Moursi}},\ }\href
  {https://doi.org/10.48550/ARXIV.1904.09043} {\bibinfo {title}
  {Differentiating through a cone program}} (\bibinfo {year}
  {2019}{\natexlab{b}})\BibitemShut {NoStop}%
\bibitem [{\citenamefont {Bottou}\ \emph {et~al.}(2018)\citenamefont {Bottou},
  \citenamefont {Curtis},\ and\ \citenamefont {Nocedal}}]{Bottou2018}%
  \BibitemOpen
  \bibfield  {author} {\bibinfo {author} {\bibfnamefont {L.}~\bibnamefont
  {Bottou}}, \bibinfo {author} {\bibfnamefont {F.~E.}\ \bibnamefont {Curtis}},\
  and\ \bibinfo {author} {\bibfnamefont {J.}~\bibnamefont {Nocedal}},\
  }\bibfield  {title} {\bibinfo {title} {Optimization methods for large-scale
  machine learning},\ }\href {https://doi.org/10.1137/16m1080173} {\bibfield
  {journal} {\bibinfo  {journal} {{SIAM} Review}\ }\textbf {\bibinfo {volume}
  {60}},\ \bibinfo {pages} {223} (\bibinfo {year} {2018})}\BibitemShut
  {NoStop}%
\bibitem [{\citenamefont {Keskar}\ \emph {et~al.}(2017)\citenamefont {Keskar},
  \citenamefont {Mudigere}, \citenamefont {Nocedal}, \citenamefont
  {Smelyanskiy},\ and\ \citenamefont {Tang}}]{keskar2017largebatch}%
  \BibitemOpen
  \bibfield  {author} {\bibinfo {author} {\bibfnamefont {N.~S.}\ \bibnamefont
  {Keskar}}, \bibinfo {author} {\bibfnamefont {D.}~\bibnamefont {Mudigere}},
  \bibinfo {author} {\bibfnamefont {J.}~\bibnamefont {Nocedal}}, \bibinfo
  {author} {\bibfnamefont {M.}~\bibnamefont {Smelyanskiy}},\ and\ \bibinfo
  {author} {\bibfnamefont {P.~T.~P.}\ \bibnamefont {Tang}},\ }\href@noop {}
  {\bibinfo {title} {On large-batch training for deep learning: Generalization
  gap and sharp minima}} (\bibinfo {year} {2017}),\ \Eprint
  {https://arxiv.org/abs/1609.04836} {arXiv:1609.04836 [cs.LG]} \BibitemShut
  {NoStop}%
\bibitem [{\citenamefont {Kübler}\ \emph {et~al.}(2021)\citenamefont
  {Kübler}, \citenamefont {Buchholz},\ and\ \citenamefont
  {Schölkopf}}]{kubler2021inductive}%
  \BibitemOpen
  \bibfield  {author} {\bibinfo {author} {\bibfnamefont {J.~M.}\ \bibnamefont
  {Kübler}}, \bibinfo {author} {\bibfnamefont {S.}~\bibnamefont {Buchholz}},\
  and\ \bibinfo {author} {\bibfnamefont {B.}~\bibnamefont {Schölkopf}},\
  }\href@noop {} {\bibinfo {title} {The inductive bias of quantum kernels}}
  (\bibinfo {year} {2021}),\ \Eprint {https://arxiv.org/abs/2106.03747}
  {arXiv:2106.03747 [quant-ph]} \BibitemShut {NoStop}%
\bibitem [{\citenamefont {McClean}\ \emph {et~al.}(2018)\citenamefont
  {McClean}, \citenamefont {Boixo}, \citenamefont {Smelyanskiy}, \citenamefont
  {Babbush},\ and\ \citenamefont {Neven}}]{McClean2018}%
  \BibitemOpen
  \bibfield  {author} {\bibinfo {author} {\bibfnamefont {J.~R.}\ \bibnamefont
  {McClean}}, \bibinfo {author} {\bibfnamefont {S.}~\bibnamefont {Boixo}},
  \bibinfo {author} {\bibfnamefont {V.~N.}\ \bibnamefont {Smelyanskiy}},
  \bibinfo {author} {\bibfnamefont {R.}~\bibnamefont {Babbush}},\ and\ \bibinfo
  {author} {\bibfnamefont {H.}~\bibnamefont {Neven}},\ }\bibfield  {title}
  {\bibinfo {title} {Barren plateaus in quantum neural network training
  landscapes},\ }\bibfield  {journal} {\bibinfo  {journal} {Nature
  Communications}\ }\textbf {\bibinfo {volume} {9}},\ \href
  {https://doi.org/10.1038/s41467-018-07090-4} {10.1038/s41467-018-07090-4}
  (\bibinfo {year} {2018})\BibitemShut {NoStop}%
\bibitem [{\citenamefont {Pesah}\ \emph {et~al.}(2021)\citenamefont {Pesah},
  \citenamefont {Cerezo}, \citenamefont {Wang}, \citenamefont {Volkoff},
  \citenamefont {Sornborger},\ and\ \citenamefont {Coles}}]{Pesah2021}%
  \BibitemOpen
  \bibfield  {author} {\bibinfo {author} {\bibfnamefont {A.}~\bibnamefont
  {Pesah}}, \bibinfo {author} {\bibfnamefont {M.}~\bibnamefont {Cerezo}},
  \bibinfo {author} {\bibfnamefont {S.}~\bibnamefont {Wang}}, \bibinfo {author}
  {\bibfnamefont {T.}~\bibnamefont {Volkoff}}, \bibinfo {author} {\bibfnamefont
  {A.~T.}\ \bibnamefont {Sornborger}},\ and\ \bibinfo {author} {\bibfnamefont
  {P.~J.}\ \bibnamefont {Coles}},\ }\bibfield  {title} {\bibinfo {title}
  {Absence of barren plateaus in quantum convolutional neural networks},\
  }\href {https://doi.org/10.1103/PhysRevX.11.041011} {\bibfield  {journal}
  {\bibinfo  {journal} {Phys. Rev. X}\ }\textbf {\bibinfo {volume} {11}},\
  \bibinfo {pages} {041011} (\bibinfo {year} {2021})}\BibitemShut {NoStop}%
\bibitem [{\citenamefont {Lloyd}\ \emph {et~al.}(2020)\citenamefont {Lloyd},
  \citenamefont {Schuld}, \citenamefont {Ijaz}, \citenamefont {Izaac},\ and\
  \citenamefont {Killoran}}]{lloyd2020quantum}%
  \BibitemOpen
  \bibfield  {author} {\bibinfo {author} {\bibfnamefont {S.}~\bibnamefont
  {Lloyd}}, \bibinfo {author} {\bibfnamefont {M.}~\bibnamefont {Schuld}},
  \bibinfo {author} {\bibfnamefont {A.}~\bibnamefont {Ijaz}}, \bibinfo {author}
  {\bibfnamefont {J.}~\bibnamefont {Izaac}},\ and\ \bibinfo {author}
  {\bibfnamefont {N.}~\bibnamefont {Killoran}},\ }\href@noop {} {\bibinfo
  {title} {Quantum embeddings for machine learning}} (\bibinfo {year} {2020}),\
  \Eprint {https://arxiv.org/abs/2001.03622} {arXiv:2001.03622 [quant-ph]}
  \BibitemShut {NoStop}%
\bibitem [{\citenamefont {Farhi}\ \emph {et~al.}(2014)\citenamefont {Farhi},
  \citenamefont {Goldstone},\ and\ \citenamefont {Gutmann}}]{farhi2014quantum}%
  \BibitemOpen
  \bibfield  {author} {\bibinfo {author} {\bibfnamefont {E.}~\bibnamefont
  {Farhi}}, \bibinfo {author} {\bibfnamefont {J.}~\bibnamefont {Goldstone}},\
  and\ \bibinfo {author} {\bibfnamefont {S.}~\bibnamefont {Gutmann}},\
  }\bibfield  {title} {\bibinfo {title} {A quantum approximate optimization
  algorithm},\ }\href@noop {} {\bibfield  {journal} {\bibinfo  {journal} {arXiv
  preprint arXiv:1411.4028}\ } (\bibinfo {year} {2014})}\BibitemShut {NoStop}%
\bibitem [{\citenamefont {Hadfield}\ \emph {et~al.}(2019)\citenamefont
  {Hadfield}, \citenamefont {Wang}, \citenamefont {O'Gorman}, \citenamefont
  {Rieffel}, \citenamefont {Venturelli},\ and\ \citenamefont
  {Biswas}}]{Hadfield2019}%
  \BibitemOpen
  \bibfield  {author} {\bibinfo {author} {\bibfnamefont {S.}~\bibnamefont
  {Hadfield}}, \bibinfo {author} {\bibfnamefont {Z.}~\bibnamefont {Wang}},
  \bibinfo {author} {\bibfnamefont {B.}~\bibnamefont {O'Gorman}}, \bibinfo
  {author} {\bibfnamefont {E.~G.}\ \bibnamefont {Rieffel}}, \bibinfo {author}
  {\bibfnamefont {D.}~\bibnamefont {Venturelli}},\ and\ \bibinfo {author}
  {\bibfnamefont {R.}~\bibnamefont {Biswas}},\ }\bibfield  {title} {\bibinfo
  {title} {From the quantum approximate optimization algorithm to a quantum
  alternating operator ansatz},\ }\href@noop {} {\bibfield  {journal} {\bibinfo
   {journal} {Algorithms}\ }\textbf {\bibinfo {volume} {12}} (\bibinfo {year}
  {2019})}\BibitemShut {NoStop}%
\bibitem [{\citenamefont {Schuld}\ \emph {et~al.}(2020)\citenamefont {Schuld},
  \citenamefont {Bocharov}, \citenamefont {Svore},\ and\ \citenamefont
  {Wiebe}}]{Schuld2020}%
  \BibitemOpen
  \bibfield  {author} {\bibinfo {author} {\bibfnamefont {M.}~\bibnamefont
  {Schuld}}, \bibinfo {author} {\bibfnamefont {A.}~\bibnamefont {Bocharov}},
  \bibinfo {author} {\bibfnamefont {K.~M.}\ \bibnamefont {Svore}},\ and\
  \bibinfo {author} {\bibfnamefont {N.}~\bibnamefont {Wiebe}},\ }\bibfield
  {title} {\bibinfo {title} {Circuit-centric quantum classifiers},\ }\href
  {https://doi.org/10.1103/PhysRevA.101.032308} {\bibfield  {journal} {\bibinfo
   {journal} {Phys. Rev. A}\ }\textbf {\bibinfo {volume} {101}},\ \bibinfo
  {pages} {032308} (\bibinfo {year} {2020})}\BibitemShut {NoStop}%
\bibitem [{\citenamefont {Welch}\ \emph {et~al.}(2014)\citenamefont {Welch},
  \citenamefont {Greenbaum}, \citenamefont {Mostame},\ and\ \citenamefont
  {Aspuru-Guzik}}]{Welch2014}%
  \BibitemOpen
  \bibfield  {author} {\bibinfo {author} {\bibfnamefont {J.}~\bibnamefont
  {Welch}}, \bibinfo {author} {\bibfnamefont {D.}~\bibnamefont {Greenbaum}},
  \bibinfo {author} {\bibfnamefont {S.}~\bibnamefont {Mostame}},\ and\ \bibinfo
  {author} {\bibfnamefont {A.}~\bibnamefont {Aspuru-Guzik}},\ }\bibfield
  {title} {\bibinfo {title} {Efficient quantum circuits for diagonal unitaries
  without ancillas},\ }\href {https://doi.org/10.1088/1367-2630/16/3/033040}
  {\bibfield  {journal} {\bibinfo  {journal} {New Journal of Physics}\ }\textbf
  {\bibinfo {volume} {16}},\ \bibinfo {pages} {033040} (\bibinfo {year}
  {2014})}\BibitemShut {NoStop}%
\bibitem [{\citenamefont {Kingma}\ and\ \citenamefont
  {Ba}(2017)}]{kingma2017adam}%
  \BibitemOpen
  \bibfield  {author} {\bibinfo {author} {\bibfnamefont {D.~P.}\ \bibnamefont
  {Kingma}}\ and\ \bibinfo {author} {\bibfnamefont {J.}~\bibnamefont {Ba}},\
  }\href@noop {} {\bibinfo {title} {Adam: A method for stochastic
  optimization}} (\bibinfo {year} {2017}),\ \Eprint
  {https://arxiv.org/abs/1412.6980} {arXiv:1412.6980 [cs.LG]} \BibitemShut
  {NoStop}%
\bibitem [{\citenamefont {O'Donoghue}\ \emph
  {et~al.}(2016{\natexlab{b}})\citenamefont {O'Donoghue}, \citenamefont {Chu},
  \citenamefont {Parikh},\ and\ \citenamefont {Boyd}}]{ocpb:16}%
  \BibitemOpen
  \bibfield  {author} {\bibinfo {author} {\bibfnamefont {B.}~\bibnamefont
  {O'Donoghue}}, \bibinfo {author} {\bibfnamefont {E.}~\bibnamefont {Chu}},
  \bibinfo {author} {\bibfnamefont {N.}~\bibnamefont {Parikh}},\ and\ \bibinfo
  {author} {\bibfnamefont {S.}~\bibnamefont {Boyd}},\ }\bibfield  {title}
  {\bibinfo {title} {Conic optimization via operator splitting and homogeneous
  self-dual embedding},\ }\href {http://stanford.edu/~boyd/papers/scs.html}
  {\bibfield  {journal} {\bibinfo  {journal} {Journal of Optimization Theory
  and Applications}\ }\textbf {\bibinfo {volume} {169}},\ \bibinfo {pages}
  {1042} (\bibinfo {year} {2016}{\natexlab{b}})}\BibitemShut {NoStop}%
\bibitem [{\citenamefont {O'Donoghue}(2021)}]{odonoghue:21}%
  \BibitemOpen
  \bibfield  {author} {\bibinfo {author} {\bibfnamefont {B.}~\bibnamefont
  {O'Donoghue}},\ }\bibfield  {title} {\bibinfo {title} {Operator splitting for
  a homogeneous embedding of the linear complementarity problem},\ }\href@noop
  {} {\bibfield  {journal} {\bibinfo  {journal} {{SIAM} Journal on
  Optimization}\ }\textbf {\bibinfo {volume} {31}},\ \bibinfo {pages} {1999}
  (\bibinfo {year} {2021})}\BibitemShut {NoStop}%
\bibitem [{sup()}]{supplement}%
  \BibitemOpen
  \href@noop {} {}\bibinfo {note} {See Supplemental Material [publisher url]
  for prelimary studies on the impact of different components of the QCC-net as
  tested on the gun-point data set referenced in the main article and for
  further explanation of how QMP was implemented.}\BibitemShut {Stop}%
\bibitem [{\citenamefont {Bowles}\ \emph {et~al.}(2023)\citenamefont {Bowles},
  \citenamefont {Wright}, \citenamefont {Farkas}, \citenamefont {Killoran},\
  and\ \citenamefont {Schuld}}]{bowles2023contextuality}%
  \BibitemOpen
  \bibfield  {author} {\bibinfo {author} {\bibfnamefont {J.}~\bibnamefont
  {Bowles}}, \bibinfo {author} {\bibfnamefont {V.~J.}\ \bibnamefont {Wright}},
  \bibinfo {author} {\bibfnamefont {M.}~\bibnamefont {Farkas}}, \bibinfo
  {author} {\bibfnamefont {N.}~\bibnamefont {Killoran}},\ and\ \bibinfo
  {author} {\bibfnamefont {M.}~\bibnamefont {Schuld}},\ }\href@noop {}
  {\bibinfo {title} {Contextuality and inductive bias in quantum machine
  learning}} (\bibinfo {year} {2023}),\ \Eprint
  {https://arxiv.org/abs/2302.01365} {arXiv:2302.01365 [quant-ph]} \BibitemShut
  {NoStop}%
\bibitem [{\citenamefont {Hastie}\ \emph {et~al.}(2009)\citenamefont {Hastie},
  \citenamefont {Tibshirani}, \citenamefont {Friedman},\ and\ \citenamefont
  {Friedman}}]{hastie2009elements}%
  \BibitemOpen
  \bibfield  {author} {\bibinfo {author} {\bibfnamefont {T.}~\bibnamefont
  {Hastie}}, \bibinfo {author} {\bibfnamefont {R.}~\bibnamefont {Tibshirani}},
  \bibinfo {author} {\bibfnamefont {J.~H.}\ \bibnamefont {Friedman}},\ and\
  \bibinfo {author} {\bibfnamefont {J.~H.}\ \bibnamefont {Friedman}},\
  }\href@noop {} {\emph {\bibinfo {title} {The elements of statistical
  learning: data mining, inference, and prediction}}},\ Vol.~\bibinfo {volume}
  {2}\ (\bibinfo  {publisher} {Springer},\ \bibinfo {year} {2009})\ p.\
  \bibinfo {pages} {134}\BibitemShut {NoStop}%
\bibitem [{\citenamefont {Cross}\ \emph {et~al.}(2019)\citenamefont {Cross},
  \citenamefont {Bishop}, \citenamefont {Sheldon}, \citenamefont {Nation},\
  and\ \citenamefont {Gambetta}}]{cross2019validating}%
  \BibitemOpen
  \bibfield  {author} {\bibinfo {author} {\bibfnamefont {A.~W.}\ \bibnamefont
  {Cross}}, \bibinfo {author} {\bibfnamefont {L.~S.}\ \bibnamefont {Bishop}},
  \bibinfo {author} {\bibfnamefont {S.}~\bibnamefont {Sheldon}}, \bibinfo
  {author} {\bibfnamefont {P.~D.}\ \bibnamefont {Nation}},\ and\ \bibinfo
  {author} {\bibfnamefont {J.~M.}\ \bibnamefont {Gambetta}},\ }\bibfield
  {title} {\bibinfo {title} {Validating quantum computers using randomized
  model circuits},\ }\href@noop {} {\bibfield  {journal} {\bibinfo  {journal}
  {Physical Review A}\ }\textbf {\bibinfo {volume} {100}},\ \bibinfo {pages}
  {032328} (\bibinfo {year} {2019})}\BibitemShut {NoStop}%
\bibitem [{\citenamefont {Ohkura}\ \emph {et~al.}(2022)\citenamefont {Ohkura},
  \citenamefont {Satoh},\ and\ \citenamefont
  {Van~Meter}}]{ohkura2022simultaneous}%
  \BibitemOpen
  \bibfield  {author} {\bibinfo {author} {\bibfnamefont {Y.}~\bibnamefont
  {Ohkura}}, \bibinfo {author} {\bibfnamefont {T.}~\bibnamefont {Satoh}},\ and\
  \bibinfo {author} {\bibfnamefont {R.}~\bibnamefont {Van~Meter}},\ }\bibfield
  {title} {\bibinfo {title} {Simultaneous execution of quantum circuits on
  current and near-future {NISQ} systems},\ }\href@noop {} {\bibfield
  {journal} {\bibinfo  {journal} {IEEE Transactions on Quantum Engineering}\ }
  (\bibinfo {year} {2022})}\BibitemShut {NoStop}%
\bibitem [{\citenamefont {Liu}\ and\ \citenamefont
  {Dou}(2021)}]{liu2021qucloud}%
  \BibitemOpen
  \bibfield  {author} {\bibinfo {author} {\bibfnamefont {L.}~\bibnamefont
  {Liu}}\ and\ \bibinfo {author} {\bibfnamefont {X.}~\bibnamefont {Dou}},\
  }\bibfield  {title} {\bibinfo {title} {{QuCloud}: A new qubit mapping
  mechanism for multi-programming quantum computing in cloud environment},\
  }in\ \href@noop {} {\emph {\bibinfo {booktitle} {2021 IEEE International
  Symposium on High-Performance Computer Architecture (HPCA)}}}\ (\bibinfo
  {organization} {IEEE},\ \bibinfo {year} {2021})\ pp.\ \bibinfo {pages}
  {167--178}\BibitemShut {NoStop}%
\bibitem [{\citenamefont {Niu}\ and\ \citenamefont
  {Todri-Sanial}(2021)}]{niu2021enabling}%
  \BibitemOpen
  \bibfield  {author} {\bibinfo {author} {\bibfnamefont {S.}~\bibnamefont
  {Niu}}\ and\ \bibinfo {author} {\bibfnamefont {A.}~\bibnamefont
  {Todri-Sanial}},\ }\bibfield  {title} {\bibinfo {title} {Enabling
  multi-programming mechanism for quantum computing in the {NISQ} era},\
  }\href@noop {} {\bibfield  {journal} {\bibinfo  {journal} {arXiv preprint
  arXiv:2102.05321}\ } (\bibinfo {year} {2021})}\BibitemShut {NoStop}%
\bibitem [{\citenamefont {Niu}\ and\ \citenamefont
  {Todri-Sanial}(2022)}]{niu2022parallel}%
  \BibitemOpen
  \bibfield  {author} {\bibinfo {author} {\bibfnamefont {S.}~\bibnamefont
  {Niu}}\ and\ \bibinfo {author} {\bibfnamefont {A.}~\bibnamefont
  {Todri-Sanial}},\ }\bibfield  {title} {\bibinfo {title} {How parallel circuit
  execution can be useful for {NISQ} computing?},\ }in\ \href@noop {} {\emph
  {\bibinfo {booktitle} {2022 Design, Automation \& Test in Europe Conference
  \& Exhibition (DATE)}}}\ (\bibinfo {organization} {IEEE},\ \bibinfo {year}
  {2022})\ pp.\ \bibinfo {pages} {1065--1070}\BibitemShut {NoStop}%
\bibitem [{\citenamefont {Park}\ \emph {et~al.}(2023)\citenamefont {Park},
  \citenamefont {Zhang}, \citenamefont {Yu},\ and\ \citenamefont
  {Korepin}}]{park2023quantum}%
  \BibitemOpen
  \bibfield  {author} {\bibinfo {author} {\bibfnamefont {G.}~\bibnamefont
  {Park}}, \bibinfo {author} {\bibfnamefont {K.}~\bibnamefont {Zhang}},
  \bibinfo {author} {\bibfnamefont {K.}~\bibnamefont {Yu}},\ and\ \bibinfo
  {author} {\bibfnamefont {V.}~\bibnamefont {Korepin}},\ }\bibfield  {title}
  {\bibinfo {title} {Quantum multi-programming for {G}rover’s search},\
  }\href@noop {} {\bibfield  {journal} {\bibinfo  {journal} {Quantum
  Information Processing}\ }\textbf {\bibinfo {volume} {22}},\ \bibinfo {pages}
  {54} (\bibinfo {year} {2023})}\BibitemShut {NoStop}%
\bibitem [{\citenamefont {Murali}\ \emph {et~al.}(2020)\citenamefont {Murali},
  \citenamefont {McKay}, \citenamefont {Martonosi},\ and\ \citenamefont
  {Javadi-Abhari}}]{murali2020software}%
  \BibitemOpen
  \bibfield  {author} {\bibinfo {author} {\bibfnamefont {P.}~\bibnamefont
  {Murali}}, \bibinfo {author} {\bibfnamefont {D.~C.}\ \bibnamefont {McKay}},
  \bibinfo {author} {\bibfnamefont {M.}~\bibnamefont {Martonosi}},\ and\
  \bibinfo {author} {\bibfnamefont {A.}~\bibnamefont {Javadi-Abhari}},\
  }\bibfield  {title} {\bibinfo {title} {Software mitigation of crosstalk on
  noisy intermediate-scale quantum computers},\ }in\ \href@noop {} {\emph
  {\bibinfo {booktitle} {Proceedings of the Twenty-Fifth International
  Conference on Architectural Support for Programming Languages and Operating
  Systems}}}\ (\bibinfo {year} {2020})\ pp.\ \bibinfo {pages}
  {1001--1016}\BibitemShut {NoStop}%
\bibitem [{\citenamefont {Ohkura}(2021)}]{ohkura2021crosstalk}%
  \BibitemOpen
  \bibfield  {author} {\bibinfo {author} {\bibfnamefont {Y.}~\bibnamefont
  {Ohkura}},\ }\bibfield  {title} {\bibinfo {title} {Crosstalk-aware {NISQ}
  {M}ulti-programming},\ }\href@noop {} {\bibfield  {journal} {\bibinfo
  {journal} {Faculty Policy Manage., Keio Univ., Tokyo, Japan}\ } (\bibinfo
  {year} {2021})}\BibitemShut {NoStop}%
\bibitem [{\citenamefont {Hubregtsen}\ \emph {et~al.}(2022)\citenamefont
  {Hubregtsen}, \citenamefont {Wierichs}, \citenamefont {Gil-Fuster},
  \citenamefont {Derks}, \citenamefont {Faehrmann},\ and\ \citenamefont
  {Meyer}}]{Hubgregson2022}%
  \BibitemOpen
  \bibfield  {author} {\bibinfo {author} {\bibfnamefont {T.}~\bibnamefont
  {Hubregtsen}}, \bibinfo {author} {\bibfnamefont {D.}~\bibnamefont
  {Wierichs}}, \bibinfo {author} {\bibfnamefont {E.}~\bibnamefont
  {Gil-Fuster}}, \bibinfo {author} {\bibfnamefont {P.-J. H.~S.}\ \bibnamefont
  {Derks}}, \bibinfo {author} {\bibfnamefont {P.~K.}\ \bibnamefont
  {Faehrmann}},\ and\ \bibinfo {author} {\bibfnamefont {J.~J.}\ \bibnamefont
  {Meyer}},\ }\bibfield  {title} {\bibinfo {title} {Training quantum embedding
  kernels on near-term quantum computers},\ }\href
  {https://doi.org/10.1103/PhysRevA.106.042431} {\bibfield  {journal} {\bibinfo
   {journal} {Phys. Rev. A}\ }\textbf {\bibinfo {volume} {106}},\ \bibinfo
  {pages} {042431} (\bibinfo {year} {2022})}\BibitemShut {NoStop}%
\bibitem [{\citenamefont {Yirka}(2021)}]{IBM_127q}%
  \BibitemOpen
  \bibfield  {author} {\bibinfo {author} {\bibfnamefont {B.}~\bibnamefont
  {Yirka}},\ }\href
  {https://phys.org/news/2021-11-ibm-qubit-quantum-processor.html} {\bibinfo
  {title} {{IBM} announces development of 127-qubit quantum processor}}
  (\bibinfo {year} {2021})\BibitemShut {NoStop}%
\bibitem [{\citenamefont {Preskill}(2018)}]{Preskill2018quantumcomputingin}%
  \BibitemOpen
  \bibfield  {author} {\bibinfo {author} {\bibfnamefont {J.}~\bibnamefont
  {Preskill}},\ }\bibfield  {title} {\bibinfo {title} {Quantum {C}omputing in
  the {NISQ} era and beyond},\ }\href
  {https://doi.org/10.22331/q-2018-08-06-79} {\bibfield  {journal} {\bibinfo
  {journal} {{Quantum}}\ }\textbf {\bibinfo {volume} {2}},\ \bibinfo {pages}
  {79} (\bibinfo {year} {2018})}\BibitemShut {NoStop}%
\bibitem [{\citenamefont {Moll}\ \emph {et~al.}(2018)\citenamefont {Moll},
  \citenamefont {Barkoutsos}, \citenamefont {Bishop}, \citenamefont {Chow},
  \citenamefont {Cross}, \citenamefont {Egger}, \citenamefont {Filipp},
  \citenamefont {Fuhrer}, \citenamefont {Gambetta}, \citenamefont {Ganzhorn}
  \emph {et~al.}}]{moll2018quantum}%
  \BibitemOpen
  \bibfield  {author} {\bibinfo {author} {\bibfnamefont {N.}~\bibnamefont
  {Moll}}, \bibinfo {author} {\bibfnamefont {P.}~\bibnamefont {Barkoutsos}},
  \bibinfo {author} {\bibfnamefont {L.~S.}\ \bibnamefont {Bishop}}, \bibinfo
  {author} {\bibfnamefont {J.~M.}\ \bibnamefont {Chow}}, \bibinfo {author}
  {\bibfnamefont {A.}~\bibnamefont {Cross}}, \bibinfo {author} {\bibfnamefont
  {D.~J.}\ \bibnamefont {Egger}}, \bibinfo {author} {\bibfnamefont
  {S.}~\bibnamefont {Filipp}}, \bibinfo {author} {\bibfnamefont
  {A.}~\bibnamefont {Fuhrer}}, \bibinfo {author} {\bibfnamefont {J.~M.}\
  \bibnamefont {Gambetta}}, \bibinfo {author} {\bibfnamefont {M.}~\bibnamefont
  {Ganzhorn}}, \emph {et~al.},\ }\bibfield  {title} {\bibinfo {title} {Quantum
  optimization using variational algorithms on near-term quantum devices},\
  }\href@noop {} {\bibfield  {journal} {\bibinfo  {journal} {Quantum Science
  and Technology}\ }\textbf {\bibinfo {volume} {3}},\ \bibinfo {pages} {030503}
  (\bibinfo {year} {2018})}\BibitemShut {NoStop}%
\bibitem [{\citenamefont {Cerezo}\ \emph {et~al.}(2021)\citenamefont {Cerezo},
  \citenamefont {Arrasmith}, \citenamefont {Babbush}, \citenamefont {Benjamin},
  \citenamefont {Endo}, \citenamefont {Fujii}, \citenamefont {McClean},
  \citenamefont {Mitarai}, \citenamefont {Yuan}, \citenamefont {Cincio} \emph
  {et~al.}}]{cerezo2021variational}%
  \BibitemOpen
  \bibfield  {author} {\bibinfo {author} {\bibfnamefont {M.}~\bibnamefont
  {Cerezo}}, \bibinfo {author} {\bibfnamefont {A.}~\bibnamefont {Arrasmith}},
  \bibinfo {author} {\bibfnamefont {R.}~\bibnamefont {Babbush}}, \bibinfo
  {author} {\bibfnamefont {S.~C.}\ \bibnamefont {Benjamin}}, \bibinfo {author}
  {\bibfnamefont {S.}~\bibnamefont {Endo}}, \bibinfo {author} {\bibfnamefont
  {K.}~\bibnamefont {Fujii}}, \bibinfo {author} {\bibfnamefont {J.~R.}\
  \bibnamefont {McClean}}, \bibinfo {author} {\bibfnamefont {K.}~\bibnamefont
  {Mitarai}}, \bibinfo {author} {\bibfnamefont {X.}~\bibnamefont {Yuan}},
  \bibinfo {author} {\bibfnamefont {L.}~\bibnamefont {Cincio}}, \emph
  {et~al.},\ }\bibfield  {title} {\bibinfo {title} {Variational quantum
  algorithms},\ }\href@noop {} {\bibfield  {journal} {\bibinfo  {journal}
  {Nature Reviews Physics}\ }\textbf {\bibinfo {volume} {3}},\ \bibinfo {pages}
  {625} (\bibinfo {year} {2021})}\BibitemShut {NoStop}%
\bibitem [{\citenamefont {Peruzzo}\ \emph {et~al.}(2014)\citenamefont
  {Peruzzo}, \citenamefont {McClean}, \citenamefont {Shadbolt}, \citenamefont
  {Yung}, \citenamefont {Zhou}, \citenamefont {Love}, \citenamefont
  {Aspuru-Guzik},\ and\ \citenamefont {O’brien}}]{peruzzo2014variational}%
  \BibitemOpen
  \bibfield  {author} {\bibinfo {author} {\bibfnamefont {A.}~\bibnamefont
  {Peruzzo}}, \bibinfo {author} {\bibfnamefont {J.}~\bibnamefont {McClean}},
  \bibinfo {author} {\bibfnamefont {P.}~\bibnamefont {Shadbolt}}, \bibinfo
  {author} {\bibfnamefont {M.-H.}\ \bibnamefont {Yung}}, \bibinfo {author}
  {\bibfnamefont {X.-Q.}\ \bibnamefont {Zhou}}, \bibinfo {author}
  {\bibfnamefont {P.~J.}\ \bibnamefont {Love}}, \bibinfo {author}
  {\bibfnamefont {A.}~\bibnamefont {Aspuru-Guzik}},\ and\ \bibinfo {author}
  {\bibfnamefont {J.~L.}\ \bibnamefont {O’brien}},\ }\bibfield  {title}
  {\bibinfo {title} {A variational eigenvalue solver on a photonic quantum
  processor},\ }\href@noop {} {\bibfield  {journal} {\bibinfo  {journal}
  {Nature communications}\ }\textbf {\bibinfo {volume} {5}},\ \bibinfo {pages}
  {1} (\bibinfo {year} {2014})}\BibitemShut {NoStop}%
\bibitem [{\citenamefont {Aaronson}\ and\ \citenamefont
  {Rall}(2020)}]{aaronson2020quantum}%
  \BibitemOpen
  \bibfield  {author} {\bibinfo {author} {\bibfnamefont {S.}~\bibnamefont
  {Aaronson}}\ and\ \bibinfo {author} {\bibfnamefont {P.}~\bibnamefont
  {Rall}},\ }\bibfield  {title} {\bibinfo {title} {Quantum approximate
  counting, simplified},\ }\href {https://doi.org/10.1137/1.9781611976014.5}
  {\bibfield  {journal} {\bibinfo  {journal} {Symposium on Simplicity in
  Algorithms}\ ,\ \bibinfo {pages} {24–32}} (\bibinfo {year}
  {2020})}\BibitemShut {NoStop}%
\bibitem [{\citenamefont {Suzuki}\ \emph {et~al.}(2020)\citenamefont {Suzuki},
  \citenamefont {Uno}, \citenamefont {Raymond}, \citenamefont {Tanaka},
  \citenamefont {Onodera},\ and\ \citenamefont
  {Yamamoto}}]{suzuki2020amplitude}%
  \BibitemOpen
  \bibfield  {author} {\bibinfo {author} {\bibfnamefont {Y.}~\bibnamefont
  {Suzuki}}, \bibinfo {author} {\bibfnamefont {S.}~\bibnamefont {Uno}},
  \bibinfo {author} {\bibfnamefont {R.}~\bibnamefont {Raymond}}, \bibinfo
  {author} {\bibfnamefont {T.}~\bibnamefont {Tanaka}}, \bibinfo {author}
  {\bibfnamefont {T.}~\bibnamefont {Onodera}},\ and\ \bibinfo {author}
  {\bibfnamefont {N.}~\bibnamefont {Yamamoto}},\ }\bibfield  {title} {\bibinfo
  {title} {Amplitude estimation without phase estimation},\ }\href@noop {}
  {\bibfield  {journal} {\bibinfo  {journal} {Quantum Information Processing}\
  }\textbf {\bibinfo {volume} {19}},\ \bibinfo {pages} {75} (\bibinfo {year}
  {2020})}\BibitemShut {NoStop}%
\bibitem [{\citenamefont {Yu}\ \emph {et~al.}(2020)\citenamefont {Yu},
  \citenamefont {Lim},\ and\ \citenamefont {Rao}}]{yu2020practical}%
  \BibitemOpen
  \bibfield  {author} {\bibinfo {author} {\bibfnamefont {K.}~\bibnamefont
  {Yu}}, \bibinfo {author} {\bibfnamefont {H.}~\bibnamefont {Lim}},\ and\
  \bibinfo {author} {\bibfnamefont {P.}~\bibnamefont {Rao}},\ }\bibfield
  {title} {\bibinfo {title} {Practical numerical integration on {NISQ}
  devices},\ }in\ \href@noop {} {\emph {\bibinfo {booktitle} {Quantum
  Information Science, Sensing, and Computation XII}}},\ Vol.\ \bibinfo
  {volume} {11391}\ (\bibinfo {organization} {SPIE},\ \bibinfo {year} {2020})\
  p.\ \bibinfo {pages} {1139106}\BibitemShut {NoStop}%
\bibitem [{\citenamefont {Grinko}\ \emph {et~al.}(2021)\citenamefont {Grinko},
  \citenamefont {Gacon}, \citenamefont {Zoufal},\ and\ \citenamefont
  {Woerner}}]{grinko2021iterative}%
  \BibitemOpen
  \bibfield  {author} {\bibinfo {author} {\bibfnamefont {D.}~\bibnamefont
  {Grinko}}, \bibinfo {author} {\bibfnamefont {J.}~\bibnamefont {Gacon}},
  \bibinfo {author} {\bibfnamefont {C.}~\bibnamefont {Zoufal}},\ and\ \bibinfo
  {author} {\bibfnamefont {S.}~\bibnamefont {Woerner}},\ }\bibfield  {title}
  {\bibinfo {title} {Iterative quantum amplitude estimation},\ }\href@noop {}
  {\bibfield  {journal} {\bibinfo  {journal} {npj Quantum Information}\
  }\textbf {\bibinfo {volume} {7}},\ \bibinfo {pages} {1} (\bibinfo {year}
  {2021})}\BibitemShut {NoStop}%
\bibitem [{\citenamefont {Rao}\ \emph {et~al.}(2020)\citenamefont {Rao},
  \citenamefont {Yu}, \citenamefont {Lim}, \citenamefont {Jin},\ and\
  \citenamefont {Choi}}]{rao2020quantum}%
  \BibitemOpen
  \bibfield  {author} {\bibinfo {author} {\bibfnamefont {P.}~\bibnamefont
  {Rao}}, \bibinfo {author} {\bibfnamefont {K.}~\bibnamefont {Yu}}, \bibinfo
  {author} {\bibfnamefont {H.}~\bibnamefont {Lim}}, \bibinfo {author}
  {\bibfnamefont {D.}~\bibnamefont {Jin}},\ and\ \bibinfo {author}
  {\bibfnamefont {D.}~\bibnamefont {Choi}},\ }\bibfield  {title} {\bibinfo
  {title} {Quantum amplitude estimation algorithms on {IBM} quantum devices},\
  }in\ \href@noop {} {\emph {\bibinfo {booktitle} {Quantum Communications and
  Quantum Imaging XVIII}}},\ Vol.\ \bibinfo {volume} {11507}\ (\bibinfo
  {organization} {SPIE},\ \bibinfo {year} {2020})\ pp.\ \bibinfo {pages}
  {49--60}\BibitemShut {NoStop}%
\bibitem [{\citenamefont {Zhang}\ and\ \citenamefont
  {Korepin}(2020)}]{zhang2020depth}%
  \BibitemOpen
  \bibfield  {author} {\bibinfo {author} {\bibfnamefont {K.}~\bibnamefont
  {Zhang}}\ and\ \bibinfo {author} {\bibfnamefont {V.~E.}\ \bibnamefont
  {Korepin}},\ }\bibfield  {title} {\bibinfo {title} {Depth optimization of
  quantum search algorithms beyond {G}rover's algorithm},\ }\href@noop {}
  {\bibfield  {journal} {\bibinfo  {journal} {Physical Review A}\ }\textbf
  {\bibinfo {volume} {101}},\ \bibinfo {pages} {032346} (\bibinfo {year}
  {2020})}\BibitemShut {NoStop}%
\bibitem [{\citenamefont {Zhang}\ \emph {et~al.}(2021)\citenamefont {Zhang},
  \citenamefont {Rao}, \citenamefont {Yu}, \citenamefont {Lim},\ and\
  \citenamefont {Korepin}}]{zhang2021implementation}%
  \BibitemOpen
  \bibfield  {author} {\bibinfo {author} {\bibfnamefont {K.}~\bibnamefont
  {Zhang}}, \bibinfo {author} {\bibfnamefont {P.}~\bibnamefont {Rao}}, \bibinfo
  {author} {\bibfnamefont {K.}~\bibnamefont {Yu}}, \bibinfo {author}
  {\bibfnamefont {H.}~\bibnamefont {Lim}},\ and\ \bibinfo {author}
  {\bibfnamefont {V.}~\bibnamefont {Korepin}},\ }\bibfield  {title} {\bibinfo
  {title} {Implementation of efficient quantum search algorithms on {NISQ}
  computers},\ }\href@noop {} {\bibfield  {journal} {\bibinfo  {journal}
  {Quantum Information Processing}\ }\textbf {\bibinfo {volume} {20}},\
  \bibinfo {pages} {1} (\bibinfo {year} {2021})}\BibitemShut {NoStop}%
\bibitem [{\citenamefont {Zhang}\ \emph {et~al.}(2022)\citenamefont {Zhang},
  \citenamefont {Yu},\ and\ \citenamefont {Korepin}}]{zhang2022quantum}%
  \BibitemOpen
  \bibfield  {author} {\bibinfo {author} {\bibfnamefont {K.}~\bibnamefont
  {Zhang}}, \bibinfo {author} {\bibfnamefont {K.}~\bibnamefont {Yu}},\ and\
  \bibinfo {author} {\bibfnamefont {V.}~\bibnamefont {Korepin}},\ }\bibfield
  {title} {\bibinfo {title} {Quantum search on noisy intermediate-scale quantum
  devices},\ }\href {https://doi.org/10.1209/0295-5075/ac90e6} {\bibfield
  {journal} {\bibinfo  {journal} {Europhysics Letters}\ }\textbf {\bibinfo
  {volume} {140}},\ \bibinfo {pages} {18002} (\bibinfo {year}
  {2022})}\BibitemShut {NoStop}%
\bibitem [{\citenamefont {Lubinski}\ \emph {et~al.}(2021)\citenamefont
  {Lubinski}, \citenamefont {Johri}, \citenamefont {Varosy}, \citenamefont
  {Coleman}, \citenamefont {Zhao}, \citenamefont {Necaise}, \citenamefont
  {Baldwin}, \citenamefont {Mayer},\ and\ \citenamefont
  {Proctor}}]{lubinski2021application}%
  \BibitemOpen
  \bibfield  {author} {\bibinfo {author} {\bibfnamefont {T.}~\bibnamefont
  {Lubinski}}, \bibinfo {author} {\bibfnamefont {S.}~\bibnamefont {Johri}},
  \bibinfo {author} {\bibfnamefont {P.}~\bibnamefont {Varosy}}, \bibinfo
  {author} {\bibfnamefont {J.}~\bibnamefont {Coleman}}, \bibinfo {author}
  {\bibfnamefont {L.}~\bibnamefont {Zhao}}, \bibinfo {author} {\bibfnamefont
  {J.}~\bibnamefont {Necaise}}, \bibinfo {author} {\bibfnamefont {C.~H.}\
  \bibnamefont {Baldwin}}, \bibinfo {author} {\bibfnamefont {K.}~\bibnamefont
  {Mayer}},\ and\ \bibinfo {author} {\bibfnamefont {T.}~\bibnamefont
  {Proctor}},\ }\bibfield  {title} {\bibinfo {title} {Application-oriented
  performance benchmarks for quantum computing},\ }\href@noop {} {\bibfield
  {journal} {\bibinfo  {journal} {arXiv preprint arXiv:2110.03137}\ } (\bibinfo
  {year} {2021})}\BibitemShut {NoStop}%
\end{thebibliography}%
\end{@fileswfalse}

\end{document}